\documentclass{aa} 
\usepackage{braket}
\usepackage{todonotes}
\usepackage{url}
\usepackage{hyperref} 
\hypersetup{
	colorlinks=true,
	citecolor = blue,
	linkcolor=blue,
	filecolor=blue,      
	urlcolor=blue,
}
\usepackage{graphicx}
\usepackage{float}
\usepackage{subcaption}
\usepackage{graphicx}
\usepackage{blindtext}
\usepackage{txfonts}
\let\oldtextbf=\textbf
\renewcommand\textbf[1]{{\boldmath\oldtextbf{#1}}}

\begin{document}

	\title{Galaxies lacking dark matter in the Illustris simulation}
	
	\author{M. Haslbauer\inst{1},
		J. Dabringhausen\inst{2},
		P. Kroupa\inst{1,2},
		B. Javanmardi\inst{3,4},
		\and 
		I. Banik\inst{1}
	}
	\authorrunning{M. Haslbauer et al.}
	\institute{Helmholtz Institut f\"ur Strahlen- und Kernphysik (HISKP), University of Bonn, Nussallee 14-16, D-53121 Bonn, 
		Germany\\
		\email{mhaslbauer@astro.uni-bonn.de}
		\and
		Charles University, Faculty of Mathematics and Physics, Astronomical Institute, V  Hole\v{s}ovi\v{c}k\'ach 2, CZ-180 00 Praha 8, Czech Republic
		\and
		School of Astronomy, Institute for Research in Fundamental Sciences (IPM), P. O. Box 19395-5531, Tehran, Iran  \and
		LESIA, Paris Observatory, PSL University, CNRS, Sorbonne University, Univ. Paris Diderot, Paris Cité Sorbonne, 5 place Jules Janssen, 92195 Meudon, France
	}
	
	\date{Received 04 July, 2018; accepted 22 April, 2019}
	
	
	\abstract
	{Any viable cosmological model in which galaxies interact predicts the existence of primordial and tidal dwarf galaxies (TDGs). In particular, in the standard model of cosmology ($\Lambda$CDM), according to the dual dwarf galaxy theorem, there must exist both primordial dark matter-dominated and dark matter-free TDGs with different radii.}
	{We study the frequency, evolution, and properties of TDGs in a $\Lambda$CDM cosmology.}
	{We use the hydrodynamical cosmological Illustris-1 simulation to identify tidal dwarf galaxy candidates (TDGCs) and study their present-day physical properties. The positions of galaxies in the radius--mass plane, depending on their nonbaryonic content, are compared with observational data and other simulations. We also present movies on the formation of a few  galaxies lacking dark matter, confirming their tidal dwarf nature. Tidal dwarf galaxy candidates can however also be formed via other mechanisms, such as from ram-pressure-stripped material or, speculatively, from cold-accreted gas.}
	{We find $97$ TDGCs with $M_{\mathrm{stellar}} >5 \times 10^{7} \, \rm{M_{\odot}}$ at redshift $z = 0$, corresponding to a co-moving number density of $2.3 \times 10^{-4} \, h^{3} \, \rm{cMpc^{-3}}$. The most massive TDGC has $M_{\mathrm{total}} = 3.1 \times 10^{9} \, \rm{M_{\odot}}$, comparable to that of the Large Magellanic Cloud. Tidal dwarf galaxy candidates are phase-space-correlated, reach high metallicities, and are typically younger than dark matter-rich dwarf galaxies.}
	{We report for the first time the verification of the dual dwarf theorem in a self-consistent $\Lambda$CDM cosmological simulation. Simulated TDGCs and dark matter-dominated galaxies populate different regions in the radius--mass diagram in disagreement with observations of early-type galaxies. The dark matter-poor galaxies formed in Illustris-1 have comparable radii to observed dwarf galaxies and to TDGs formed in other galaxy-encounter simulations. In Illustris-1, only $0.17$~percent of all selected galaxies with $M_{\mathrm{stellar}} = 5 \times 10^{7}-10^{9} \, \rm{M_{\odot}}$ are TDGCs or dark matter-poor dwarf galaxies. The occurrence of NGC 1052-DF2-type objects is discussed.}
	

	\maketitle 
	%
	\section{Introduction}
	The current standard model of cosmology is based on Einstein's general relativity and requires the existence of cold dark matter (CDM) and a cosmological constant ($\Lambda$) in Einstein's gravitational field equations. This $\Lambda$CDM model is a much-used description of the large-scale structure of the Universe, but fundamental problems, not only on galactic and galaxy-group scales, remain unsolved \citep[e.g.,][]{Kroupa_2010, Famaey_2012, Pawlowski_2014,Kroupa_2012,Kroupa_2015,Mueller_2018}.
	
	In the $\Lambda$CDM framework, the dual dwarf theorem has to be valid, according to which primordial and tidal dwarf galaxies (TDGs) must exist \citep{Kroupa_2010, Kroupa_2012}. These two types of dwarf galaxies are characterized by different formation scenarios and differ mainly by their amounts of nonbaryonic dark matter.
	
	Primordial galaxies are formed by the collapse of cold dark matter particles into halos. These structures become gravitationally bound and their deep gravitational potentials act on the baryonic matter, which streams and condenses into the halos. Thus, each primordial galaxy has to be dark matter-dominated \citep{Bournaud_2006,Bournaud_2008b,Ploeckinger_2018}.
	
	In the hierarchical $\Lambda$CDM cosmology, the formation of dwarf galaxies can also be triggered by interactions of gas-rich galaxies. Galaxy encounters create tidal forces, which distort the galactic disk and cause the expulsion of gas and stars. The ejected stars and gas form tidal tails and arms, which surround and orbit around the host galaxy. Overdensities within tidal arms collapse and grow continually in mass \citep{Barnes_1992,Bournaud_2006, Wetzstein_2007, Bournaud_2008a,Bournaud_2008b,Fouquet_2012,Ploeckinger_2014,Ploeckinger_2015}. These substructures reach stellar masses between $  10^{6} \, \rm{M_{\odot}}$ and $  10^{9} \, \rm{M_{\odot}}$ and are called TDGs. The high velocity dispersion of dark matter particles and the relatively shallow gravitational potential compared to their host galaxy prevent TDGs from capturing a significant amount of dark matter \citep{Barnes_1992, Wetzstein_2007,Bournaud_2008a,Bournaud_2008b,Fouquet_2012,Yang_2014,Ploeckinger_2018}. Consequently, TDGs are not dark matter-dominated \citep{Kroupa_2012}. 
	The small amount of dark matter also has implications for the survival time of TDGs. Since the dynamical friction force depends linearly on the density of the surrounding matter field and on the square of the mass of the dwarf galaxy, dark matter-dominated dwarf galaxies have a faster orbital decay with respect to their host galaxy \citep{Angus_2011}. Therefore, in spite of the vicinity of TDGs to a larger host galaxy, it has been shown that especially low-mass TDGs have survival times comparable with the Hubble time  \citep{Kroupa_1997, Recchi_2007,Casas_2012,Ploeckinger_2014,Ploeckinger_2015}. Observational constraints also show that TDGs survive for many gigayears \citep{Duc_2014}. Ram-pressure stripping, interactions with their host galaxy, star formation, and evolution can deplete the gas reservoir of TDGs over cosmic time. Therefore, long-lived and gas-poor TDGs can potentially resemble dwarf elliptical galaxies (dEs) \citep{Dabringhausen_2013}, and models suggest that the Large and Small Magellanic Clouds can also be TDGs \citep{Fouquet_2012}. Estimates based on the merger tree in the CDM cosmological model have shown that TDGs can probably account for the observed number density of dEs \citep{Okazaki_2000}. Because of the different formation scenarios, TDGs should typically be phase-space-correlated while primordial dwarfs should be spheroidally distributed in phase-space around their host \citep{Kroupa_2005,Pawlowski_2011,Kroupa_2012,Pawlowski_2018}. In the local Universe, phase-space correlations (a clustering of the direction of the orbital angular momentum vectors of dwarf galaxies) are observed around the majority of the nearest ($\lesssim 4 \, \rm{Mpc}$) major galaxies, namely M31 \citep{Metz_2007b,Ibata_2013}, the Milky Way \citep{Pawlowksi_2013,Pawlowski_2018}, and Centaurus A \citep{Mueller_2018}. Observing the phase-space distribution of distant satellite galaxies is currently very difficult, but a significant excess of observed co-rotating satellite pairs over that expected in a $\Lambda$CDM universe has been found \citep{Ibata_2014}. Disks of satellites thus appear to be the rule rather than the exception. Phase-space-correlated satellite systems may however be destroyed if the host galaxy suffers another encounter. The observed high incidence of disk-of-satellite systems thus suggests that such encounters, let alone mergers, cannot be frequent. 
	
	Several observations of interacting galaxies have confirmed the existence of gaseous tidal tails, arms, and TDGs in the Universe \citep[e.g.,][]{Mirabel_1992, Duc_2000, Mendes_2001,Weilbacher_2002,Martinez_2010,Kaviraj_2012,Lee_2012,Duc_2014}. Since primordial dwarf galaxies form in the dark matter halo while tidal dwarf galaxies form naked under their own self-gravity, the latter are expected to have systematically smaller radii if dark matter exists \citep{Kroupa_2012}. \cite{Dabringhausen_2013} studied the position of early-type galaxies and ultra compact dwarf galaxies (UCDs) in the radius--mass plane. These latter authors conclude that no significant difference in the radius--mass plane between observed dEs and observed TDGs can be found, which is in conflict with the current standard model of cosmology. However, the data they used are from different observations \citep{Bender_1992,Bender_1993,Ferrarese_2006,Misgeld_2008, Misgeld_2009, Misgeld_2011,Miralles_2012}. Moreover, UCDs and globular clusters (GCs) are clearly separated from dEs and TDGs in the radius--mass plane \citep{Gilmore_2007, Dabringhausen_2013}. Until now, no self-consistent study exists of formation in a cosmological context  quantifying the expected differences between TDGs and primordial dwarf galaxies.
	
	The recently observed ultra-diffuse galaxy NGC 1052-DF2 with a dark matter mass 400 times smaller than theoretically expected based on an internal velocity dispersion of $\sigma_{\mathrm{intr}}=3.2_{+5.5}^{-3.2} \, \rm{km \, s^{-1}}$, seems to support the existence of dark matter-free galaxies in our Universe \citep{vDokkum_2018b}. \citet{vDokkuma_2018} derived a revised internal velocity dispersion of $\sigma_{\mathrm{intr}}=7.8_{+5.2}^{-2.2} \, \rm{km \, s^{-1}}$ using ten GCs surrounding this galaxy. \citet{Danieli_2019} measured a stellar velocity dispersion of $\sigma_{\mathrm{stars}}=8.5_{+2.3}^{-3.1} \, \rm{km \, s^{-1}}$ with the Keck Cosmic Web Imager (KCWI). The high relative velocity to the nearby massive elliptical galaxy NGC 1052 underpins the theory that this observed dark matter-lacking galaxy is indeed a TDG. However, \cite{Martin_2018} revised the internal velocity of NGC 1052-DF2 to a $90$~percent upper limit of $17.3 \, \rm{km \, s^{-1}}$ corresponding to a mass-to-light ratio of $M/L_{\mathrm{V}}<8.1 \, \rm{\Upsilon_{\odot}}$, consistent with many Local Group dwarf galaxies. \citet{Emsellem_2018} obtain $M/L_{\mathrm{V}}$ in the range $3.5-3.9 (\pm1.8) \, \rm{\Upsilon_{\odot}}$  using the Jeans model if located at $D = 20 \, \rm{Mpc}$. This result would be close to the $2 \sigma$ upper limit of the study from \citet{Martin_2018}.       The lack of dark matter and the unusual high luminosity of ten globular cluster-like objects surrounding this galaxy only holds if NGC 1052-DF2 is located at a distance of around $20 \, \rm{Mpc}$ \citep{vDokkum_2018b}. \citet{Danieli_2019} confirmed that DF2 is dark matter deficient and concluded that it is an outlier to dwarf galaxies of the Local Group. In contrast to that, \cite{Trujillo_2019} derived a revised distance to NGC 1052-DF2 of $D = 13.0 \pm 0.4 \, \rm{Mpc}$ based on five redshift-independent measurements including the tip of the red giant branch and the surface brightness fluctuation method. Thus, NGC 1052-DF2 would be a dwarf galaxy with an ordinary dark matter content $M_{\mathrm{halo}}/M_{\mathrm{stellar}} > 20$ and a normal globular cluster population. Meanwhile, \citet{vDokkum_2019} reported that the dwarf galaxy NGC 1052-DF4 also lacks dark matter and is found at a distance of $D = 20 \, \rm{Mpc}$.
	
	In this paper we investigate dark matter-free galaxies in the Illustris simulation, which is currently one of  the most advanced cosmological computations. We analyze their physical properties and qualitatively estimate the probability of finding NGC 1052-DF2-like galaxies in a $\Lambda$CDM Universe at
	redshift $z=0$ assuming that this observed ultra-diffuse galaxy is indeed free of dark matter. High-resolution runs of modern cosmological hydrodynamical simulations such as EAGLE \citep{McAlpine_2016} and Illustris \citep{Vogelsberger_2014b} allow the analysis of TDGs in a self-consistent cosmological $\Lambda$CDM framework. The formation of TDGs in the EAGLE simulation has been studied by \cite{Ploeckinger_2018}. The formation of TDGs in individual galaxy--galaxy encounters in the $\Lambda$CDM context is well established \citep{Wetzstein_2007,Bournaud_2008a,Bournaud_2008b}.
	
	The layout of the paper is as follows. In Section~\ref{sec:Methods}, we introduce the Illustris simulation and the selection criteria for dark matter-free galaxies. Section~\ref{sec:Results} presents the results, in particular we study different physical properties of dark matter-free galaxies and we plot the radius--mass relation. The results are compared with observational data. The evolution of dark matter-free galaxies over cosmic time is shown. The results are discussed in Section~\ref{sec:Discussion}. We finally summarize and conclude with Section~\ref{sec:Conclusion}. Throughout this paper co-moving distances are marked with the prefix ``c'' (i.e., cpc, ckpc, cMpc). 
	We note that at redshift $z=0$, the scale factor $a(t)$ becomes unity and by definition proper and co-moving distances become the same.
	
	\section{Methods} \label{sec:Methods}
	
	We use the cosmological hydrodynamical Illustris simulation to study the evolution and physical properties of dark matter-free galaxies. This section introduces the Illustris project by describing the cosmological and numerical parameters and the implemented physics of galaxy-formation models. The selection criteria for primordial and tidal dwarf galaxy candidates (TDGCs) are stated. Movies on the formation and evolution of TDGCs are attached in the supplementary material. 
	
	\subsection{Illustris simulation}
	The Illustris simulation project\footnote{\url{http://www.illustris-project.org}} is a set of cosmological hydrodynamical and dark matter-only simulations at different resolutions performed with the moving-mesh code $\textsf{AREPO}$ \citep{Springel_2010}. The simulations assume a flat $\Lambda$CDM cosmology based on the Wilkinson Microwave Anisotropy Probe (WMAP)-9 measurements with the values of the cosmological parameters at the present time being $\Omega_{\mathrm{m,0}}=0.2726$, $\Omega_{\mathrm{\Lambda,0}}=0.7274$, $\Omega_{\mathrm{b,0}}=0.0456$, $\sigma_8=0.809$, $n_{\mathrm{s}}=0.963$, and $H_0=100 \, h^{-1} \, \rm{km \, s^{-1} \, Mpc^{-1}}$ with $h = 0.704$ \citep{Hinshaw_2013}. The main simulations cover a co-moving volume of $(75 \, h^{-1} \, \rm{cMpc})^{3}$ and start at redshift $z=127$. The evolution of dark matter particles, gas cells, passive gas tracers, stars and stellar wind particles, and supermassive black holes (SMBHs) are followed up to redshift $z=0$ \citep{Nelson_2015}.
	
	Dark matter halos are identified with the standard friends-of-friends (FOF) algorithm \citep{Davis_1985} with a linking length of $0.2$ times the mean particle separation. The minimum particle number of each halo is 32. Subhalos within halos are identified with the Subfind algorithm \citep{Springel_2001, Dolag_2009} and have a unique identification number (ID) within each snapshot. The particle with the minimum gravitational potential energy defines the spatial position of the subhalo (halo) within the periodic box, and the total mass of a subhalo (halo) is defined as the sum of the individual masses of particles (cells) connected to the subhalo (halo). The physical properties of FOF and Subfind objects for each snapshot are listed in the group catalogs, which can be downloaded from the Illustris webpage \citep{Vogelsberger_2014b,Genel_2014}.
	
	Throughout this paper, we use the highest-resolution run (Illustris-1) with a dark matter mass resolution of $6.26 \times 10^{6} \, \rm{M_{\odot}}$ (the mass of one particle) and an initial baryonic mass resolution of $1.26 \times 10^{6} \, \rm{M_{\odot}}$ (the mass of one particle). The gravitational softening lengths of dark matter and baryonic particles are $1420 \, \rm{cpc}$ and $710 \, \rm{cpc}$ in co-moving length scale, respectively \citep{Vogelsberger_2014a, Nelson_2015}. 
	
	\cite{Torrey_2015} provide images for subhalos with $M_{\mathrm{stellar}} > 10^{10} \, \rm{M_{\odot}}$ at redshift $z = 0$, which are produced with the radiative transfer code \texttt{SUNRISE} \citep{Jonsson_2006,Jonsson_2010}. These galaxy PNG images and fits files can be downloaded with the web-based search tool Illustris Galaxy Observatory from the Illustris webpage.\footnote{\url{http://www.illustris-project.org/galaxy_obs/}}
	
	In addition, the Illustris team supplies an online tool called ``The Illustris Explorer'' which visualizes a slice with a depth of $15 \, h^{-1} \rm{Mpc}$ in projection of the Illustris-1 simulation box at redshift $z = 0$. This deep zoom map interface allows one to visualize, for example, the gas temperatures and densities, the dark matter densities, and the stellar luminosities in Johnson/SDSS filters.\footnote{\url{http://www.illustris-project.org/explorer/}}
	
	\subsection{Galaxy-formation models}
	A detailed galaxy formation model for simulating astrophysical processes is implemented in the Illustris simulation. The model includes a stochastic star formation description in dense gas, stellar evolution with mass loss and chemical enrichment, cooling and heating mechanisms of the ISM, AGN feedback, and the growth and evolution of SMBHs. The implemented physical models and a comparison with observations can be found in detail in \citet{Vogelsberger_2013} and \citet{Torrey_2014}. We point out that in the Illustris simulation the mass loading and wind velocity are scaled with the local dark matter velocity dispersion \citep{Vogelsberger_2013}. This is in contradiction with the standard view of cold dark matter theory, which assumes weak interactions between nonbaryonic and baryonic matter. With this recipe, more massive halos produce stronger baryonic feedback.
	
	\subsection{Selection criteria for dark matter-containing and dark matter-free stellar objects}
	\label{sec:Methods_DMF_subhalos}
	We select two different kinds of stellar objects based on their baryonic and dark matter masses. We identify subhalos with a stellar mass $M_{\mathrm{stellar}}> 0$ and a nonzero dark matter mass and refer to them as dark matter-containing (DMC) stellar objects. Dark matter-free (DMF) stellar objects are defined as subhalos with a stellar mass $M_{\mathrm{stellar}}>0$ and a dark matter mass of $M_{\mathrm{dm}}=0 $. These selection criteria give us $304 \, 302$ DMC and $3484$ DMF stellar objects at redshift $z=0$.
	
	\subsection{Selection criteria for DMC dwarf galaxies and tidal dwarf galaxy candidates} \label{sec:Methods_distancecriteria}
	The  selection criteria stated above for DMF and DMC stellar objects are independent of the environment. In fact, DMF and DMC stellar objects can be substructures which are embedded in the galactic disk of their host galaxies rather than real physical galaxies \citep{Ploeckinger_2018,Graus_2018}. 
	Therefore we divide stellar objects based on the separation, $s$, to their next host galaxy.\footnote{The separation, s, between two subhalos is defined as the distance between the particles with the minimum gravitational potential energy in each subhalo.} A host galaxy is defined as the closest subhalo with $M_{\mathrm{stellar}}>10^{9} \, \rm{M_{\odot}}$ and a stellar mass at least  ten times larger than the considered stellar object. A stellar object with a separation to its host halo smaller than or equal to ten times the stellar half-mass radius of the host ($\leq 10 \times R_{\mathrm{0.5 \, stellar}}^{\mathrm{host}}$) is defined as a substructure within a galaxy such as a massive GC or a numerical artifact. Dark matter-free or dark matter-containing stellar objects beyond the distance criterion of $10 \times R_{\mathrm{0.5 \, stellar}}^{\mathrm{host}}$ and within $100 \times R_{\mathrm{0.5 \, stellar}}^{\mathrm{host}}$ are identified as TDGCs or dark matter-containing dwarf galaxies (DMC DGs), respectively. We label these dark matter-free objects explicitly as TDG ``candidates'' in order to emphasis that apart from galactic interactions (tidal forces) such objects can also be formed in other scenarios, such as for example ram-pressure disruption or perhaps cold accretion.
	
	The minimum separation criterion is motivated by the fraction of the separation between the Milky Way (MW) galaxy and the Large Magellanic Cloud (LMC) \citep[$s_{\mathrm{MW-LMC}} \approx 50 \, \rm{kpc}$,][]{Pietrzynski_2013} to the 3D deprojected half-light radius of the MW \citep[$R_{\mathrm{0.5 \, light}}^{\mathrm{MW}} \approx 4.8 \, \rm{kpc}$,][]{Koda_2015,Wolf_2010}. 
	A maximum separation limit is used because the catalog of observed early-type galaxies from \cite{Dabringhausen_2016} only includes dwarf galaxies which are found in dense galactic regions. Ignoring this criterion the most distant TDGC has a separation of $987 \, \rm{kpc}$ to its host and was probably expelled by a galaxy--galaxy interaction.
	
	Furthermore, we restrict our main analysis to TDGCs with $M_{\mathrm{stellar}}>5 \times 10^{7} \, \rm{M_{\odot}}$ (hereafter TDGC sample A) and DMC DGs within the $5 \times 10^{7} - 10^{9} \, \rm{M_{\odot}}$ stellar mass regime. The minimum stellar mass ensures that these subhalos are resolved with at least $50$ stellar particles. Using these selection criteria we find $97$ TDGCs corresponding to a co-moving number density of $2.3 \times 10^{-4} \, h^{3} \, \rm{cMpc^{-3}}$ at redshift $z=0$.
	
\begin{figure}
		\centering
		\includegraphics[width=\columnwidth,trim={0cm 0.0cm 0.0cm 0.0cm},clip]{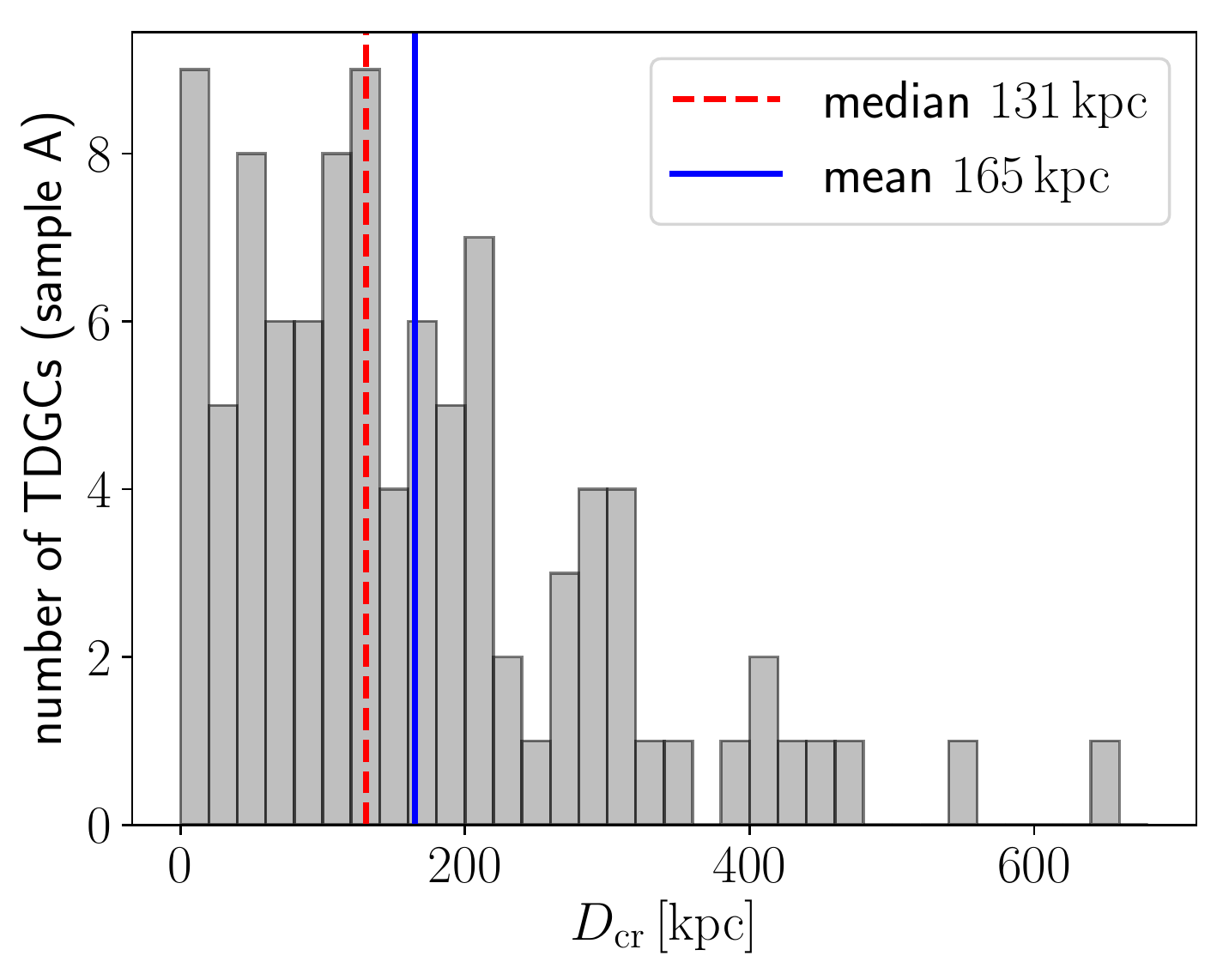}
		\caption{Distribution of the 3D distance-criterion parameter, $D_{\mathrm{cr}}$ (Eq. \ref{equation_distance_criterion_parameter}), for TDGCs of sample A at redshift $z = 0$. The dashed red and the solid blue lines highlight the median and the mean of the distribution, respectively.}
		\label{histogram_distance_subhalo}
	\end{figure}

	In order to study the separation of TDGCs to their host galaxies we introduce the 3D distance-criterion parameter $D_{\mathrm{cr}}$,
	\begin{equation}
	\begin{aligned}
	D_{\mathrm{cr}} \equiv s_{\mathrm{TDGC-host}}-10 \times R_{\mathrm{0.5 \, stellar}}^{\mathrm{host}} \, ,
	\end{aligned}
	\label{equation_distance_criterion_parameter}
	\end{equation}
	where $s_{\mathrm{TDGC-host}}$ is the 3D separation between the TDGC and its host galaxy, and $R_{\mathrm{0.5 \, stellar}}^{\mathrm{host}}$ is the stellar half-mass radius of the host galaxy as already defined in the text above. The distribution of the $D_{\mathrm{cr}}$ parameter for TDGCs of sample A is shown in Fig.~\ref{histogram_distance_subhalo}. This plot and Table~\ref{tab:TDGCs} point out that most of the TDGCs are located in the vicinity of a larger galaxy, which is theoretically expected from the formation theory of TDGs. In contrast to that, a significant number of DMC objects are also beyond the chosen maximum separation limit of $100 \times R_{\mathrm{0.5 \, stellar}}^{\mathrm{host}}$ (i.e., $D_{\mathrm{cr,max}}=90 \times R_{\mathrm{0.5 \, stellar}}^{\mathrm{host}}$) as summarized in Table~\ref{tab:DMCDGs}. DMC DGs with less dark matter than baryonic mass are found mostly close to their host galaxies suggesting that TDGCs can in principle capture dark matter particles. About $0.35$~percent of all galaxies with $M_{\mathrm{stellar}} = 5 \times 10^{7}-10^{9} \, \rm{M_{\odot}}$ and within the applied distance criteria are TDGCs or DM-poor DGs. This reduces to $0.17$~ percent when ignoring the maximum separation limit on the samples.
	
	\begin{table*}
		\centering
		\caption{Number of TDGCs for different selection criteria identified at redshift $z=0$. }
		\label{tab:TDGCs}
		\begin{tabular}{lllllllll} \hline 
			TDGCs & $M_{\mathrm{gas}} \, [\rm{M_{\odot}}]$ & & $M_{\mathrm{stellar}} \,  \, [\rm{M_{\odot}}]$ & $> 5 \times R_{\mathrm{0.5 \, stellar}}^{\mathrm{host}}$ & $ > 10 \times R_{\mathrm{0.5 \, stellar}}^{\mathrm{host}}$ & $(10-50) \times R_{\mathrm{0.5 \, stellar}}^{\mathrm{host}}$ & $(10-100) \times R_{\mathrm{0.5 \, stellar}}^{\mathrm{host}}$ \\
			\hline \hline
			sample A & $\geq 0$ & $\land$    & $> 5 \times 10^{7}$  & $119$ & $98$ & $76$ & $97$ \\ 
			sample B & $> 5 \times 10^{7}$ & $\land$ & $> 0$   & $987$ & $575$ & $317$ & $416$ \\ 
			sample C & $> 5 \times 10^{7}$ & $\land$  & $> 5 \times 10^{7}$ & $15$ & $10$ & $9$ & $10$ \\
			\hline 
		\end{tabular}
		\tablefoot{TDGCs of our main samples have to fulfill the $(10-100) \times R_{\mathrm{0.5 \, stellar}}^{\mathrm{host}}-$ distance criterion (last column, eq. \ref{equation_distance_criterion_parameter}) and have $M_{\mathrm{dm}} = 0$.}
	\end{table*}
	
	\begin{table*}
		\centering
		\caption{Number of DMC DGs for different selection criteria identified at redshift $z=0$.}
		\label{tab:DMCDGs}
		\begin{tabular}{lllllllll} \hline 
			DMC DGs & $M_{\mathrm{dm}}/M_{\mathrm{baryonic}}$ &  $> 5 \times R_{\mathrm{0.5 \, stellar}}^{\mathrm{host}}$ & $ > 10 \times R_{\mathrm{0.5 \, stellar}}^{\mathrm{host}}$ & $(10-50) \times R_{\mathrm{0.5 \, stellar}}^{\mathrm{host}}$ & $(10-100) \times R_{\mathrm{0.5 \, stellar}}^{\mathrm{host}}$ \\ \hline \hline
			all        & $> 0$ & $67 \, 585$   & $65 \, 815$  & $17 \, 682$ & $32 \, 055$ \\ 
			DM-rich & $\geq 1$ & $67 \, 560$   & $65 \, 799$  & $17 \, 668$ & $32 \, 040$\\ 
			DM-poor & $< 1$ & $25$ & $16$     & $14$    & $15$\\ 
			\hline 
		\end{tabular}
		\tablefoot{DMC DGs of our main samples have to fulfill the $(10-100) \times R_{\mathrm{0.5 \, stellar}}^{\mathrm{host}}-$ distance criterion (last column).}
	\end{table*}
	
	The criteria applied here for TDGCs are independent of the gas half-mass radius of the host galaxy, $R_{\mathrm{0.5 \, gas}}^{\mathrm{host}}$, in contrast to \cite{Ploeckinger_2018}. In particular, \cite{Ploeckinger_2018} consider TDG candidates with $M_{\mathrm{gas}} > 10^{7} \, \rm{M_{\odot}}$ and $M_{\mathrm{stellar}} > 2.26 \times 10^{5} \, \rm{M_{\odot}}$ which are located beyond $2 \times R_{\mathrm{0.5 \, gas}}^{\mathrm{host}}$ and within a proper radius of $\rm{200 \, \rm{kpc}}$ or $<20 \times R_{\mathrm{0.5 \, gas}}^{\mathrm{host}}$ (i.e., $s_{\mathrm{max}} = \mathrm{min}[200 \, \rm{kpc}$,$ \, 20 \times R_{\mathrm{0.5 \, gas}}^{\mathrm{host}}$]). The host galaxy is defined as a galaxy with $M_{\mathrm{gas}}>10^{9} \, \rm{M_{\odot}}$ or a galaxy that has a gas content at least  ten times higher than the considered TDGC.
	We therefore define another sample, TDGCs sample B, which includes TDGCs with $M_{\mathrm{gas}} > 5 \times 10^{7} \, \rm{M_{\odot}}$ and at least one stellar particle (see Table \ref{tab:TDGCs}). The different described samples and where they are described are summarized in Table~\ref{tab:samples}.
	
	\begin{table}
		\centering
		\caption{Listed are the different defined samples and where we discuss them.}
		\label{tab:samples}
		\begin{tabular}{ll} \hline
			sample & relevant sections/Tables \\ \hline \hline 
			DMC \& DMF stellar objects & Section~\ref{sec:Methods_DMF_subhalos} \\ 
			DMC \& DMF substructures & Section~\ref{sec:Methods_distancecriteria} \\
			TDGCs (sample A) & see Table~\ref{tab:TDGCs}   \\ 
			TDGCs (sample B) & see Table~\ref{tab:TDGCs}   \\ 
			DMC DGs & see Table~\ref{tab:DMCDGs}  \\ 
			DM-rich DGs & see Table~\ref{tab:DMCDGs}   \\ 
			DM-poor DGs & see Table~\ref{tab:DMCDGs}   \\ 
			\hline 
		\end{tabular}
	\end{table}

	\subsection{Formation scenarios of TDGCs}
	\label{sec:formation_TDGCs}
	
	In order to confirm the tidal nature of TDGCs, we present a series of snapshots of the formation and evolution of some DM-poor DGs and TDGCs by plotting 2D histograms of the gas distribution at different time steps. The corresponding movies can be found as supplementary material. TDGCs and DM-poor objects are identified at redshift $z=0$ and are then backtraced by following their individual stellar particle IDs found in the Subfind subhalos at different time steps (excepted are the subhalos of their potentially host galaxies). The backtracing algorithm developed here stops when stellar particles can no longer be detected in a potential progenitor of the considered object. 
	
\begin{figure*}
		\centering
		\includegraphics[width=91mm,trim={0.5cm 1.0cm 3.0cm 2.0cm},clip]{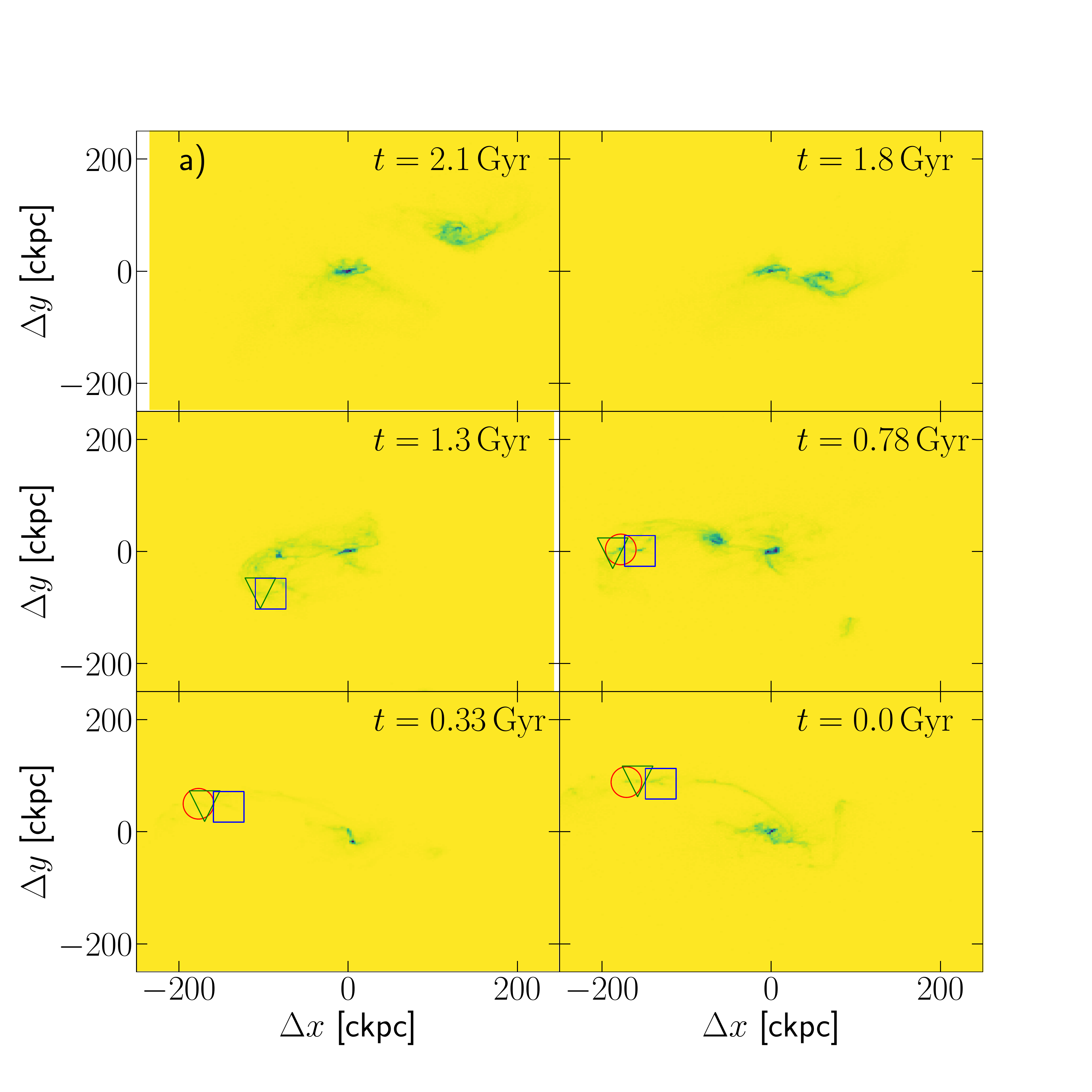} 
		\includegraphics[width=91mm,trim={0.5cm 1.0cm 3.0cm 2.0cm},clip]{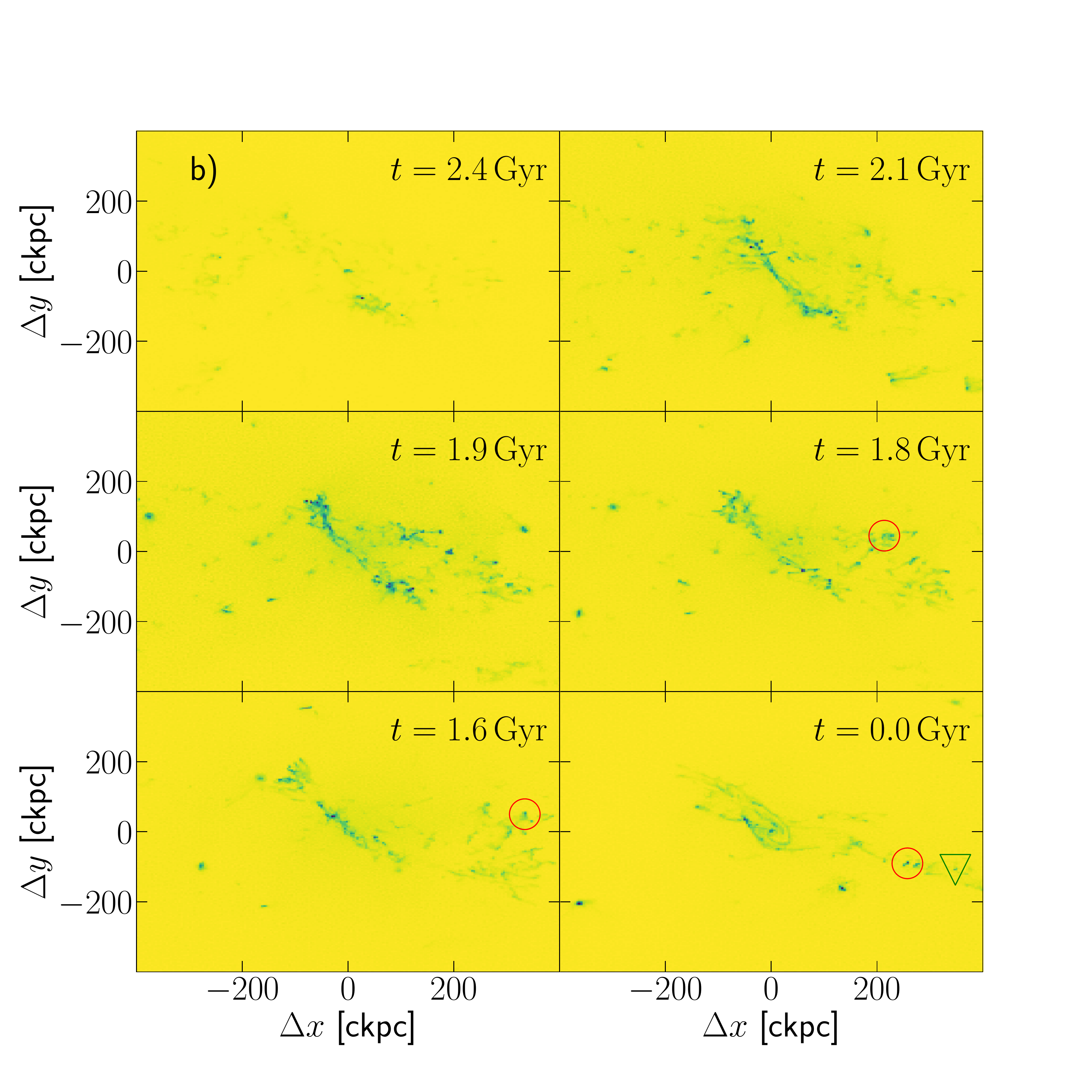} 
		\includegraphics[width=91mm,trim={0.5cm 1.0cm 3.0cm 2.0cm},clip]{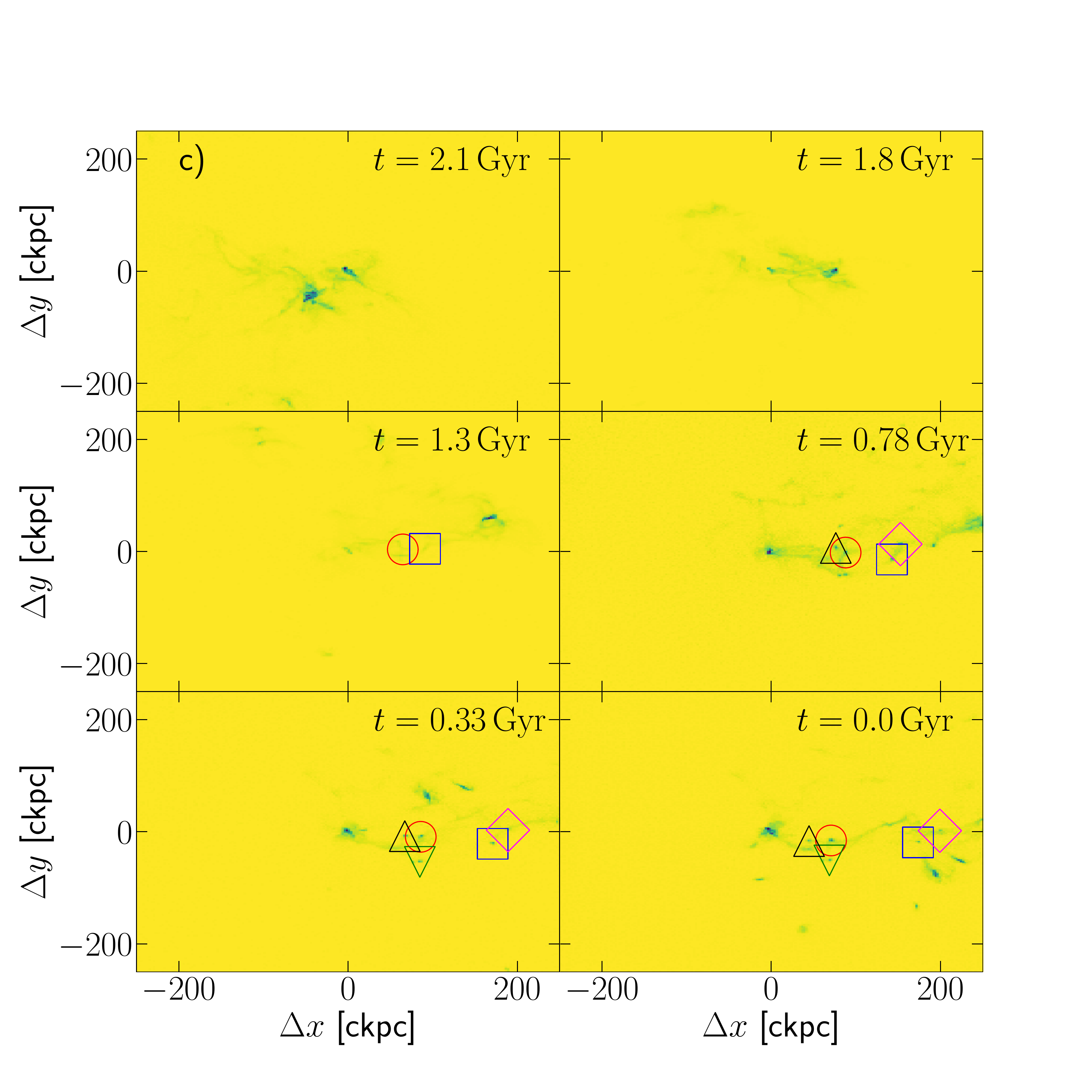} 
		\includegraphics[width=91mm,trim={0.5cm 1.0cm 3.0cm 2.0cm},clip]{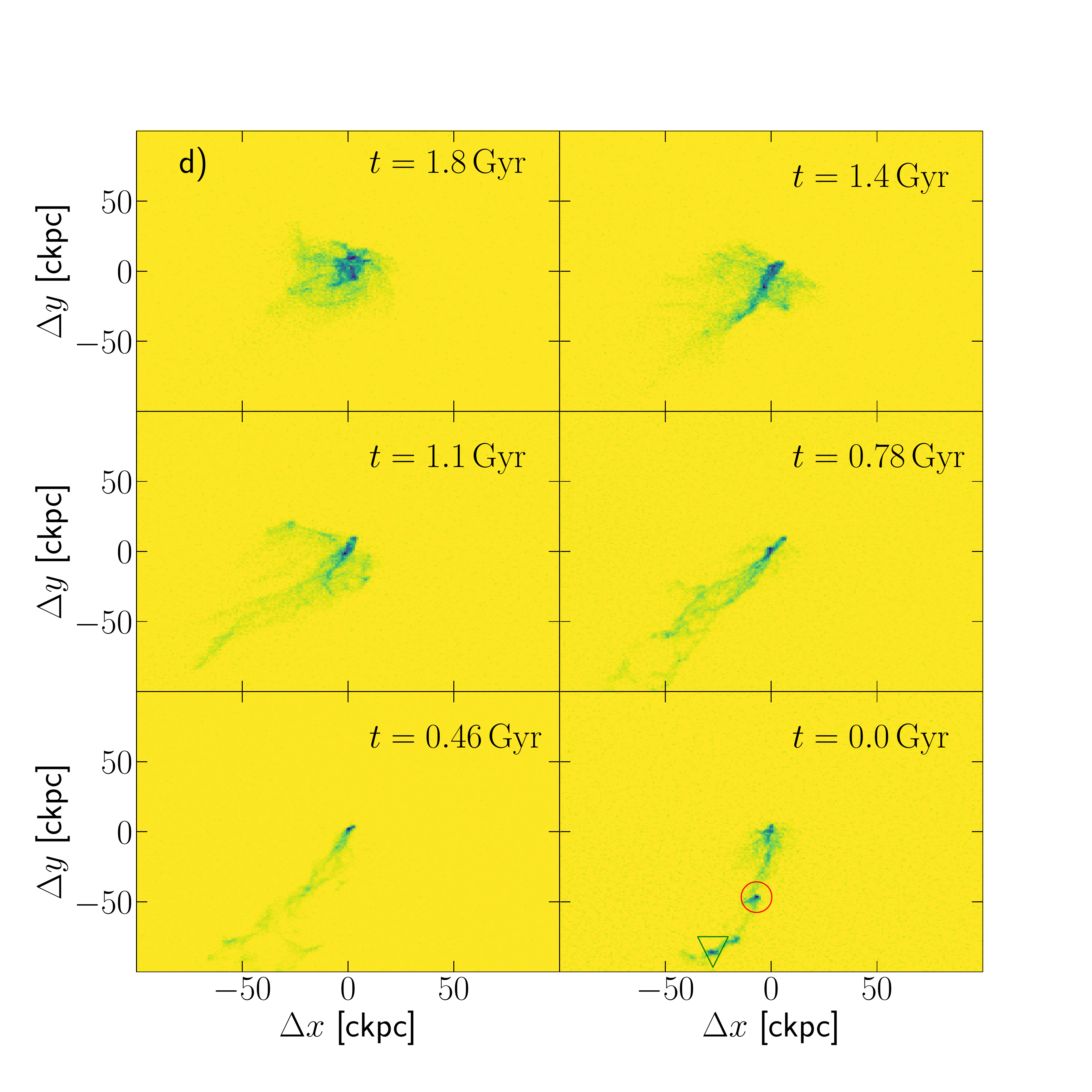} 
		\begingroup 
		\phantomsubcaption\label{fig:time_evolution_TDGCs_a}
		\phantomsubcaption\label{fig:time_evolution_TDGCs_b}
		\phantomsubcaption\label{fig:time_evolution_TDGCs_c}
		\phantomsubcaption\label{fig:time_evolution_TDGCs_d}
		\endgroup
		\vspace{-0.2cm}
		\caption{Time evolution of the gas distribution weighted by the logarithm of the gas cell mass and with position relative to the subhalo center of host galaxies. The TDGC and DM-poor objects identified at $z=0$ are being backtraced by their individual stellar particle IDs and are highlighted in the panels until stellar particles can no longer be found in their subhalo. The lookback time of the corresponding snapshot is given in the upper-right corner of the panels.
			a) Host galaxy ID $404871$ with the TDGC of sample B ID $404882$ (red circle), DM-poor substructure ID $404873$ (blue square), and the subhalo ID $404879$ (green down-pointing triangle) being identified at $z = 0$  (see also Fig.~\ref{fig:appendix_Images_TDGCs_DMCDGs} in Appendix~\ref{sec:appendix_Images_TDGCs_DMCDGs} and the movie ``ID404871.mp4'' in the supplementary information). The subhalo ID $404879$ has $M_{\mathrm{stellar}} = 2.2 \times 10^{7} \, \rm{M_{\odot}}$ and thus does not fulfill our criteria for a DM-poor DG (see Table \ref{tab:DMCDGs}). A close encounter of two galaxies happens at a lookback time of about $1.6 \, \rm{Gyr}$ creating a large extended tidal arm in which these dark matter-lacking subhalos are identified.
			b) Host galaxy ID $73663$ with the TDGCs ID $74010$ (DF2-like; red circle) and ID $74810$ (green down-pointing triangle) being identified at $z = 0$ (both of sample A; see also Section~\ref{sec:Results_NGC1052-DF2} and the movies ``ID73663.mp4'' and ``ID73663$\_$zoom.mp4''). A galaxy merger occurs at a lookback time of around $1.9 \, \rm{Gyr}$.
			c) Host galaxy ID $150872$ with the TDGCs of sample B ID $151014$ (red circle), ID $151271$ (blue square), ID $151299$  (black up-pointing triangle), ID $151878$  (magenta diamond), and ID $151332$  (green down-pointing triangle) formed by an interaction around $1.9 \, \rm{Gyr}$ ago (see the movie ``ID150872.mp4'').
			d) The host galaxy ID $138$ with the DMF substructures ID $878$ (red circle) and ID $1683$ (green down-pointing triangle) being identified at $z = 0$ (see also Fig.~\ref{fig:appendix_Images_TDGCs_DMCDGs} in Appendix~\ref{sec:appendix_Images_TDGCs_DMCDGs} and the movies ``ID138.mp4''). These are not TDGs because they form from gas ram-pressure stripped from the host ID $138$. Ram-pressure stripping can be observed at a lookback time $ \lesssim 1 \, \rm{Gyr}$.} 
		\label{fig:time_evolution_TDGCs}
	\end{figure*}

	First, we study in Fig.~\ref{fig:time_evolution_TDGCs_a} the evolution of the host galaxy with the identification number ID $404871$ at redshift $z = 0$ (see also Fig.~\ref{fig:appendix_Images_TDGCs_DMCDGs} in Appendix~\ref{sec:appendix_Images_TDGCs_DMCDGs} and the movie ``ID404871.mp4'') by following its main progenitor branch \citep{Rodriguez_2015}. This galaxy hosts a TDGC in a gaseous tidal arm, which was formed by a close encounter with another massive galaxy around $1.6 \, \rm{Gyr}$ ago. A similar formation process of the TDGCs ID $78410$ and ID $74010$ (both of sample A) is seen in Fig.~\ref{fig:time_evolution_TDGCs_b}. A galaxy merger at a lookback time of around $1.9 \, \rm{Gyr}$ creates tidal debris. The first stellar particles in the subhalos of both identified TDGCs at $z=0$ appear at about $0.1 \, \rm{Gyr}$ (ID $74010$) and $0.5 \, \rm{Gyr}$ (ID $74810$) after the merger, allowing us to estimate their ages to be about $1.8 \, \rm{Gyr}$ and $1.4 \, \rm{Gyr}$, respectively. At present,  ID $74810$ and ID $74010$ have $63$ and $200$ stellar particles, respectively. ID $74010$ has similar properties to the observed NGC 1052-DF2 galaxy by \cite{vDokkum_2018b} (see also Section~\ref{sec:Results_NGC1052-DF2} and the movies ``ID73663.mp4'' and ``ID73663$\_$zoom.mp4''). Figure~\ref{fig:time_evolution_TDGCs_c} shows the host galaxy ID $150872$ with the TDGCs of sample B IDs $151014$, $151271$, $151299$, $151878$, and $151132$, which were formed again through a galaxy--galaxy encounter around $1.9 \, \rm{Gyr}$ ago (see also the movie ``ID150872.mp4'').
	
	Finally, by tracing the host galaxy ID $138$ back in time, a different formation process of dark matter-lacking subhalos compared to the above discussed examples can be observed in Fig.~\ref{fig:time_evolution_TDGCs_d} (see also Fig.~\ref{fig:appendix_Images_TDGCs_DMCDGs} in Appendix~\ref{sec:appendix_Images_TDGCs_DMCDGs} and the movie ``ID138.mp4''). At a lookback time $ \lesssim 1 \, \rm{Gyr}$ this galaxy undergoes ram-pressure stripping. This is an example of how baryon-dominated dwarf galaxies can form from material stripped from a host galaxy through ram-pressure (the ``type B dwarfs'' of \citealt{Kroupa_2012} and ``fireballs'' observed by \citealt{Yoshida_2008,Yagi_2010}). Recent observations have shown that enhanced star formation can appear in the ram-pressure stripped tails of jellyfish galaxies \citep{Vulcani_2018}. The DMF subhalos around the host galaxy ID $138$ have $M_{\mathrm{stellar}}>5 \times 10^{7} \, \rm{M_{\odot}}$ but are located within $10 \times R_{\mathrm{0.5 \, stellar}}^{\mathrm{host}}$ and are defined as DMF substructures of their host and therefore are not counted as TDGCs in this work. This example also demonstrates that we have applied a very stringent minimum separation criterion in order to avoid a misidentification of DMF substructures. In other words, we expect to have several false negatives but accept this in order to minimize false positives.
	
\begin{figure}
		\centering
		\includegraphics[width=\linewidth,trim={0.0cm 0.0cm 0.0cm 0.0cm},clip]{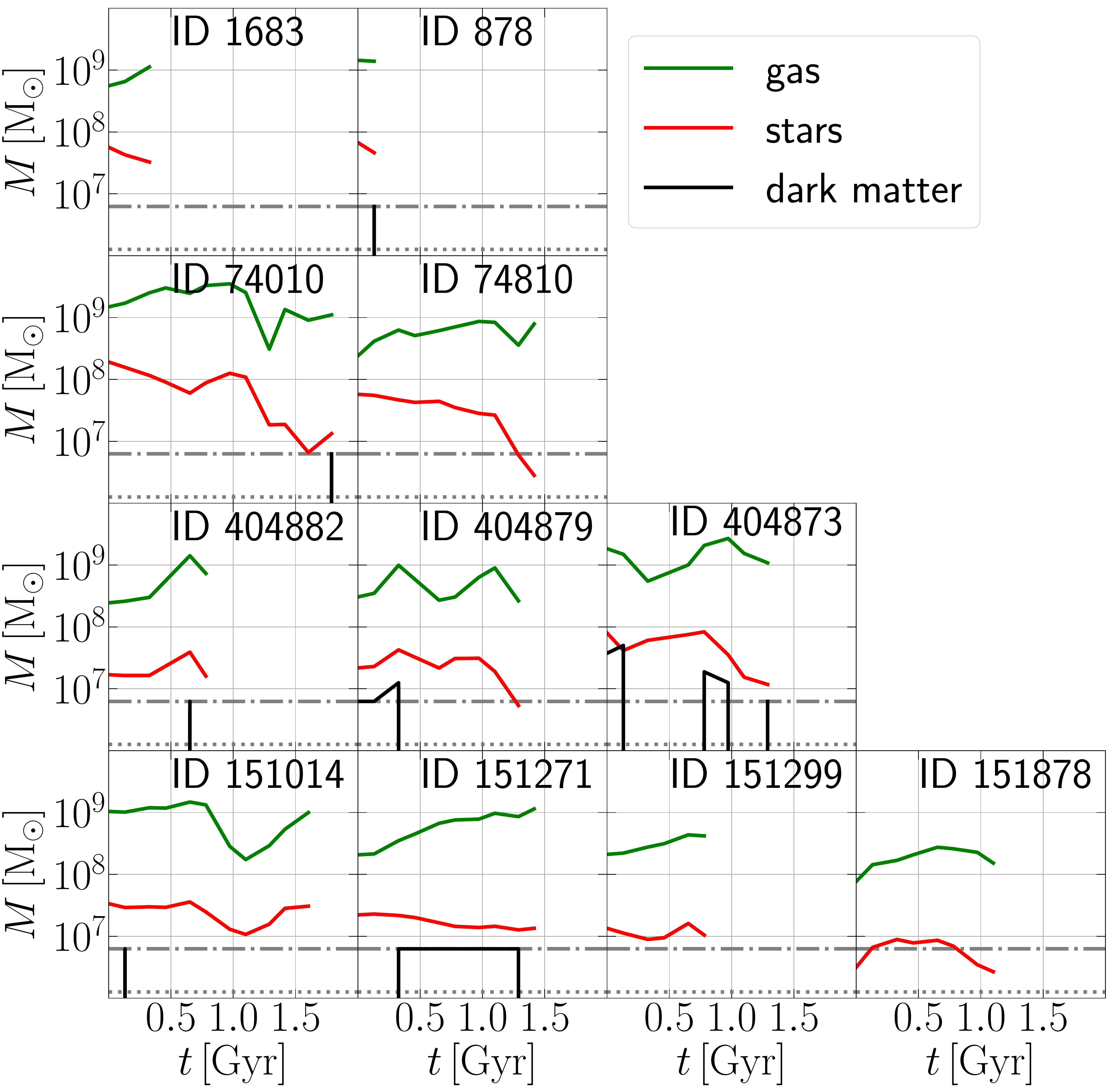}
		\caption{Gas (green), stellar (red), and dark matter (black) mass evolution of TDGCs and DM-poor objects shown in Fig.~\ref{fig:time_evolution_TDGCs} starting from the first time step at which the first stellar particles in their subhalos appeared. Their ID numbers are given in the upper-right corner of the panels and the discussed subhalos of each row belong to the same host galaxy. The subhalo ID $404879$ has $M_{\mathrm{stellar}} = 2.2 \times 10^{7} \, \rm{M_{\odot}}$, $M_{\mathrm{gas}} = 3.0 \times 10^{8} \, \rm{M_{\odot}}$, and $M_{\mathrm{dm}} = 6.3 \times 10^{6} \, \rm{M_{\odot}}$ and is thus not included in the main sample of DMC (-poor) DGs (see Table \ref{tab:DMCDGs}). The dashed and long-dashed horizontal lines indicate the initial baryonic ($1.26 \times 10^{6} \, \rm{M_{\odot}}$) and dark matter mass ($6.26 \times 10^{6} \, \rm{M_{\odot}}$) of a particle. The dark matter content is short lived and is due to individual dark matter particles crossing the objects.}
		\label{fig:mass_evolution}
	\end{figure}

	In Fig.~\ref{fig:mass_evolution} we address the gas, stellar, and dark matter mass evolution of the objects discussed here. Each of these subhalos has at most one dark matter particle at the time when their first stellar particle was identified. Given the high velocity dispersion of dark matter particles, their presence in the objects could simply be transients detected by the Subfind algorithm. Moreover, the objects are always baryon-dominated. 
	
	\subsection{The orbital angular momentum of dwarf galaxies}
	\label{sec:Methods_orbital_angular_momentum}
	
	The different formation scenarios of galaxies with and without dark matter cause differences in their phase-space distributions. In particular, TDGs can be significantly correlated in phase space \citep{Kroupa_2012,Ploeckinger_2015}. 
	The specific orbital angular momenta of dwarf galaxies with respect to their host galaxies are calculated by, 
	\begin{equation}
	\begin{aligned}
	\vec{L}_{\mathrm{orbit}} = (\vec{r}_{\mathrm{DG}}-\vec{r}_{\mathrm{host}}) \times (\vec{v}_{\mathrm{DG}}-\vec{v}_{\mathrm{host}}) \, ,
	\end{aligned}
	\label{eq:orbital_angular_momentum}
	\end{equation}
	where $\vec{r}_{\mathrm{DG}}$, and $\vec{r}_{\mathrm{host}}$, and $\vec{v}_{\mathrm{DG}}$, and $\vec{v}_{\mathrm{host}}$ are the position and velocity vectors of the dwarf galaxy and host galaxy, respectively.
	
	The degree of the phase-space correlation of a system with more than two TDGCs or DMC DGs is then determined by
	
	\begin{equation}
	\begin{aligned}
	\sigma_{\mathrm{orbit}} = \sqrt{\mathrm{var}(l_{\mathrm{orbit, \, x}})+\mathrm{var}(l_{\mathrm{orbit, \, y}})+\mathrm{var}(l_{\mathrm{orbit, \, z}})} \, ,
	\end{aligned}
	\label{eq:dregree_phase_space_correlation}
	\end{equation}
	with $\mathrm{var}(l_{\mathrm{orbit,x}})$, $\mathrm{var}(l_{\mathrm{orbit,y}})$, and $\mathrm{var}(l_{\mathrm{orbit,z}})$ being the variances of the x, y, and z components of the normalized specific orbital angular momenta given by Eq. \ref{eq:orbital_angular_momentum}; for example,
	
	\begin{equation}
	\begin{aligned}
	\mathrm{var}(l_{\mathrm{orbit, \, x}}) = \frac{1}{N} \sum_{i=1}^{N} (l_{\mathrm{orbit,x},i} - \bar{l}_{\mathrm{orbit,x}})^{2} \, ,
	\end{aligned}
	\label{eq:dregree_phase_space_correlation_var}
	\end{equation}
	where $N$ is the number of dwarf galaxies around a host galaxy and $\bar{l}_{\mathrm{orbit,x}}$ is the mean of all x-components of the normalized specific orbital angular momenta. 
	
	This method is independent of the coordinate system. In the case of a purely spherical distribution of the angular momenta the degree of the phase-space correlation becomes $\sigma_{\mathrm{orbit}} = 1$.
	
	\subsection{Dispersion- and rotation-dominated galaxies} 
	\label{sec:Methods_kappa_morphology}
\begin{figure}
		\centering
		\includegraphics[width=44mm]{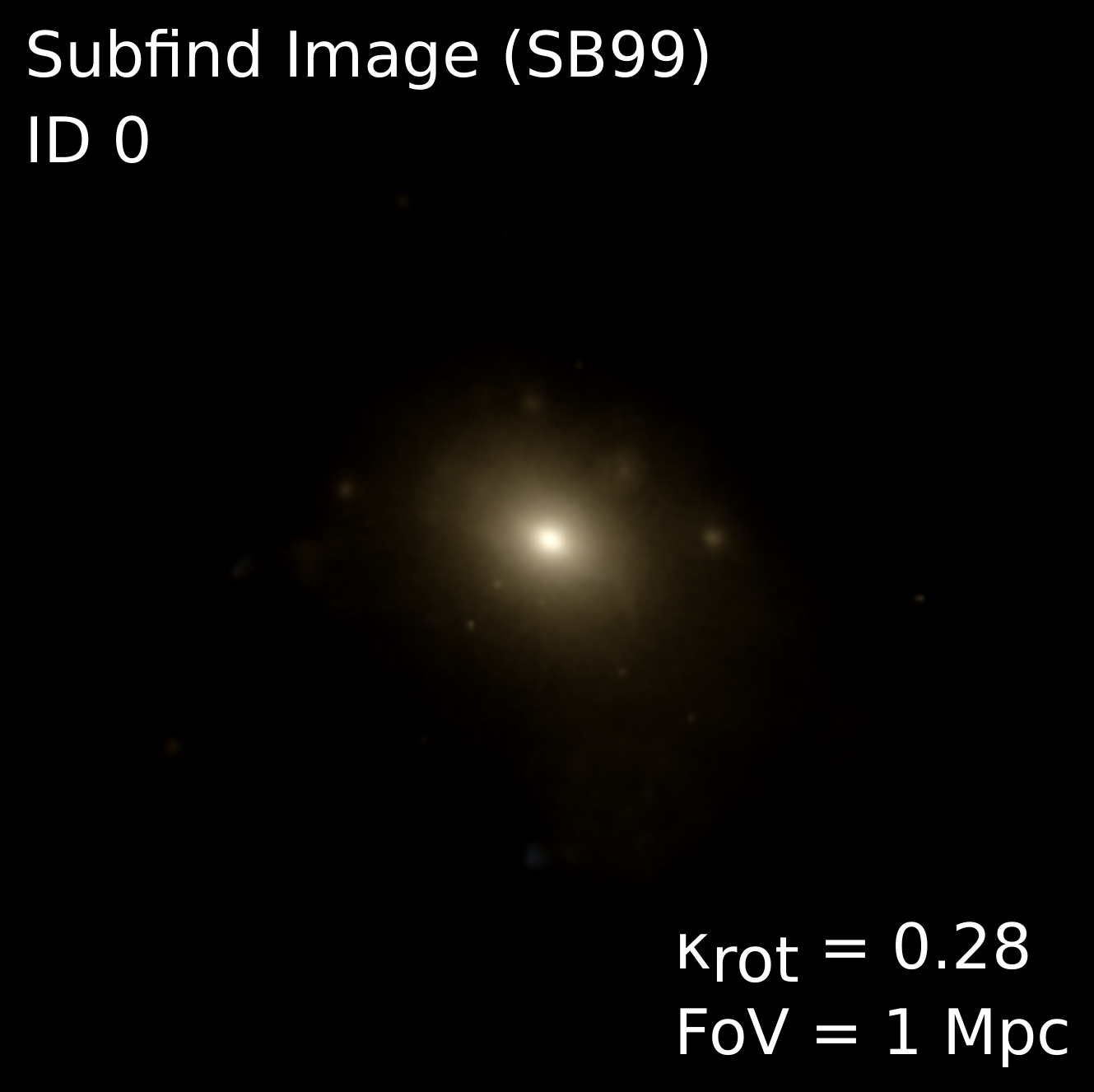}
		\includegraphics[width=44mm]{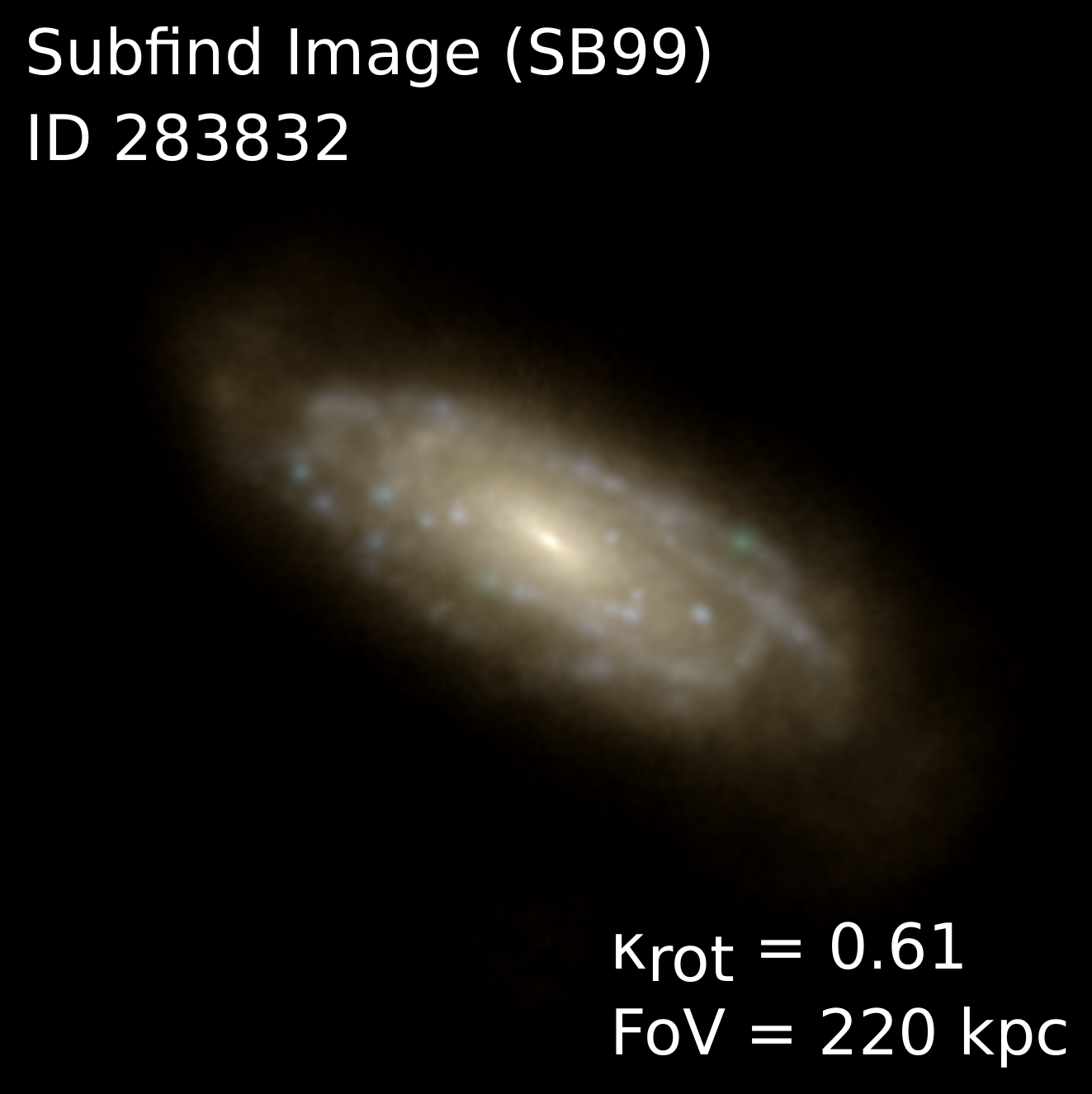}
		\caption{Examples of Subfind Starburst 99 (SB99) images of  the most massive dispersion- (left; ID 0, $\kappa_{\mathrm{rot}} = 0.28$, $M_{\mathrm{stellar+gas}}= 2.9 \times 10^{13} \, \rm{M_{\odot}}$, $M_{\mathrm{dm}}= 2.9 \times 10^{14} \, \rm{M_{\odot}}$) and most massive rotation-dominated (right; ID 283832, $\kappa_{\mathrm{rot}} = 0.61$, $M_{\mathrm{stellar+gas}}= 4.9 \times 10^{11} \, \rm{M_{\odot}}$, $M_{\mathrm{dm}}= 6.7 \times 10^{12} \, \rm{M_{\odot}}$) galaxy in the Illustris-1 simulation at redshift $z = 0$. The image field of view (FoV) is ten times the stellar half-mass radius, $R_{\mathrm{0.5 \, stellar}}$, of the shown galaxy. Credit: Illustris Galaxy Observatory \url{http://www.illustris-project.org/galaxy_obs/} [25.08.2018]}
		\label{fig:morphology}
	\end{figure}

	Determining the morphology of simulated galaxies and the comparison thereof with observations is a nontrivial task. Here, we use the $\kappa_{\mathrm{rot}}$ morphological parameter in order to separate them in dispersion- and rotation-dominated systems, which was already studied by \cite{Sales_2012} and \cite{Rodriguez_2017}. The $\kappa_{\mathrm{rot}}$ parameter is defined as the fraction of the rotational energy, $K_{\mathrm{rot}}$, to the kinetic energy, $K$, of all stellar particles in the considered subhalo. The morphological parameter, $\kappa_{\mathrm{rot}}$, is
	
	\begin{equation}
	\begin{aligned}
	\kappa_{\mathrm{rot}} \equiv \frac{K_{\mathrm{rot}}}{K} = \frac{1}{K} \sum_{i} \frac{1}{2} m_{i} \bigg( \frac{\vec{\hat{h}} \cdot \vec{h}_{i}}{R_{i}} \bigg)^2 \, ,
	\end{aligned}
	\label{equation_rotational_parameter}
	\end{equation}
	where $\vec{\hat{h}}$ is a unit vector proportional to the total stellar angular momentum of the galactic system, $\vec{h}_{i}$ is the specific angular momentum vector, $m_{i}$ is the mass, and $R_{i}$ is the projected radius of the i-th stellar particle. The positions and velocities of the stellar particles are calculated with respect to the center of mass of the subhalo. According to Eq. \ref{equation_rotational_parameter}, the $\kappa_{\mathrm{rot}}$ parameter can range between $0$ and $1$ meaning that in the latter case all stellar particles move on circular orbits with respect to the total stellar angular momentum. Subhalos with $\kappa_{\mathrm{rot}}$ smaller or larger than $0.5$ are dispersion- or rotation-dominated systems, respectively. Images of the most massive dispersion- and rotation-dominated Illustris galaxies identified at redshift $z = 0$ are presented in Fig.~\ref{fig:morphology}. An interesting discussion about the properties of the $\kappa_{\mathrm{rot}}$ morphological parameter for dynamical systems is found in the Appendix A of \cite{Rodriguez_2017}.
	
	\section{Results} \label{sec:Results}
	
	We present the physical properties of TDGCs and DMC DGs and their positions in the radius-mass plane at redshift $z=0$. The results are compared with observational data from \cite{Dabringhausen_2016} and \cite{Mieske_2008, Mieske_2013}. The metallicities of TDGCs and DMC DGs are studied in Appendix \ref{sec:appendix_metallicity}. In addition, a discussion about the internal structures and kinematics of TDGCs including a $\sigma$-clipping scheme as a $6$D phase-space halo finder applied on gas-free Subfind TDGCs of sample A can be found in Appendix \ref{sec:appendix_energy} where the gravitationally bound nature of these simulated objects is also discussed.

	\subsection{Phase-space correlation of TDGCs and DMC DGs}
	\label{Results:angularmomentum}
	
\begin{figure}
		\centering
		\includegraphics[width=\columnwidth,trim={0cm 0.0cm 0 0.0cm},clip]{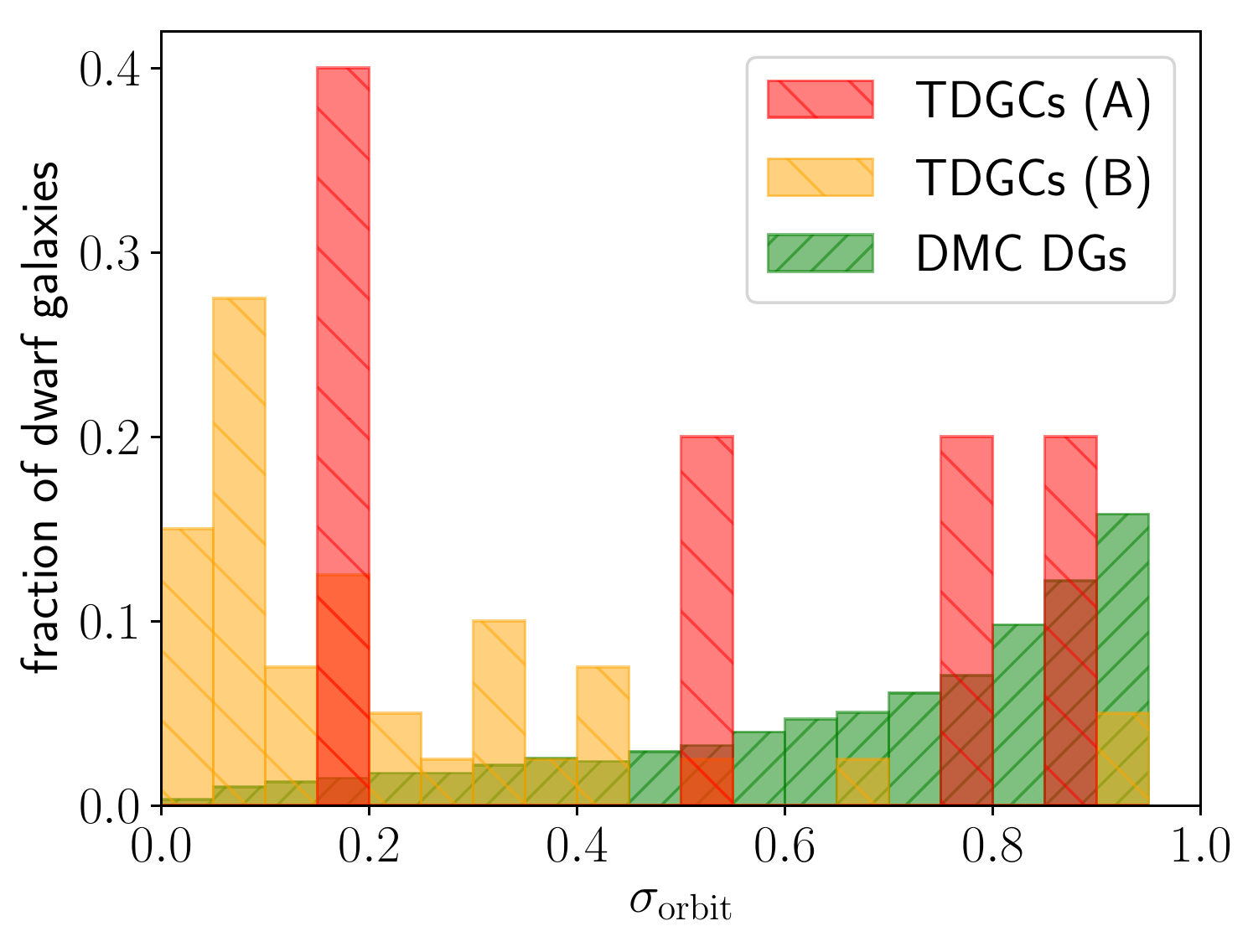}
		\caption{Degree of the phase-space correlation, $\sigma_{\mathrm{orbit}}$ (eq. \ref{eq:dregree_phase_space_correlation}), for TDGCs of sample A (red) and sample B (orange) and DMC DGs (green). The histograms have a bin width of $\Delta \sigma_{\mathrm{orbit}} = 0.05$.}
		\label{fig:phasespace_correlation}
	\end{figure}

	We quantify the degree of the phase-space correlation, $\sigma_{\mathrm{orbit}}$ (Eq.~\ref{eq:dregree_phase_space_correlation}), for all galactic systems with more than one TDGC or DMC DG. Considering all TDGCs with $M_{\mathrm{stellar}}>5 \times 10^{7} \, \rm{M_{\odot}}$ (sample A) gives only five galactic systems that host more than one such TDGC. Therefore, we determine the phase-space correlation for both samples A and B. The results are listed in Table~\ref{tab:phasespace_correlation} and the distributions of the degree of the phase-space correlation, $\sigma_{\mathrm{orbit}}$, for TDGCs and DMC DGs are shown in Fig.~\ref{fig:phasespace_correlation}. Tidal dwarf galaxy candidates from sample B are significantly more phase-space-correlated than DMC DGs. These results are discussed in Section~\ref{sec:discussion_DMF_galaxies}.
	
	\begin{table}
		\centering
		\caption{Degree of the phase-space correlation, $\sigma_{\mathrm{orbit}}$ (eq. \ref{eq:dregree_phase_space_correlation}), for TDGC samples and DMC DGs at redshift $z=0$. }
		\label{tab:phasespace_correlation}
		\begin{tabular}{llll} \hline
			sample & counts & mean & median  \\ \hline \hline
			TDGCs (sample A) & $5$ & $0.52$ & $0.54$ \\
			TDGCs (sample B) & $40$ & $0.22$ & $0.15$ \\ 
			DMC DGs & $7810$ & $0.73$ & $0.82$ \\
			\hline 
		\end{tabular}
		\tablefoot{Listed are the number of galactic systems with more than one TDGC or DMC DG, the mean, and the median of the degree of the phase-space correlation for each dwarf galaxy sample.}
	\end{table}
	
	\subsection{NGC 1052-DF2-like galaxies in the Illustris-1 simulation} \label{sec:Results_NGC1052-DF2}
	
	The ultra-diffuse galaxy NGC 1052–DF2 has $M_{\mathrm{stellar}} \approx 2 \times 10^{8} \, \rm{M_{\odot}}$ with a dark matter mass $400$ times smaller than theoretically predicted and has an effective radius along the major axis of $R_{\mathrm{e}} = 2.2 \, \rm{kpc}$, assuming it is at a distance of $20 \, \rm{Mpc}$ \citep{vDokkum_2018b}. \cite{Wolf_2010} derived a scaling relation between the 2D projected half-light radius, $R_{\mathrm{e}}$, and the 3D deprojected half-light radius, $R_{\mathrm{0.5 \, light}}$, for stellar systems. These latter authors showed that the relation,
	\begin{equation}
	\begin{aligned}
	R_{\mathrm{0.5 \, light}} \approx \frac{4}{3} \times R_{\mathrm{e}} \, , 
	\end{aligned}
	\label{equation_radii_scaling_relation_2D_3D}
	\end{equation}
	is accurate for most surface brightness profiles of spherical stellar systems with S\'{e}rsic indices in the range $0.10 \leq n^{-1} \leq 2.0$ \cite[see Appendix B in][]{Wolf_2010}.
	Applying this scaling relation to the effective radius of NGC 1052-DF2  and taking into account its axis ratio being $0.85$ gives a 3D stellar half-light radius of $2.7 \, \rm{kpc}$.
	
	Using the Illustris-1 simulation we found no single TDGC fulfilling a minimum stellar mass criterion of $2 \times 10^{8} \, \rm{M_{\odot}}$ and a minimum stellar half-mass radius criterion of  $2.7 \, \rm{kpc}$ at the same time. 
	Choosing instead $20$~percent reduced lower limits of $M_{\mathrm{stellar}} = 0.8 \times \left(2 \times 10^{8} \right) \, \rm{M_{\odot}}$ and $R_{\mathrm{0.5 \, stellar}}=0.8 \times 2.7 \, \rm{kpc}$ gives only one TDGC at redshift $z = 0$ (ID 74010). The probability of finding such a NGC1502-DF2-like galaxy among all TDGCs of sample A is around $1.0 \times 10^{-2}$. In particular, this TDGC has $R_{\mathrm{0.5 \, stellar}} = 2.4 \, \rm{kpc}$, $M_{\mathrm{stellar}} = 1.9 \times 10^{8} \, \rm{M_{\odot}}$, $M_{\mathrm{gas}} = 1.5 \times 10^{9} \, \rm{M_{\odot}}$, and $\kappa_{\mathrm{rot}} = 0.46$. The separation to its host galaxy (ID 73679) is about $219 \, \rm{kpc}$, which is roughly consistent with the observed NGC 1052-DF2 galaxy found in the vicinity of the massive elliptical galaxy NGC 1052.\footnote{The statistically expected 3D separation between the observed NGC 1052 and NGC 1052-DF2 galaxies is about $100 \, \rm{kpc}$, which is $\sqrt{3/2}$ times its projected separation, assuming NGC 1052-DF2 is at a distance from us comparable to that of NGC 1052 \citep[$20 \, \rm{Mpc}$,][]{vDokkum_2018b}. However, this distance may be revised \citep{Trujillo_2019}.} The simulated host galaxy is dispersion-dominated and has $M_{\mathbf{stellar}} = 6.4 \times 10^{10} \, \rm{M_{\odot}}$ and $M_{\mathrm{dm}} = 4.4 \times 10^{11} \, \rm{M_{\odot}}$. Interestingly, this galaxy hosts a second gas-rich TDGC (ID 74810) at $150 \, \rm{kpc}$. As seen in a series of snapshots in Section~\ref{sec:formation_TDGCs} (see also the movies ``ID73663.mp4'' and ``ID73663$\_$zoom.mp4'') these TDGCs were formed from the gas expelled by tidal forces from massive interacting galaxies. 
	
	\begin{table}
		\centering
		\tiny
		\caption{Probability of finding a NGC 1052-DF2-like galaxy in the Illustris-1 simulation at redshift $z=0$. }
		\label{table_NGC1052-DF2_probability}
		\begin{tabular}{lllll} \hline
			& TDGCs & TDGCs & TDGCs  \\
			\hline \hline
			$R_{\mathrm{0.5 \, stellar}} \, [\rm{kpc}]$  & $\geq 2.7$ & $\geq 0.8 \times 2.7 $ &  $ \geq 0.6 \times2.7 $  \\
			$M_{\mathrm{stellar}} \, [\rm{M_{\odot}}]$ & $\geq 2 \times 10^{8}$ & $\geq 0.8 \times \left( 2 \times 10^{8} \right)$  & $\geq 0.6 \times \left( 2 \times 10^{8} \right)$  \\ \hline \hline 
			Number & $0$ & $1$ & $6$    \\ \hline 
			Probability & $0.0$ & $1.0 \times 10^{-2}$ & $6.2 \times 10^{-2}$ \\
			wrt. sample A &  &  &  \\ \hline \hline
			gas free & TDGCs & TDGCs & TDGCs  \\
			\hline \hline
			$R_{\mathrm{0.5 \, stellar}} \, [\rm{kpc}]$  & $\geq 2.7$ & $\geq 0.8 \times 2.7$ &  $\geq 0.6 \times 2.7$  \\
			$M_{\mathrm{stellar}} \, [\rm{M_{\odot}}]$ & $\geq 2 \times 10^{8}$ & $\geq 0.8 \times \left( 2 \times 10^{8} \right)$  & $\geq 0.6 \times \left( 2 \times 10^{8} \right)$  \\ \hline \hline 
			Number & $0$ & $0$ & $5$    \\ \hline 
			Probability & $0.0$ & $0.0$ & $5.2 \times 10^{-2}$ \\
			wrt. sample A &  &  &  \\
			\hline 
		\end{tabular}
		\tablefoot{The second part of the table only refers to gas-free NGC 1052-DF2-like galaxies. The probabilities are calculated by dividing the number of selected TDGCs by the number of all TDGCs of sample A ($97$) at redshift $z=0$.}
	\end{table}
	
	Summing up, there is no TDGC in the Illustris-1 simulations at redshift $z = 0$ which has a stellar mass and a stellar-half mass radius equal to or larger than the observed NGC 1052–DF2 at the same time. However, relaxing the lower mass limits of the selection criteria by $20$ and $40$~percent yields one and six TDGCs, respectively. But invoking the condition $M_{\mathrm{gas}} = 0$ because NGC 1052-DF2 is gas-free \citep{Chowdhury_2019,Sardone_2019} and choosing lower limits of $M_{\mathrm{stellar}} = 0.8 \times \left( 2 \times 10^{8} \right) \, \rm{M_{\odot}}$ and $R_{\mathrm{0.5 \, stellar}}= 0.8 \times 2.7 \, \rm{kpc}$ lead to no similar dwarf galaxies existing in the Illustris-1 simulation. A parameter study of different selection criteria is given in Table~\ref{table_NGC1052-DF2_probability}. Regardless of the exact definition, finding a NGC 1052-DF2-like galaxy at redshift $z=0$ in the Illustris-1 simulation is extremely rare. This analysis does not include a comparison of the peculiar velocity of the observed NGC 1052-DF2 with simulated analogs. 
	
	However, the observed velocity dispersion is rather uncertain and allows for a significant dark matter content \citep{Martin_2018}. In addition, \citet{Trujillo_2019} concluded that NGC 1052-DF2 is at a distance of $13.0 \pm 0.4 \, \rm{Mpc}$ from Earth and is not an outlier to dwarf galaxies of the Local Group. \footnote{These calculations include only completely dark matter-free galaxies. In a further analysis subhalos with the ratio $M_{\mathrm{halo}}/M_{\mathrm{stellar}}$ at least $400$ times lower than theoretically expected can be included. This analysis would be an interesting extension to the present work.}
	
	\subsection{Physical properties of TDGCs and DMC DGs}
	\label{sec:Physical properties of TDGCs and DMC DGs}
	
\begin{figure}
		\centering
		\includegraphics[width=\columnwidth,trim={0.0cm 0.0cm 0.0cm 0.0cm},clip]{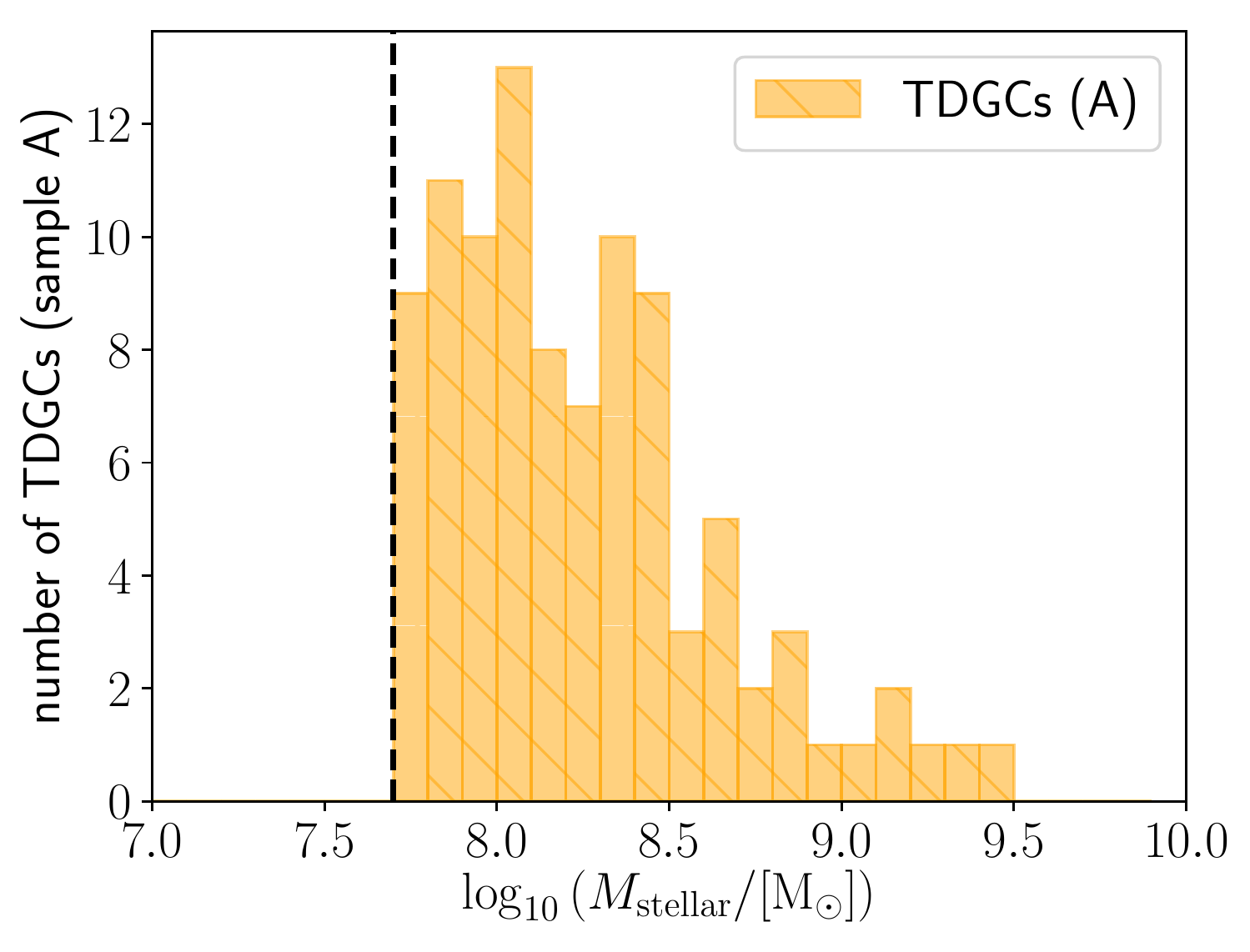}
		\includegraphics[width=\columnwidth,trim={0.0cm 0.0cm 0.0cm 0.0cm},clip]{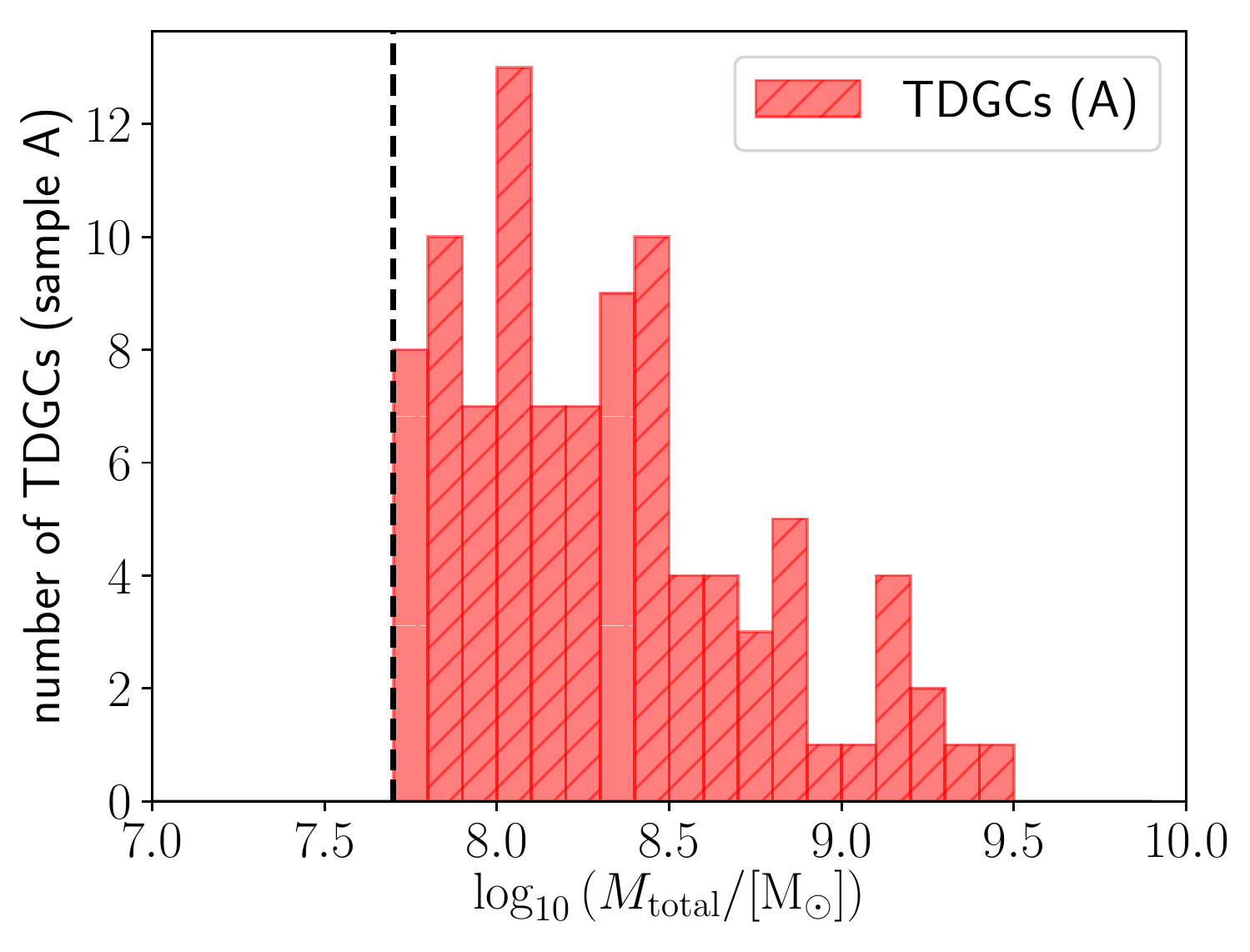}
		\caption{Stellar mass, $M_{\mathrm{stellar}}$, (top) and total mass, $M_{\mathrm{total}}$, (bottom) distributions of TDGCs (sample A) at redshift $z = 0$. The dashed vertical line illustrates the minimum stellar mass criterion of $5 \times 10^{7} \, \rm{M_{\odot}}$. The histograms have a bin width of $\log_{10}(\Delta M_{\mathrm{stellar}} / [\rm{M_{\odot}}]) = 0.10$ and $\log_{10}(\Delta M_{\mathrm{total}} / [\rm{M_{\odot}}]) = 0.10$.}
		\label{fig:histogram_stellar_total_TDGCs}
	\end{figure}

	Figure~\ref{fig:histogram_stellar_total_TDGCs} shows the stellar (top) and total (bottom) mass distributions of TDGCs with $M_{\mathrm{stellar}}>5 \times 10^{7} \, \rm{M_{\odot}}$ (sample A). As expected, TDGCs have typically small masses whereby the most massive TDGC has $M_{\mathrm{total}} = 3.1 \times 10^{9} \, \rm{M_{\odot}}$. In high-resolution simulations of merging galaxies with dark matter, the most massive TDGs have been reported to have baryonic masses in the range of $10^{8} \, \rm{M_{\odot}}$ to $10^{9} \, \rm{M_{\odot}}$ such that also the Large and Small Magellanic Clouds can be TDGs \citep{Bournaud_2008a,Bournaud_2008b,Fouquet_2012}.
	
	The applied selection criteria for TDGCs of sample A identify dwarf galaxies with low amounts of gas. In particular, around $89$~percent of all TDGCs are completely gas-free and also have no star formation. If we apply similar selection criteria for TDGCs as in the work of \cite{Ploeckinger_2018} (sample B), we obtain many more TDGCs than in sample A (see Table~\ref{tab:TDGCs}).\footnote{Here, we refer to the TDGC sample B, which includes DMF stellar objects beyond $10 \times R_{\mathrm{0.5 \, stellar}}^{\mathrm{host}}$ and within $100 \times R_{\mathrm{0.5 \, stellar}}^{\mathrm{host}}$ with $M_{\mathrm{gas}}>5 \times 10^{7} \, \rm{M_{\odot}}$ containing at least one stellar particle.} This can be understood by the formation scenario of TDGCs, which are formed in gas-rich tidal tails expelled from their host galaxies triggered by galactic interactions. The median (mean) of the stellar mass of TDGCs belonging to sample B is about $1.7$ ($5.6$) higher than the median (mean) of the TDGCs from \cite{Ploeckinger_2018}. Furthermore, we report a median (mean) of the gas mass that is 2.4 (1.6) times higher than the sample of \cite{Ploeckinger_2018}.
	These discrepancies could be caused by the different selection criteria for TDGCs and the use of different cosmological simulations. \cite{Ploeckinger_2018} set a minimum gas mass limit of $10^{7} \, \rm{M_{\odot}}$ because of the higher resolution of baryonic and dark matter particles in the EAGLE simulations they used compared to the Illustris-1 run employed here. Moreover, \cite{Ploeckinger_2018} select TDGCs within $z \leq 2.0$.

	\begin{figure}
		\centering
		\includegraphics[width=\columnwidth,trim={0cm 0.0cm 0 0.0cm},clip]{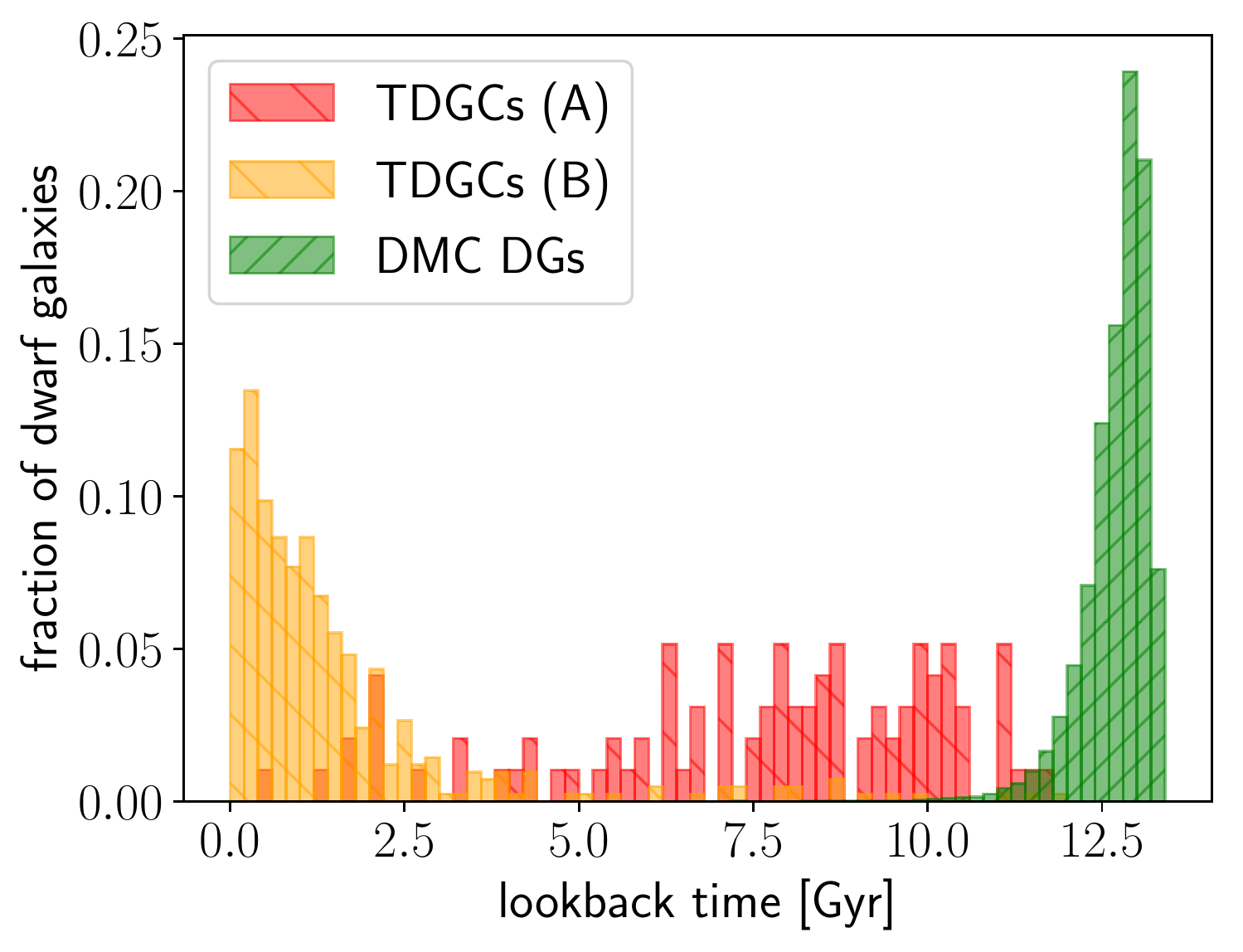}
		\caption{Distribution of the age of the oldest stellar particles within dwarf galaxies identified at redshift $z=0$. The histograms have a bin width of $\Delta t = 0.2 \, \rm{Gyr}$. The statistics of the distributions shown here are listed in Table~\ref{table_physical_propertiers}.}
		\label{fig:age_dwarfgalaxies}
	\end{figure}
	
	The age of dwarf galaxies is estimated here by the formation time of the oldest stellar particle within a subhalo identified at redshift $z=0$. The derived age distribution of different dwarf galaxy samples are studied in Fig.~\ref{fig:age_dwarfgalaxies} and analyzed in more detail in Table~\ref{table_physical_propertiers}. The mean age of DM-rich DGs is $12.7 \, \rm{Gyr}$, which is significantly higher than for DM-poor DGs ($8.9 \, \rm{Gyr}$) and TDGCs and underlines that DM-rich DGs are formed in early stages of the Universe. The mean ages of the TDGCs of samples A and B are $7.6 \, \rm{Gyr}$ and $1.5 \, \rm{Gyr}$, respectively. Therefore TDGCs with a vanishing gas content are older objects, which have already consumed or lost their gas reservoir via ram-pressure stripping and interactions. 
	
\begin{figure}
		\centering
		\includegraphics[width=\linewidth,trim={0cm 0.0cm 0 0.0cm},clip]{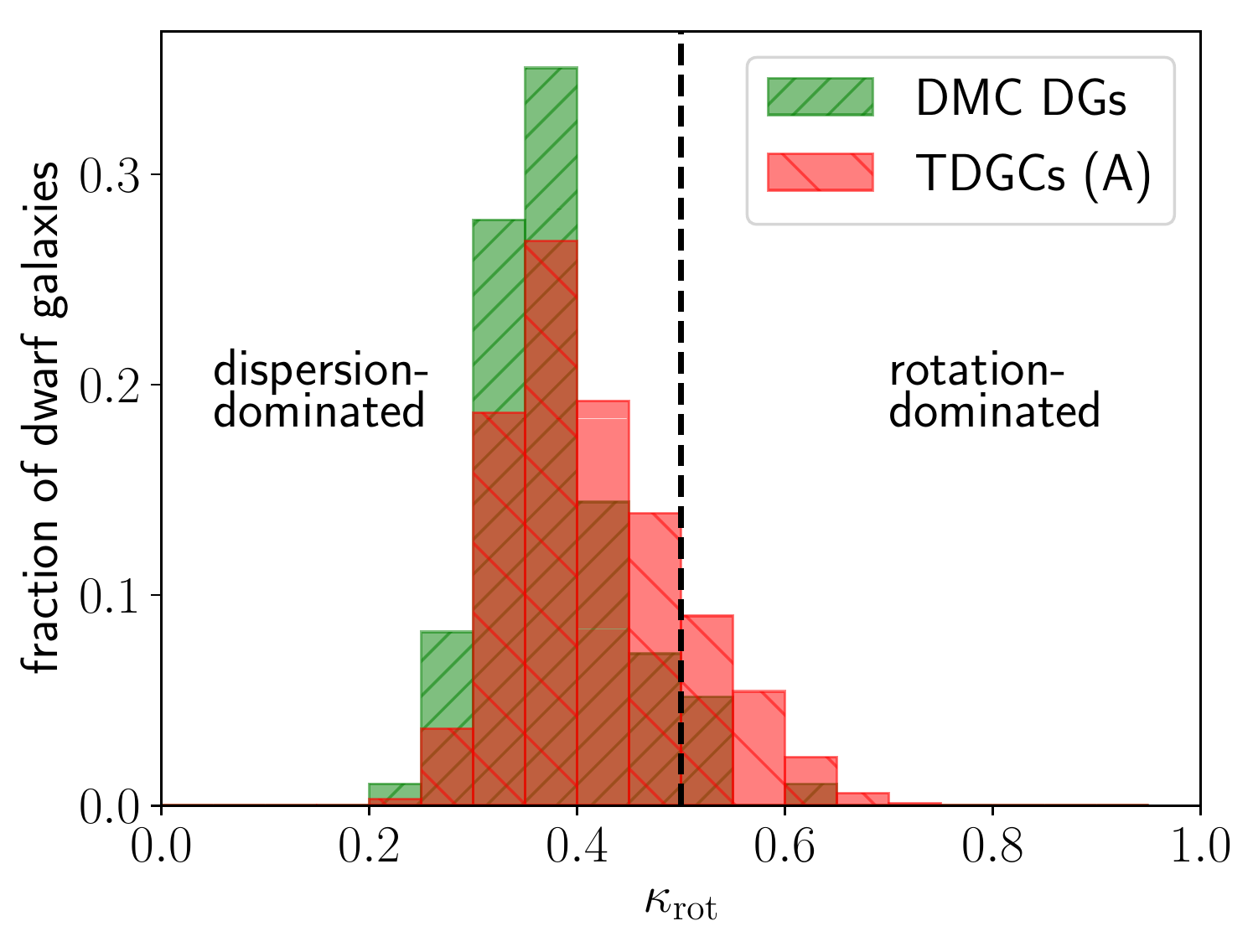}
		\caption{Distribution of the $\kappa_{\mathrm{rot}}$ morphological parameter for DMC DGs (green) and TDGCs (red) with $M_{\mathrm{stellar}}>5 \times 10^{7} \, \rm{M_{\odot}}$ (sample A). DMC DGs and TDGCs are divided into rotation- and dispersion-dominated objects at $\kappa_{\mathrm{rot}} = 0.5$ (dashed black line). The histograms have a bin width of $\Delta \kappa_{\mathrm{rot}} = 0.05$.}
		\label{fig:kapp_DMF_TDGCs}
	\end{figure}

	The distribution of the $\kappa_{\mathrm{rot}}$ morphology parameter of TDGCs (sample A) and DMC DGs is presented in Fig.~\ref{fig:kapp_DMF_TDGCs}, which states that around $94$~percent of all TDGCs with $M_{\mathrm{stellar}}> 5 \times 10^{7} \, \rm{M_{\odot}}$ (sample A) are dispersion-dominated ($\kappa_{\mathrm{rot}}<0.5$) at redshift $z=0$. The high fraction of dispersion-dominated TDGCs is unexpected, because high-resolution simulations of galaxy encounters by \cite{Bournaud_2008a,Bournaud_2008b} have shown that the most massive stellar TDGs in the mass range of $10^{8}-10^{9} \, \rm{M_{\odot}}$ are dominated by rotation. However, the TDGs in their simulations are young and gas-rich while most of the observed satellite galaxies surrounding the MW are old and typically dispersion-dominated. The simulations by \cite{Bournaud_2008a,Bournaud_2008b} suggest that feedback processes such as SN explosions transform them into gas-poor DGs which suffer from a loss of angular momentum. Therefore, TDGs can be transformed into dwarf spheroidal satellite galaxies within a Hubble time \citep{Metz_2007, Dabringhausen_2013}. 
	The medians and means of the distribution of the $\kappa_{\mathrm{rot}}$ parameter for simulated TDGCs (sample A) and DMC DGs are almost the same (see Table~\ref{table_physical_propertiers}). 
	Nevertheless, calculating the $\kappa_{\mathrm{rot}}$ parameter for objects with a small number of stellar particles is insecure and the present results should be treated with caution. 
	
	Finally, we study the 1D velocity dispersion of simulated dwarf galaxies which is calculated by all particles (cells) belonging to the considered subhalo. The medians and means of the 1D velocity dispersion for different dwarf galaxy samples reveal information about the properties of baryonic and dark matter particles. Dwarf galaxies with a small amount of dark matter have significantly lower velocity dispersions than dark matter-dominated objects, as theoretically expected. The medians  of the 1D velocity dispersion of simulated TDGCs (sample A) and DM-poor DGs are $7.8 \, \rm{km \, s^{-1}}$ and $9.7 \, \rm{km \, s^{-1}}$, respectively. The intrinsic velocity dispersion of NGC 1052-DF2 derived by observing ten GCs is $\sigma_{\mathrm{intr}} = 7.8_{-2.2}^{+5.2} \, \rm{km \, s^{-1}}$ \citep{vDokkuma_2018}. 
	
	The stellar and gas metallicities of TDGCs belonging to sample A are shown and discussed in Appendix \ref{sec:appendix_metallicity} and in Section~\ref{sec:discussion_gas_sfr}, respectively. The physical properties of different TDGCs and DMC DGs samples at redshift $z=0$ are summarized in Table~\ref{table_physical_propertiers}.
	
\begin{figure}
		\centering
		\includegraphics[width=\columnwidth,trim={0cm 0.00cm 0 0.0cm},clip]{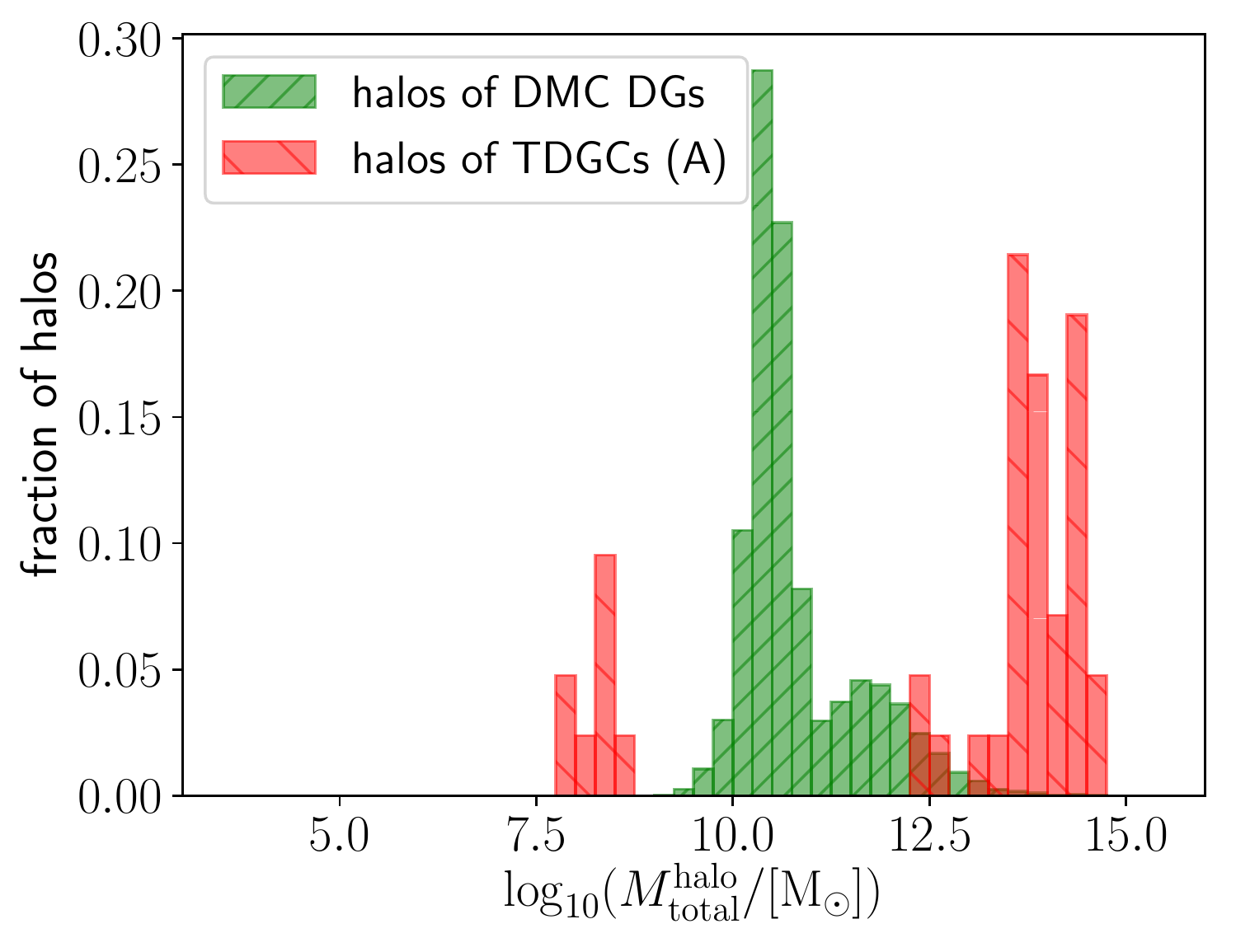}
		\includegraphics[width=\columnwidth,trim={0cm 0.00cm 0 0.2cm},clip]{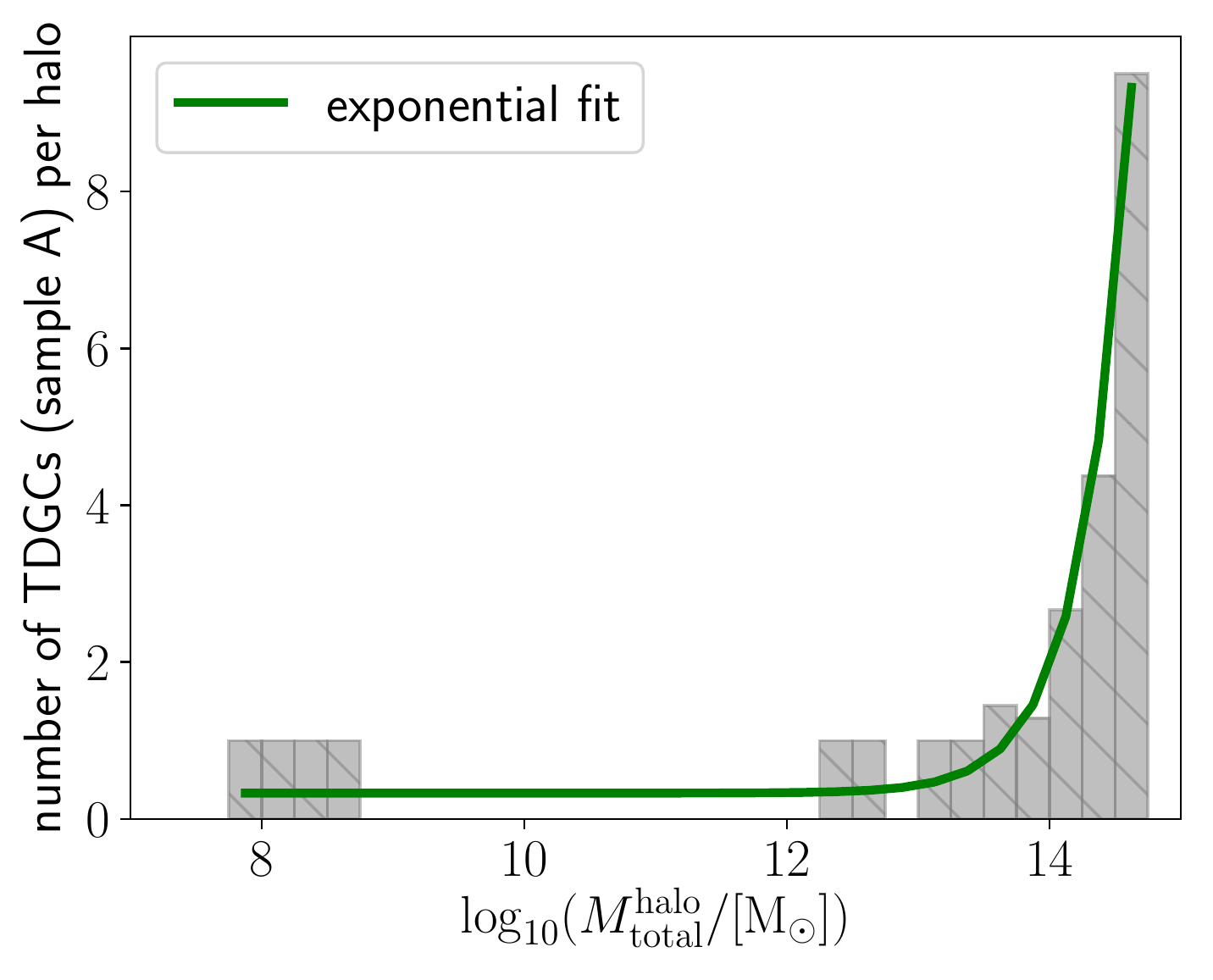}
		\caption{
			Top: Total host halo mass, $M_{\mathrm{total}}^{\mathrm{halo}}$, distribution of host halos in which at least one DMC DG (green) is embedded is shown in green and in which at least one TDGC of sample A is embedded is shown in red for redshift $z=0$. \newline
			Bottom: Distribution of the number of TDGCs (sample A) per host halo at redshift $z=0$. The histogram is fitted with an exponential function (solid green line) given by Eq. \ref{equation_halo_power_fit}. The fitting parameters are listed in the text. The histograms have a bin width of $\log_{10}(\Delta M_{\mathrm{total}}^{\mathrm{halo}}/[\rm{M_{\odot}}]) = 0.25$.}
		\label{histogram_totalmass_halo}
	\end{figure}

	The host halos of TDGCs (sample A) are studied in Fig.~\ref{histogram_totalmass_halo}, which shows the total host halo mass distribution of host halos which contain at least one DMC DG (top; green) and/ or at least one TDGC (top; red) at redshift $z=0$. Most of the TDGCs appear in host halos with a total halo mass range of $M_{\mathrm{total}}^{\mathrm{halo}} \approx 10^{12} - 4.6 \times 10^{14} \, \rm{M_{\odot}}$, but a small number of TDGCs can also be found in the $7.8 \times 10^{7} - 2.7 \times 10^{8} \, \rm{M_{\odot}}$ total halo mass regime. In fact, $97$ TDGCs belong to $42$ different host halos and the most massive host halo ($M_{\mathrm{total}}^{\mathrm{halo}}=~4.6 \times 10^{14} \, \rm{M_{\odot}}$) possesses the highest number of TDGCs ($n_{\mathrm{TDGCs}}=~12$). Only a very small number of TDGCs are found in low-mass host halos. The number of TDGCs per number of host halos within a given mass bin is shown in Fig.~\ref{histogram_totalmass_halo} (bottom). The number of TDGCs per host halo increases with the total host halo mass. The distribution is fitted with an exponential function of the form, 
	
	\begin{equation}
	\begin{aligned}
	\tilde{\rho}_{\mathrm{TDGCs}} \bigg( \log_{10} \bigg( \frac{M_{\mathrm{total}}^{\mathrm{halo}}}{\rm{M_{\odot}}} \bigg) \bigg) =~ a + b ^{\bigg[ \log_{10} \bigg( \frac{M_{\mathrm{total}}^{\mathrm{halo}}}{\rm{M_{\odot}}} \bigg) - c \bigg]} \, ,
	\end{aligned}
	\label{equation_halo_power_fit}
	\end{equation}
	with the fitting parameters
	\begin{equation*}
		\begin{aligned}
			&a = 0.331 \pm 0.098, \\
			&b = 16.0 \pm 4.5, \\
			&\log_{10} \bigg( \frac{c}{\rm{M_{\odot}}} \bigg) = 13.832 \pm 0.077,
		\end{aligned}
		\label{parameters_halo_power_fit}
	\end{equation*}
	where $\tilde{\rho}_{\mathrm{TDGCs}}(\log_{10}(M_{\mathrm{total}}^{\mathrm{halo}})) d \log_{10}(M_{\mathrm{total}}^{\mathrm{halo}})=d \tilde{N}$ is the number of TDGCs per host halo with a mass in the range $\log_{10}(M_{\mathrm{total}}^{\mathrm{halo}})$ to $\log_{10}(M_{\mathrm{total}}^{\mathrm{halo}})+d\log_{10}(M_{\mathrm{total}}^{\mathrm{halo}})$.
	
	The higher probability for galactic interactions and mergers in massive host halos can explain the increase of TDGCs per host halo with total host halo mass, consistent with these dark matter-free galaxies indeed being TDGs. This is qualitatively consistent with the analysis by \cite{Okazaki_2000}.

	\begin{table*}
		\centering
		\caption{Physical properties of DMC DG and TDGC samples at redshift $z=0$.}
		\label{table_physical_propertiers}
		\begin{tabular}{lllllllll} \hline 
			properties & sample & $M_{\mathrm{dm}}/M_{\mathrm{baryonic}}$ & minimum & maximum & median & mean \\ \hline \hline 
			$M_{\mathrm{stellar}} \, [\rm{M_{\odot}}]$ & DMC DGs & $> 0$ & $5.0 \times 10^{7}$ & $1.0 \times 10^{9}$ & $1.4 \times 10^{8}$ & $2.4 \times 10^{8}$ \\
			& DM-rich DGs & $\geq 1$ & $5.0 \times 10^{7}$ & $1.0 \times 10^{9}$ & $1.4 \times 10^{8}$ & $2.4 \times 10^{8}$ \\
			& DM-poor DGs & $< 1$ & $5.3 \times 10^{7}$ & $8.9 \times 10^{9}$ & $1.4 \times 10^{8}$ & $3.6 \times 10^{8}$ \\ 
			& TDGCs (A)  & $0$ & $5.0 \times 10^{7}$ & $3.1 \times 10^{9}$ & $1.5 \times 10^{8}$  & $3.1 \times 10^{8}$ \\ 
			& TDGCs (B) & $0$ & $5.1 \times 10^{5}$ & $4.4 \times 10^{8}$ & $2.9 \times 10^{6}$  & $1.0 \times 10^{7}$ \\ 
			& TDGCs (C) & $0$ & $5.7 \times 10^{7}$ & $4.3 \times 10^{8}$ & $1.1 \times 10^{8}$  & $1.6 \times 10^{8}$ \\\hline 
			$ M_{\mathrm{gas}} \, [\rm{M_{\odot}}]$ & DMC DGs & $> 0$ & $0.0$ & $2.2 \times 10^{10}$ & $1.8 \times 10^{9}$ & $2.6 \times 10^{9}$ \\
			& DM-rich DGs & $\geq 1$ & $0.0$ & $2.2 \times 10^{10}$ & $1.8 \times 10^{9}$ & $2.6 \times 10^{9}$ \\
			& DM-poor DGs & $< 1$ & $0.0$ & $8.4 \times 10^{8}$ & $0.0$ & $1.0 \times 10^{8}$ \\
			& TDGCs (A)  & $0$ & $0.0$ & $1.5 \times 10^{9}$ & $0.0$  & $6.2 \times 10^{7}$ \\ 
			& TDGCs (B) & $0$ & $5.4 \times 10^{7}$ & $1.5 \times 10^{9}$ & $2.0 \times 10^{8}$  & $2.5 \times 10^{8}$\\
			& TDGCs (C) & $0$ & $1.1 \times 10^{8}$ & $1.5 \times 10^{9}$ & $4.6 \times 10^{8}$  & $5.9 \times 10^{8}$ \\\hline 
			$\kappa_{\mathrm{rot}}$ & DMC DGs & $> 0$ & $0.19$ & $0.75$ & $0.40$ & $0.42$ \\
			& DM-rich DGs & $\geq 1$ & $0.19$ & $0.75$ & $0.40$ & $0.42$ \\
			& DM-poor DGs & $< 1$ & $0.28$ & $0.46$ & $0.34$ & $0.35$\\ 
			& TDGCs (A)  & $0$ & $0.23$ & $0.60$ & $0.37$  & $0.38$\\
			& TDGCs (B) & $0$ & - & - & -  & -\\
			& TDGCs (C) & $0$ & $0.35$ & $0.55$ & $0.43$  & $0.44$ \\\hline 
			$ v_{\mathrm{disp}} \, [\rm{km \, s^{-1}}]$ & DMC DGs & $> 0$ & $4.0$ & $52$ & $27$ & $26$ \\
			& DM-rich DGs & $\geq 1$ & $6.0$ & $52$ & $27$ & $26$ \\
			& DM-poor DGs & $< 1$ & $4.0$ & $19$ & $9.7$ & $10$\\
			& TDGCs (A)  & $0$ & $3.7$ & $35$ & $7.8$ & $9.6$ \\
			& TDGCs (B) & $0$ & $1.3$ & $26$ & $5.4$ & $5.8$ \\
			& TDGCs (C) & $0$ & $7.2$ & $26$ & $13$  & $14$ \\\hline 
			$\psi_{\mathrm{sfr}} \, [\rm{M_{\odot}\, yr^{-1}}]$ & DMC DGs & $> 0$ & $0.0$ & $0.59$ & $0.0031$ & $0.013$\\
			& DM-rich DGs & $\geq 1$ & $0.0$ & $0.59$ & $0.0031$ & $0.013$\\
			& DM-poor DGs & $< 1$ & $0.0$ & $0.049$ & $0.0$ & $0.0048$\\ 
			& TDGCs (A)  & $0$ & $0.0$ & $0.95$ & $0.0$  & $0.027$\\ 
			& TDGCs (B) & $0$ & $0.0$ & $0.95$ & $0.00066$  & $0.011$ \\
			& TDGCs (C) & $0$ & $0.0028$ & $0.95$ & $0.20$  & $0.26$ \\\hline 
			$ Z_{\mathrm{stellar}}$ & DMC DGs & $> 0$   & $0.00040$ & $0.052$ & $0.0016$ & $0.0021$\\
			& DM-rich DGs & $\geq 1$ & $0.00040$ & $0.029$ & $0.0016$ & $0.0021$\\
			& DM-poor DGs & $< 1$ & $0.0084$ & $0.052$ & $0.028$ & $0.029$\\ 
			& TDGCs (A)  & $0$ & $0.0089$ & $0.052$ & $0.019$  & $0.022$\\ 
			& TDGCs (B) & $0$ & $0.0$ & $0.045$ & $0.0030$  & $0.0047$ \\
			& TDGCs (C) & $0$ & $0.0090$ & $0.045$ & $0.015$  & $0.021$ \\\hline 
			$Z_{\mathrm{gas}}$ & DMC DGs & $> 0$ & $0.0$ & $0.025$ & $0.0018$ & $0.0022$\\
			& DM-rich DGs & $\geq 1$ & $0.0$ & $0.025$ & $0.0018$ & $0.0022$ \\
			& DM-poor DGs & $< 1$ & $0.0$ & $0.018$ & $0.0$ & $0.0028$ \\ 
			& TDGCs (A) & $0$ & $0.0$ & $0.053$ & $0.0$  & $0.0030$\\
			& TDGCs (B) & $0$ & $0.0$ & $0.053$ & $0.0033$  & $0.0052$\\
			& TDGCs (C) & $0$ & $0.010$ & $0.053$ & $0.022$  & $0.027$ \\\hline 
			$t_{\text{age}} \, [\rm{Gyr}]$ & DMC DGs & $> 0$ & - & - & - & -\\
			& DM-rich DGs & $\geq 1$ & $7.29$ & $13.5$ & $12.8$ & $12.7$ \\
			& DM-poor DGs & $< 1$ & $1.8$ & $13.3$ & $9.2$ & $8.9$ \\ 
			& TDGCs (A) & $0$ & $0.47$ & $11.7$ & $8.0$  & $7.6$\\
			& TDGCs (B) & $0$ & $0.0$ & $12.0$ & $1.0$  & $1.5$\\
			& TDGCs (C) & $0$ & - & - & -  & - \\\hline 
		\end{tabular}
		\tablefoot{Listed are the stellar mass, $M_{\mathrm{stellar}}$, gas mass, $M_{\mathrm{gas}}$, kinematical morphological parameter, $\kappa_{\mathrm{rot}}$, 1D velocity dispersion of all the member particles/cells, $v_{\mathrm{disp}}$,  star formation rate, $\psi_{\mathrm{sfr}}$, and the stellar and gas mass-weighted average metallicities, $Z_{\mathrm{stellar}}= (M_{\mathrm{>He}}/M_{\mathrm{tot}})_{\mathrm{stellar}}$ and $Z_{\mathrm{gas}}= (M_{\mathrm{>He}}/M_{\mathrm{tot}})_{\mathrm{gas}}$, where $M_{\mathrm{>He}}$ is the mass of all elements above Helium (only cells within twice the stellar half-mass radius are considered), and the age, $t_{\mathrm{age}}$, of the oldest stellar particle within a dwarf galaxy identified at redshift $z=0$. The metallicities of TDGCs are analyzed in more detail in Appendix \ref{sec:appendix_metallicity}.}
	\end{table*}
	
	\subsection{Radius-mass relation}
	\label{sec:MR_BTF}
	
\begin{figure}
		\centering
		\includegraphics[width=\linewidth]{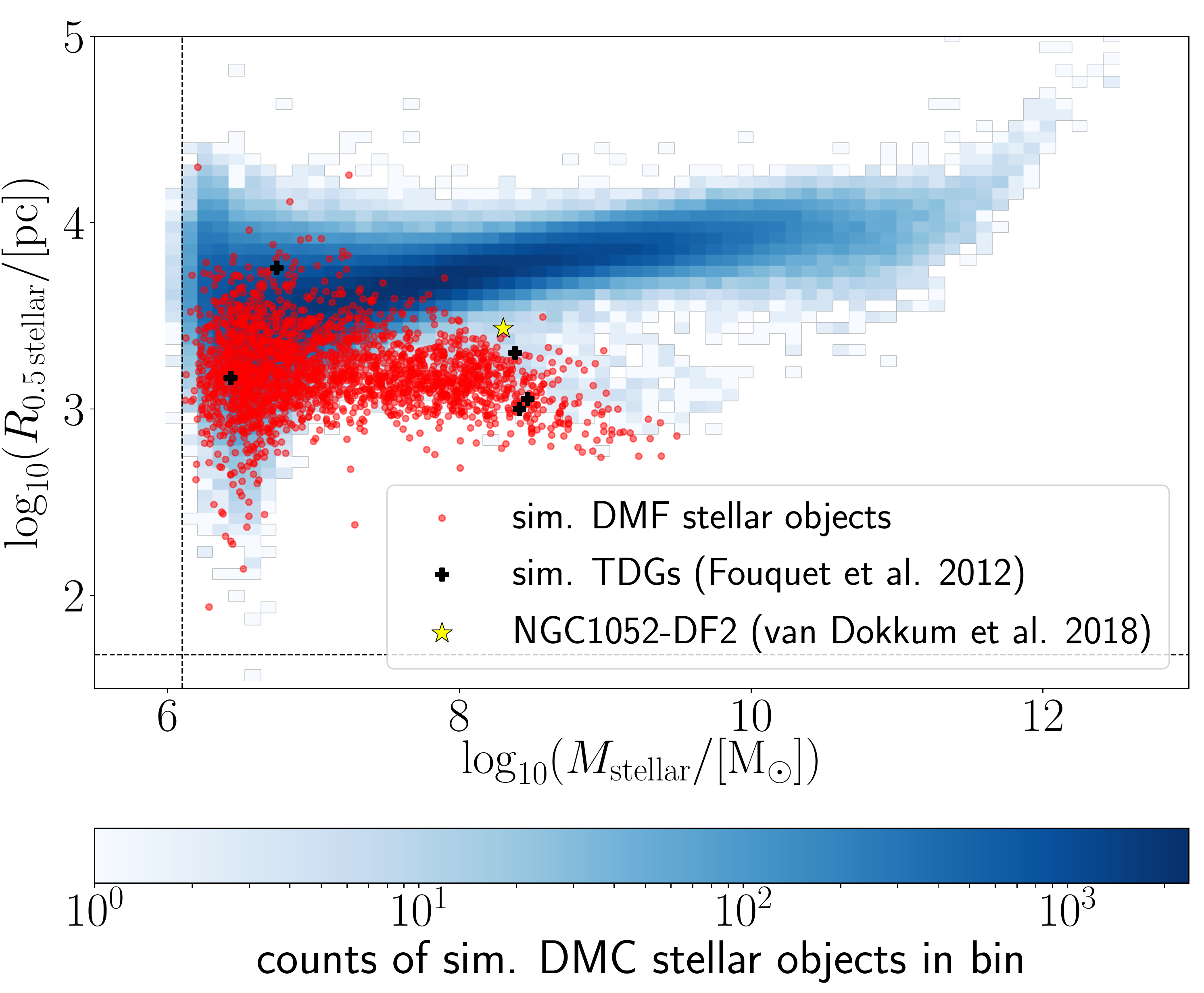} 
		\caption{Proper radius containing half of the stellar mass, $R_{\mathrm{0.5 \, stellar}}$, as a function of the stellar mass, $M_{\mathrm{stellar}}$, of simulated stellar objects at redshift $z=0$. Blue  bins are DMC and red dots are DMF stellar objects. The stellar masses and the total half-mass radii of simulated TDGs by \cite{Fouquet_2012} are shown as black crosses. A few DMC stellar objects can also be found in the regions of DMF stellar objects. The properties of these DMC stellar objects are studied in Fig.~\ref{figure_ratio_Mdm_to_Mstellar}. The dashed vertical and horizontal lines indicate the initial baryonic matter mass of a particle ($1.26 \times 10^{6} \, \rm{M_{\odot}}$) and the smallest fiducial cell size ($48 \, \rm{pc}$) of the Illustris-1 run, respectively. Subhalos with a stellar half-mass radius below the cell resolution are not shown in the plots. The yellow star shows the position of NGC 1052-DF2 with $M_{\mathrm{stellar}} = 2 \times 10^{8} \, \rm{M_{\odot}}$ and a 3D deprojected half-light radius of $2.7 \, \rm{kpc}$ \citep{vDokkum_2018b}.}
		\label{S_massinhalfrad_stellar_vs_S_halfmassrad_stellar_simulation}
	\end{figure}

\begin{figure}
		\centering
		\includegraphics[width=\linewidth]{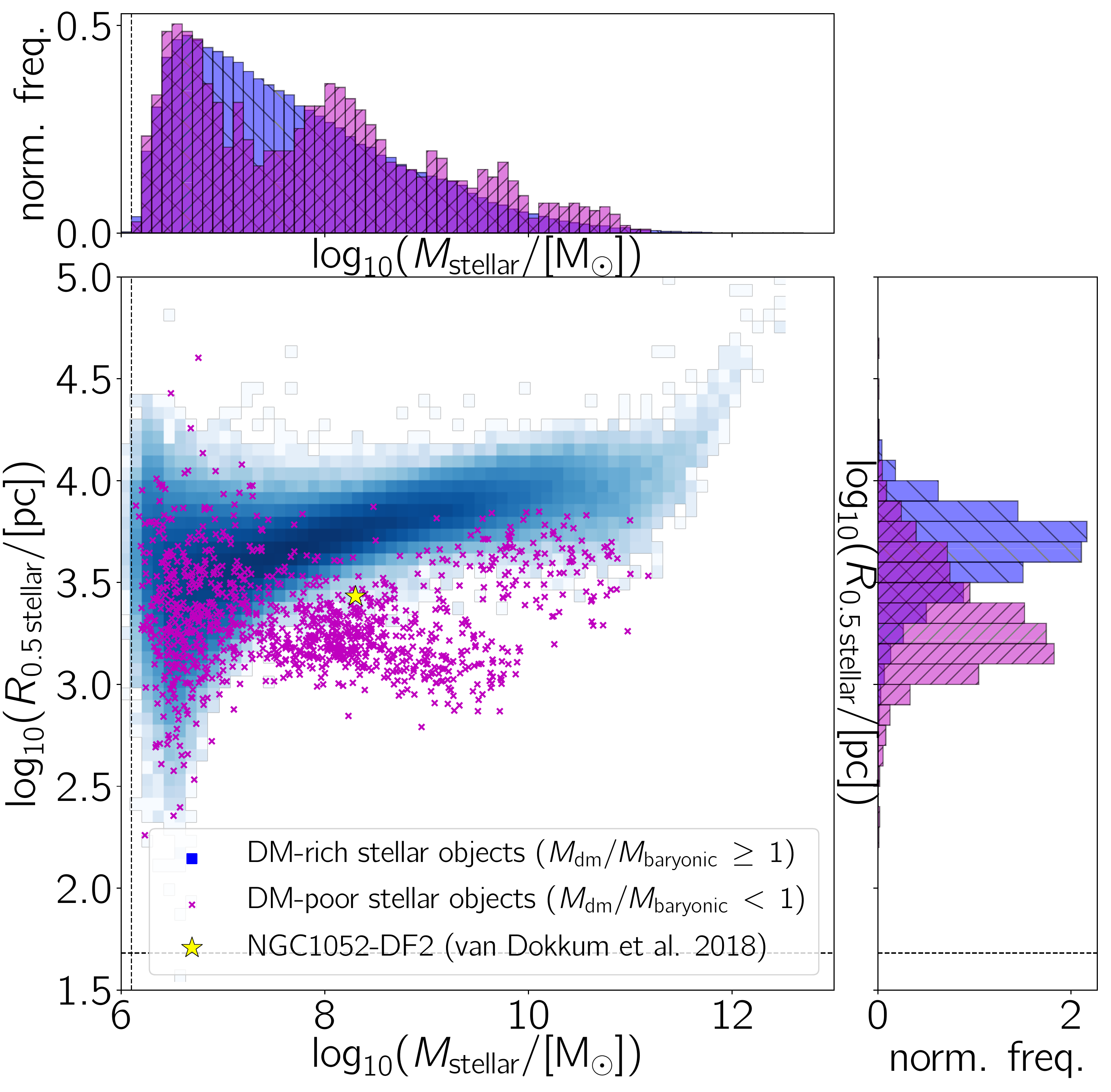}
		\caption{Proper radius containing half of the stellar mass, $R_{\mathrm{0.5 \, stellar}}$, as a function of the stellar mass, $M_{\mathrm{stellar}}$, of simulated DMC stellar objects at redshift $z=0$. Dark matter-containing stellar objects are separated in dark matter-rich ($M_{\mathrm{dm}}/M_{\mathrm{baryonic}} \geq 1$; blue bins) and dark matter-poor ($M_{\mathrm{dm}}/M_{\mathrm{baryonic}}<1$; purple crosses) types. The yellow star shows the position of NGC 1052-DF2 with $M_{\mathrm{stellar}} = 2 \times 10^{8} \, \rm{M_{\odot}}$ and a 3D deprojected half-light radius of $2.7 \, \rm{kpc}$ \citep{vDokkum_2018b}. The dashed vertical and horizontal lines indicate the initial baryonic matter mass of a particle ($1.26 \times 10^{6} \, \rm{M_{\odot}}$) and the smallest fiducial cell size ($48 \, \rm{pc}$). Subhalos with a stellar half-mass radius below the cell resolution are not shown in the plots. The histograms are normalized such that the total area is equal to 1.0 and such that they have bin widths of $\log_{10}(\Delta M_{\mathrm{stellar}}/[\mathrm{M_{\odot}}])=0.10$ and $\log_{10}(\Delta R_{\mathrm{0.5 \, stellar}}/[\mathrm{pc}])=0.10$.}
		\label{figure_ratio_Mdm_to_Mstellar}
	\end{figure}

	According to the dual dwarf theorem, two types of dwarf galaxies must exist and they can be distinguished based on their stellar masses and radii \citep{Kroupa_2010, Kroupa_2012,Dabringhausen_2013}. In order to verify these predictions in the $\Lambda$CDM cosmological Illustris simulation, we show first the positions of DMC and DMF stellar objects in the radius--mass plane in Fig.~\ref{S_massinhalfrad_stellar_vs_S_halfmassrad_stellar_simulation}. Due to the cell resolution of the Illustris-1 simulation, a significant number of DMF stellar objects have a stellar half-mass radius below the resolution limit and in this sense are consistent with a radius equal to zero.\footnote{In the Illustris-1 simulation the smallest fiducial cell size $r_{\mathrm{cell}}^{\mathrm{min}}$ is $48 \, \rm{pc}$ and the minimum mass $m_{\mathrm{cell}}^{\mathrm{min}}$ of a cell is $0.15 \times 10^{5} \, \rm{M_{\odot}}$ \citep{Vogelsberger_2014a}.} These $1240$ subhalos are removed from the diagram; they could be UCDs \citep{Hilker_2007, Baumgardt_2008}. Dark matter-free and DMC stellar objects are clearly distributed differently in the radius--mass diagram. 
	In general, DMF stellar objects have smaller stellar half-mass radii than most of the galaxies with a nonzero dark matter component, confirming the prediction by \cite{Kroupa_2012}. However, a few objects with a nonvanishing dark matter mass and with stellar masses between $10^{7} \, \rm{M_{\odot}}$ and $10^{10} \, \rm{M_{\odot}}$ can also be found in the region of DMF stellar objects. The properties of these DMC stellar objects are discussed in Fig.~\ref{figure_ratio_Mdm_to_Mstellar}, wherein the populations of dark matter-poor and dark matter-rich DMC stellar objects in the radius--mass plane are plotted. Most DMC stellar objects in the region of DMF stellar objects have a dark matter-to-baryonic mass ratio $M_{\mathrm{dm}}/M_{\mathrm{baryonic}}<1$ and are thus dark matter-poor. 
	
\begin{figure*}
		\centering
		
		
		
		\includegraphics[width=155mm,trim={0cm 8.4cm 0 3.0cm},clip]{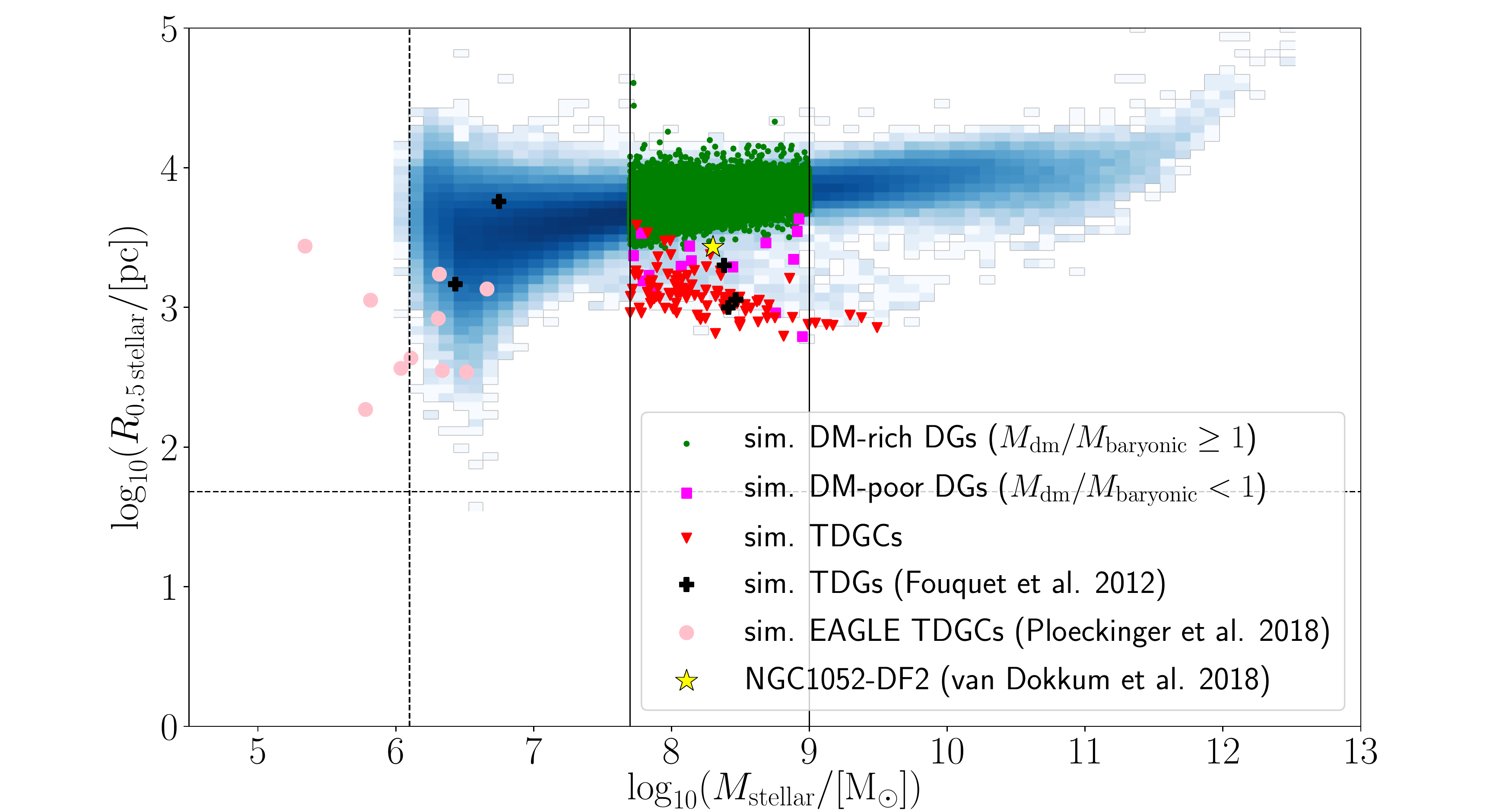}
		
		
		\includegraphics[width=155mm,trim={0cm 3.5cm 0 2.0cm},clip]{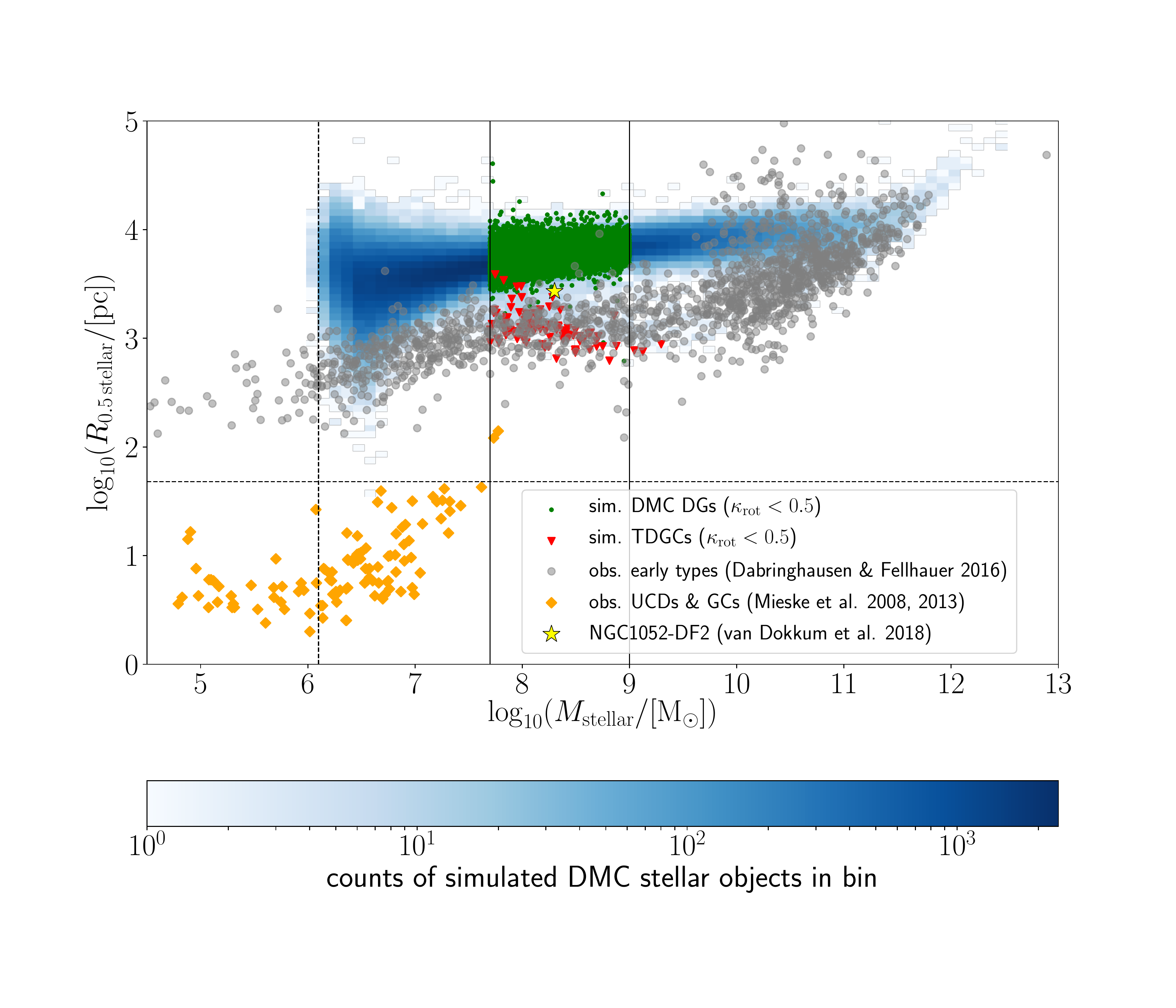}
		
		\vspace{+0.2cm}
		
		\caption{Top: Proper radius containing half of the stellar mass, $R_{\mathrm{0.5 \, stellar}}$, as a function of the stellar mass, $M_{\mathrm{stellar}}$, of simulated stellar objects at redshift $z=0$. Blue bins are DMC stellar objects, green dots are DM-rich DGs ($M_{\mathrm{dm}}/M_{\mathrm{baryonic}} \geq 1$), magenta squares are DM-poor DGs ($M_{\mathrm{dm}}/M_{\mathrm{baryonic}}<1$), and red triangles are TDGCs. The stellar masses and the total half-mass radii of simulated TDGs by \cite{Fouquet_2012} are shown as black crosses. Pink circles are simulated TDGCs identified in the EAGLE simulations by \cite{Ploeckinger_2018}. Bottom: Masses and radii of the simulated stellar objects compared with the 3D deprojected half-light radii of observed galaxies. The colors here refer to the same objects as in the top panel, but only dispersion-dominated ($\kappa_{\mathrm{rot}}<0.5$) TDGCs and DMC DGs are shown. Gray circles are early-type galaxies from faint dwarf spheroidals to giant ellipticals taken from \cite{Dabringhausen_2016}. Orange diamonds are ultra compact dwarf galaxies (UCDs) and globular clusters (GCs) taken from \cite{Mieske_2008, Mieske_2013}. The stellar masses of UCDs and GCs are calculated by assuming a constant stellar mass-to-light ratio in the V-band of $2.5 \, \rm{M_{\odot}/L_{\mathrm{\odot}}^{\mathrm{V}}}$. The yellow star shows the position in the radius--mass plane of the ultra-diffuse galaxy NGC 1052-DF2, which has $M_{\mathrm{stellar}} = 2 \times 10^{8} \, \rm{M_{\odot}}$ and a 3D deprojected half-light radius of $2.7 \, \rm{kpc}$ \citep{vDokkum_2018b}. 
			The dashed vertical and horizontal lines indicate the initial baryonic matter mass of a particle ($1.26 \times 10^{6} \, \rm{M_{\odot}}$) and the smallest fiducial cell size ($48 \, \rm{pc}$). Subhalos with a stellar half-mass radius below the cell resolution are not shown in the plots.
			The KS-test is applied for dwarf galaxies with stellar masses between $5 \times 10^{7} \, \rm{M_{\odot}}$ and $10^{9} \, \rm{M_{\odot}}$ marked by the two solid vertical black lines.}
		\label{S_massinhalfrad_stellar_vs_S_halfmassrad_stellar_DMCDGs_TDGCs}
	\end{figure*}

	Some of the stellar objects shown here are substructures with a separation to their host galaxies smaller than $10 \times R_{\mathrm{0.5 \, stellar}}^{\mathrm{host}}$ according to  Tables \ref{tab:TDGCs} and \ref{tab:DMCDGs}. 
	Therefore, we discuss the radius--mass diagram for TDGCs of sample A\footnote{In this section we only refer to TDGCs of sample A.} and DMC DGs in Fig.~\ref{S_massinhalfrad_stellar_vs_S_halfmassrad_stellar_DMCDGs_TDGCs}, which shows for the first time that the dual dwarf theorem is valid in a self-consistent $\Lambda$CDM simulation. It is worth noting that the independent simulations of galaxy-galaxy encounters (in a dark matter Universe) by \cite{Fouquet_2012} of TDG formation show these to have radii consistent with the DMF stellar objects and TDGCs in the Illustris simulation (upper panel of Fig.~\ref{S_massinhalfrad_stellar_vs_S_halfmassrad_stellar_DMCDGs_TDGCs}).
	
\begin{figure}
		\centering
		\includegraphics[width=\linewidth]{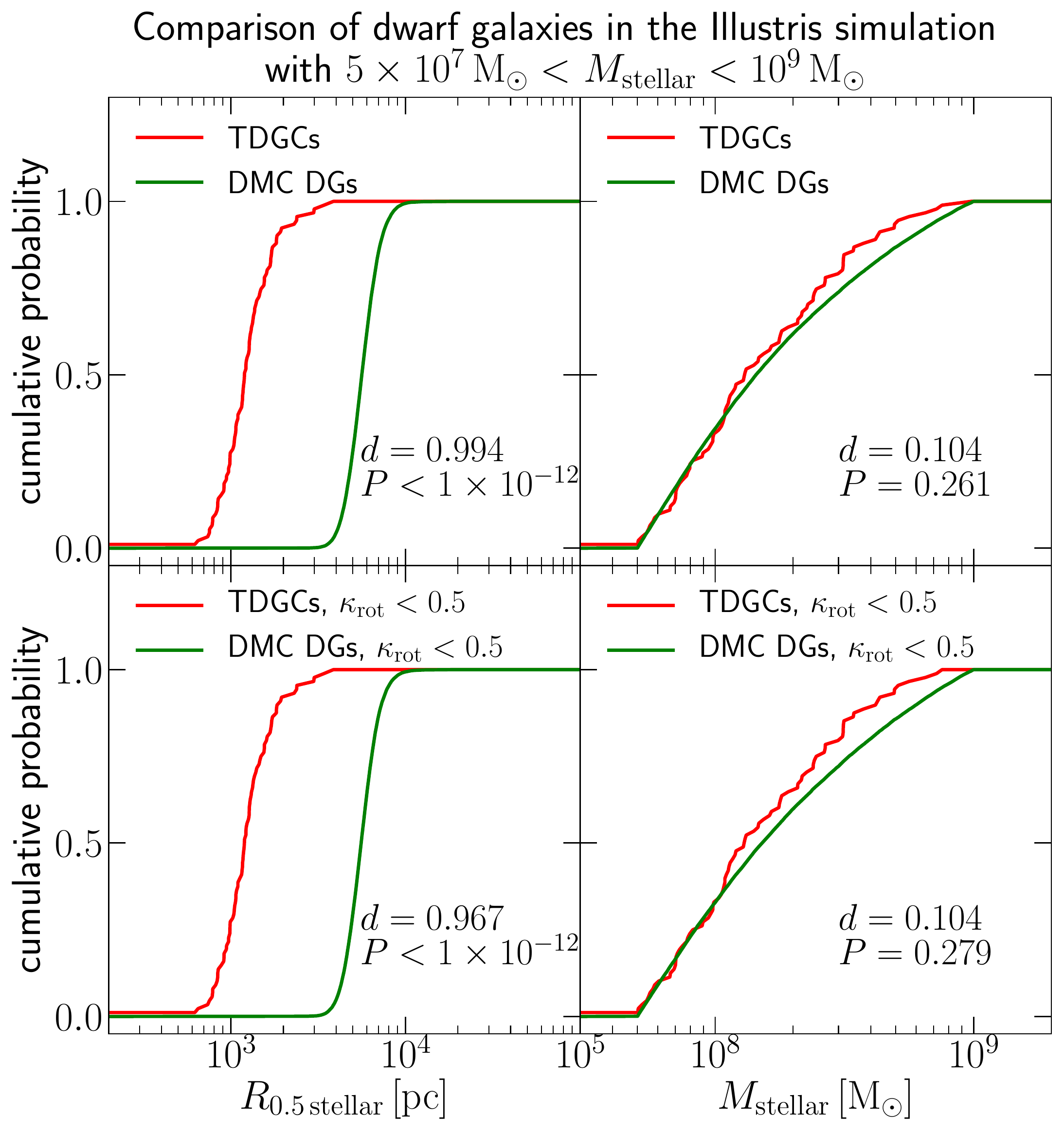}
		\caption{KS test for all and dispersion-dominated ($\kappa_{\mathrm{rot}} < 0.5$) simulated DMC DGs (green) and TDGCs (red) in the $5 \times 10^{7} - 10^{9} \, \rm{M_{\odot}}$ stellar mass regime at redshift $z = 0$.}
		\label{fig:KS_illustris}
	\end{figure}

	The Kolmogorov-Smirnov (KS) test is applied in Fig.~\ref{fig:KS_illustris} in order to decipher whether or not DMC DGs and TDGCs follow the same stellar mass and stellar half-mass radius distribution. 
	We only include simulated dwarf galaxies with stellar masses between $5 \times 10^{7} \, \rm{M_{\odot}}$ and $10^{9} \, \rm{M_{\odot}}$. The lower mass limit ensures that only well-resolved galaxies with a significant number of stellar particles are included in the statistical analysis. The P-values for the stellar half-mass radii are $<10^{-12}$, which quantitatively confirms the dual dwarf theorem. 
	Moreover, we find only $15$ DM-poor DGs that are possibly TDGs that captured dark matter particles from their host galaxy. These DM-poor DGs typically reside in the radius--mass plot between the DMC DG and TDGC branches. The probability of such a capture is very small because of the high velocity dispersion of dark matter particles and the shallow gravitational potential of TDGs, consistent with the small number of such DM-poor DGs.
	
	The simulated dispersion-dominated ($\kappa_{\mathrm{rot}} < 0.5$) galaxies formed in a $\Lambda$CDM framework are compared with observational data from \cite{Dabringhausen_2016} (early type galaxies) and \cite{Mieske_2008, Mieske_2013} (UCDs and GCs). The data for UCDs and GCs only include dynamical masses but not stellar masses. Therefore, we estimate the stellar mass by assuming a constant stellar mass-to-light ratio of $2.5 \, \rm{M_{\odot}/L_{\odot}^{\mathrm{V}}}$ in the V-band. The 2D effective radii, $R_{\mathrm{e}}$, are converted into 3D deprojected half-light radii, $R_{\mathrm{0.5 \, light}}$, by multiplying them by a factor of $4/3$ \citep[see Appendix B in][]{Wolf_2010}. 
	
\begin{figure}
		\centering
		\includegraphics[width=\linewidth]{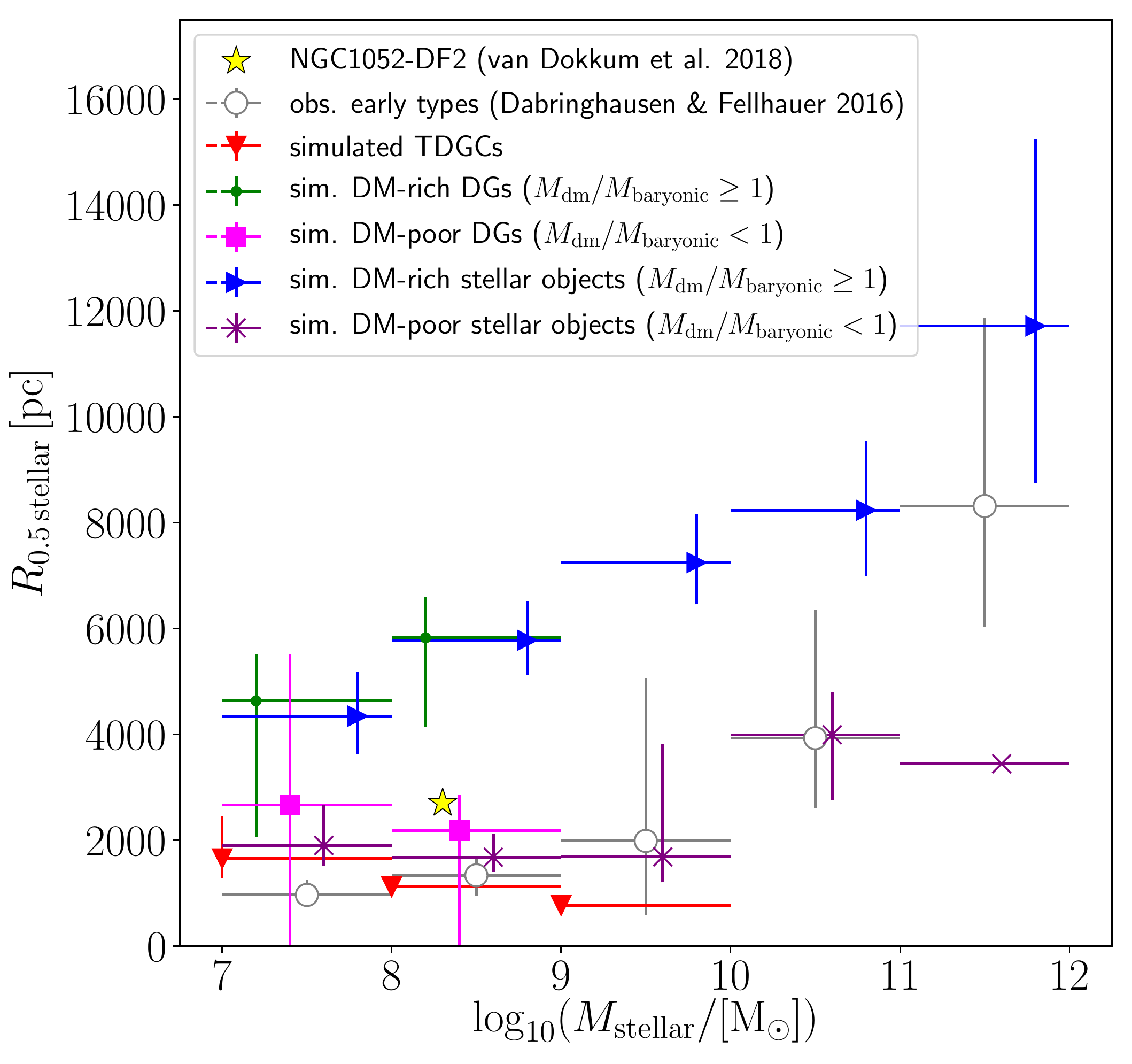}
		\caption{Median, first, and third quartile of simulated stellar half-mass radii, $R_{\mathrm{0.5 \, stellar}}$, of dispersion-dominated ($\kappa_{\mathrm{rot}}<0.5$) objects and 3D deprojected half-light radii, $R_{\mathrm{0.5 \, light}}$, of observed early type galaxies for different stellar mass bins ($10^{7}-10^{8} \, \rm{M_{\odot}}$, $10^{8}-10^{9} \, \rm{M_{\odot}}$, $10^{9}-10^{10} \, \rm{M_{\odot}}$, $10^{10}-10^{11} \, \rm{M_{\odot}}$, and $10^{11}-10^{12} \, \rm{M_{\odot}}$). 
			Blue right triangles are DM-rich stellar objects, purple crosses are DM-poor stellar objects, green dots are DM-rich DGs, magenta squares are DM-poor DGs, and red triangles are TDGCs (see Table~\ref{tab:samples}). Gray open circles are observed early-type galaxies taken from \citet{Dabringhausen_2016}. The yellow star shows the position of NGC 1052-DF2 with $M_{\mathrm{stellar}} = 2 \times 10^{8} \, \rm{M_{\odot}}$ and a 3D deprojected half-light radius of $2.7 \, \rm{kpc}$ \citep{vDokkum_2018b}. The medians of simulated and observed galaxies for different mass ranges are listed in Table~\ref{table_massradius_simulated_observed}.}
		\label{figure_statistics}
	\end{figure}

	The bottom panel of Fig.~\ref{S_massinhalfrad_stellar_vs_S_halfmassrad_stellar_DMCDGs_TDGCs} demonstrates that UCDs and GCs are separated from early-type galaxies in the radius--mass diagram and are found below the spatial resolution limit of the Illustris-1 run. The yellow star in the radius--mass diagram represents the ultra-diffuse galaxy NGC 1052-DF2, which has $M_{\mathrm{stellar}} = 2 \times 10^{8} \,\rm{M_{\odot}}$ and an effective radius along the major axis of $R_{\mathrm{e}}=2.2 \, \rm{kpc}$ \citep{vDokkum_2018b} corresponding to a 3D deprojected half-light radius of $2.7 \, \rm{kpc}$. Especially remarkable is the large effective radius of NGC 1052-DF2 compared to the sample of observed early-type galaxies from \cite{Dabringhausen_2016}. 
	The median values of simulated stellar half-mass radii and 3D deprojected half-light radii of observed galaxies for different stellar mass ranges are listed in Table~\ref{table_massradius_simulated_observed} and are shown in Fig.~\ref{figure_statistics}. The median of simulated TDGCs is within the first and third quartiles of observed half-light radii for galaxies with stellar masses between $10^{8} \, \rm{M_{\odot}}$ and $10^{10} \, \rm{M_{\odot}}$. However, the observed galaxy NGC 1052-DF2 is not within the first and third quartiles of simulated TDGCs and DM-rich DGs. 
	
	A series of KS tests are performed to decipher whether or not the stellar masses and radii of observed galaxies follow the same distribution as simulated dwarf galaxies. The full sample from \cite{Dabringhausen_2016} is observationally biased, such that different types of galaxies can be over- or under-represented resulting in an incorrect stellar mass function of galaxies. In contrast to that, the sample from the Illustris simulation includes all formed galaxies without any mass, luminosity, or radius restrictions except for the resolution limits. Therefore, we choose a statistically fair subsample from the catalog by \cite{Dabringhausen_2016}, which includes all galaxies of the Fornax, the Hydra, and the Centaurus cluster catalogs with $M_{\mathrm{stellar}}>5 \times 10^{7} \, \rm{M_{\odot}}$. These catalogs include dwarf galaxies as well as large galaxies such that these catalogs sample the observed galaxy luminosity and stellar mass function of galaxies over a wide range.
	
\begin{figure}
		\centering
		\includegraphics[width=\linewidth]{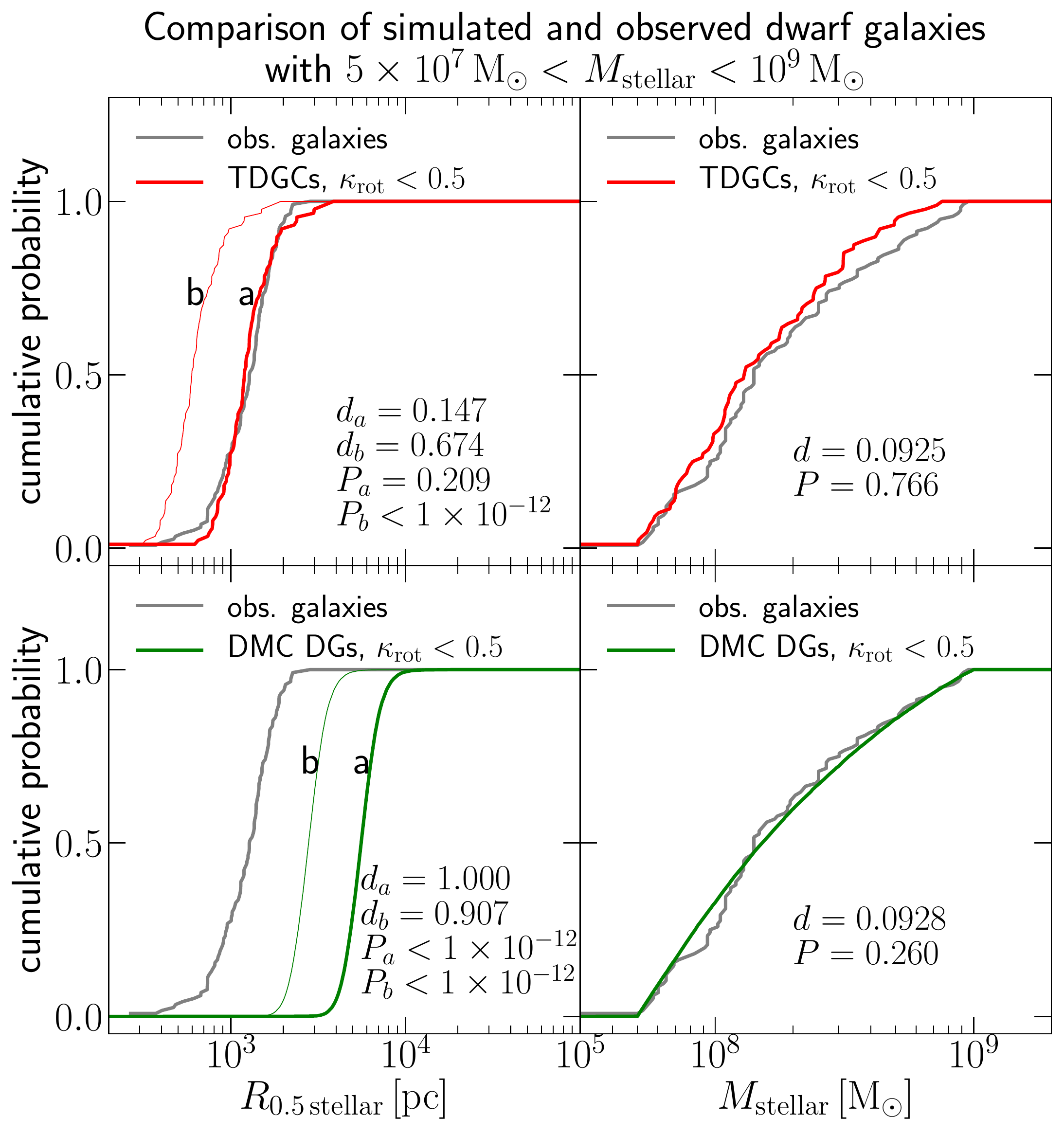}
		\caption{KS test for observed late-type galaxies (gray) and dispersion-dominated ($\kappa_{\mathrm{rot}} < 0.5$) simulated (red, green) DGs with stellar masses between $5 \times 10^{7} \, \rm{M_{\odot}}$ and $10^{9} \, \rm{M_{\odot}}$. The thick lines and the displayed $d_{\mathrm{a}}$- and $P_{\mathrm{a}}$-values refer to the real stellar half-mass radius distribution of the Illustris-1 simulation. The thin lines and the $d_{\mathrm{b}}$- and $P_{\mathrm{b}}$-values refer to a distribution in which all radii in the Illustris-1 simulation are divided by two (see text). The observational data are a subset of the catalog from \cite{Dabringhausen_2016} including all galaxies from the Fornax, Hydra, and Centaurus cluster catalog with stellar masses between  $5 \times 10^{7} \, \rm{M_{\odot}}$ and $10^{9} \, \rm{M_{\odot}}$.}
		\label{fig:KS_illustris_observations}
	\end{figure}

	Since galaxy cluster surveys almost always include the central parts of the clusters where the massive galaxies tend to gather, dwarf galaxies can be under-represented in the observational sample. In order to remove this bias towards high stellar masses for observed galaxies we restrict the KS test to dwarfs with stellar masses between $5 \times 10^{7} \, \rm{M_{\odot}}$ and $10^{9} \, \rm{M_{\odot}}$ as shown in Fig.~\ref{fig:KS_illustris_observations}. We find that the stellar mass distribution for all simulated DMC DGs and TDGCs fits the observed stellar mass distribution with a P-value of $0.260$ and $0.766$, respectively. The P-value for the stellar half-mass radius distribution for DMC DGs is $<10^{-12}$, which means that it is virtually impossible that the simulated and observed radii can be described with the same distribution function. In contrast to that, the P-value obtained by comparing the stellar half-mass radius distributions of the observed dwarf galaxies with simulated TDGCs is $0.209$. This means that if the treatment of baryonic physics in the Illustris-1 simulations is a reasonable approximation of reality, then the observed (real) dE galaxies ought to be TDGs. This conclusion was reached independently by \citet{Okazaki_2000}.
	
	\cite{Pillepich_2018} compared the galaxy sizes in the Illustris simulation and in the Illustris TNG (The Next Generation) simulation.\footnote{\url{http://www.tng-project.org}}
	\footnote{The data of the TNG simulation project are not
		yet public available [02.06.2018].} These latter authors concluded that the TNG simulation produces stellar half-mass radii two times smaller than in the Illustris simulation for galaxies $M_{\mathrm{stellar}} < 10^{10} \, \rm{M_{\odot}}$, which is caused by a modification of the treatment of galactic winds. Although the new galaxy physics model improves the simulated galaxy sizes, a mismatch between stellar half-mass radii is still present in Illustris TNG. Therefore, we not only compare the observed radius distribution with the radius distributions directly from the Illustris simulation, but also with the distributions that follow when every radius is divided by two. The P-values of the KS test in Fig.~\ref{fig:KS_illustris_observations} (red and green thin lines) for galaxies that are twice as compact as the original Illustris data are $<10^{-12}$ for both DMC DGs and TDGCs. Interestingly, simulated TDGCs become more compact than the observed ones when their radii are divided by two.  However, since the Illustris TNG data are not yet publicly available, we do not know at present whether the dark matter-poor and dark matter-free galaxies are indeed also more compact in the Illustris TNG simulation than in the Illustris simulation. Here we assume that all galaxies in the Illustris TNG simulation are more compact than in the Illustris simulation  by a
	factor of two. Nevertheless, the vast majority of the simulated galaxies (i.e., the dark matter-dominated galaxies) would still have radii that are  too large to be consistent with the observed radius distribution. 
	
	Summarizing, the observations do not clearly show different populations of galaxies based on their masses and radii, which was already reported by \cite{Dabringhausen_2013} using a sample they consider to be TDGs (TDG candidates) as discussed in their Section $2.2.1$. This is in disagreement with the Illustris simulation, which predicts two populations of dwarf galaxies in the radius--mass plane. The possible implications of this for $\Lambda$CDM cosmology are discussed in Section~\ref{sec:Discussion_Mass_radius_relation}. 
	
	\begin{table*}
		\centering
		\caption{Medians of simulated stellar half-mass radii, $R_{\mathrm{0.5 \, stellar}}$, for dispersion-dominated ($\kappa_{\mathrm{rot}} < 0.5$) DMC stellar objects, DMC DGs, and TDGCs samples for different stellar mass ranges.}
		\label{table_massradius_simulated_observed}
		\begin{tabular}{llllllllll} \hline
			sample & $M_{\mathrm{dm}}/M_{\mathrm{baryonic}}$  & $M_{\mathrm{stellar}} \, [\rm{M_{\odot}}]:$ & $10^{7}-10^{8}$ & $10^{8}-10^{9}$  & $10^{9}-10^{10}$ & $10^{10}-10^{11}$ & $10^{11}-10^{12}$  \\ \hline \hline
			DMC stellar objects & $> 0$ & $\braket{R_{\mathrm{0.5 \, stellar}}} \, [\rm{pc}]:$ &  $4342$ & $5770$ & $7232$ & $8213$ & $11 \,697$  \\
			DM-rich stellar objects & $\geq 1$ & & $4346$ & $5779$ & $7246$ & $8235$ & $11 \, 717$ \\
			DM-poor stellar objects & < 1& & 1903 & 1681 & 1687 & 3994 & 3445 \\ \hline 
			DMC DGs  & $> 0$ & & $4633$ & $582$ & -- & -- & -- \\
			DM-rich DGs & $\geq 1$ & & $4635$ & $5826$ & -- & -- & --\\ 
			DM-poor DGs & $< 1$ & & $2667$ & $2185$ & -- & -- & --\\ \hline 
			TDGCs & $0$ & & $1657$ & $1125$ & $773$ & -- & -- \\ \hline 
			observed & -- & $\braket{R_{\mathrm{0.5 \, light}}} \, [\rm{pc}]:$ &  $969$ & $1339$ & $1988$ & $3930$ & $8315$ \\ \hline 
		\end{tabular}
		\tablefoot{The medians of the 3D deprojected half-light radii, $R_{\mathrm{0.5 \, light}}$, of observed early type galaxies for different stellar mass ranges are given in the last row \citep{Dabringhausen_2016}. The statistical properties of simulated and observed galaxies are visualized in Fig.~\ref{figure_statistics}.}
	\end{table*}
	
	\subsection{Evolution of the number density of TDGCs across cosmic time}  \label{sec:Results_evolution}
\begin{figure}
		\centering
		\includegraphics[width=\columnwidth]{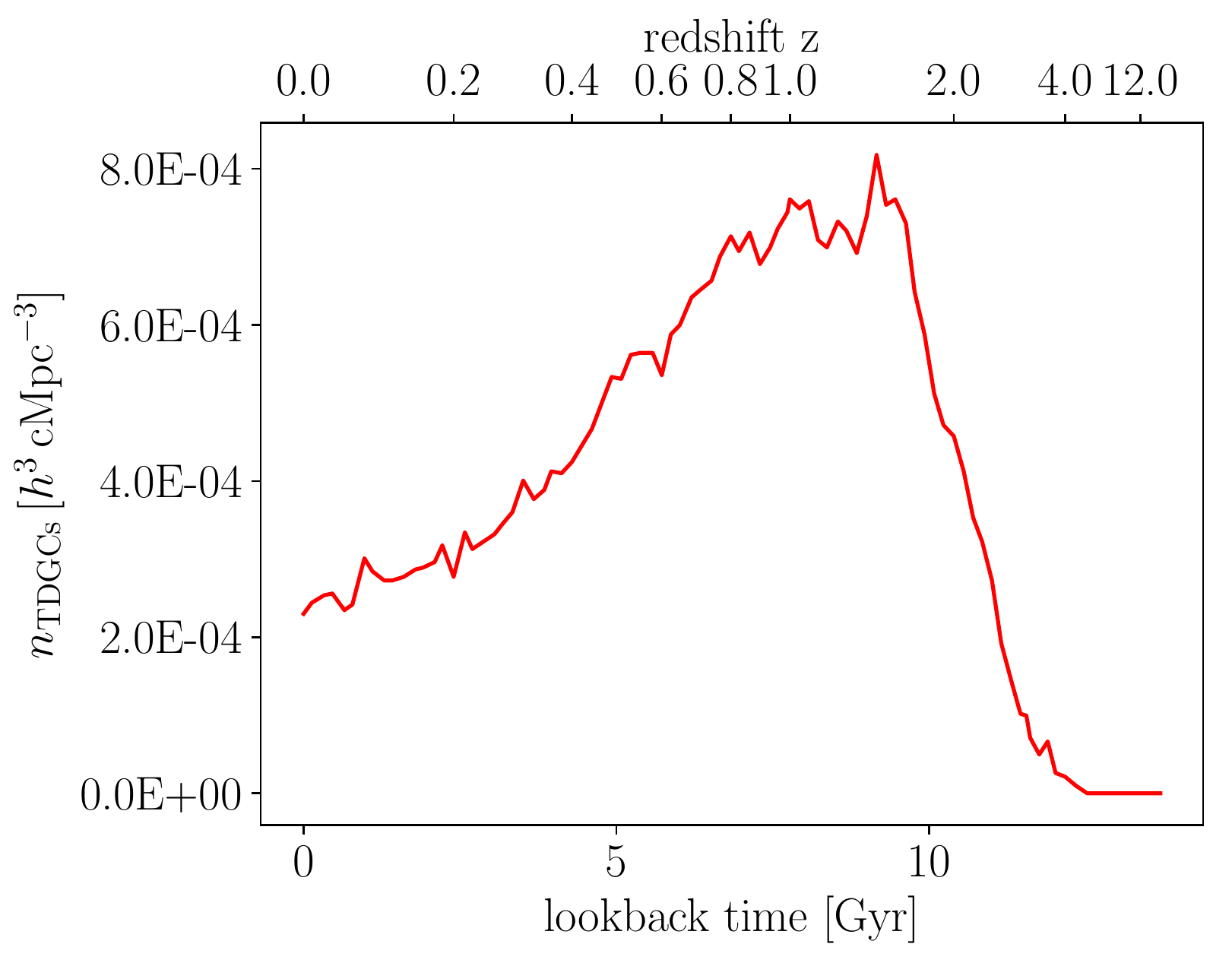}
		\caption{Time evolution of the co-moving number density of TDGCs,  $n_{\mathrm{TDGCs}}$, identified with the same selection criteria as for sample A. The $x$-axis shows the lookback time in gigayears (i.e., $0 \, \rm{Gyr}$ corresponds to the present time) and redshift $z$.}
		\label{time_evolution}
	\end{figure}

	Figure~\ref{time_evolution} shows the evolution in the co-moving number density, $n_{\mathrm{TDGCs}}$, of simulated TDGCs (sample A) over cosmic time. These TDGCs with $M_{\mathrm{stellar}} >5 \times 10^{7} \, \rm{M_{\odot}}$ are identified by the searching algorithm for the first time at redshift $z=4.7$ and therefore appear $0.752 \, \rm{Gyr}$ later than DMC stellar objects with a stellar mass of at least $5 \times 10^{7} \, \rm{M_{\odot}}$. This may indicate that the formation of TDGCs is triggered by the encounters of DMC galaxies once these DMC galaxies have grown sufficiently in mass through mergers to spawn TDGCs above a stellar mass threshold of $5 \times 10^{7} \, \rm{M_{\odot}}$. Less-massive TDGCs are most likely formed earlier, but cannot be resolved in the Illustris simulation. The number density of TDGCs increases up to redshift $z=1.4$, where a global maximum of $n_{\mathrm{TDGCs}}(z=1.4) = 8.2 \times 10^{-4} \, h^{3} \, \rm{cMpc^{-3}}$ is reached. Later on, the number density of TDGCs decreases in time. Since galaxies at higher redshifts were more gas-rich, metal-poor, and more dynamically active, TDGCs are formed efficiently through galaxy interactions resulting in an increase of the co-moving number density with decreasing redshift.
	
	\section{Discussion} \label{sec:Discussion}
	
	In this section we discuss the properties of TDGCs and DMC DGs in the Illustris simulation. The dual dwarf theorem and its implications for $\Lambda$CDM cosmology are considered.
	
	\subsection{Formation and evolution of TDGCs} \label{sec:discussion_DMF_galaxies}
	The highest-resolution run of the Illustris suite allows us to study the formation and evolution of TDGCs. We consider that baryonic substructures may be spurious objects or fragments within a galaxy and that TDGCs may be formed out of the gas of a disk galaxy during galactic interactions. Mergers of rotationally supported galaxies in dark matter halos occur in the Illustris simulation. Previous work has shown that TDGs form in such encounters \citep{Barnes_1992, Bournaud_2006,Wetzstein_2007,Fouquet_2012,Yang_2014} and thus it can be expected that they would also form in the self-consistent cosmological Illustris simulation. The verification of this theory would require following the merger tree of all TDGCs over cosmic time. However, none of the TDGCs at redshift $z=0$ are included in the merger trees provided by the Illustris team \citep{Rodriguez_2015} and backtracing all TDGCs by their particle data is very resource consuming. Therefore we have shown for some TDGCs that these subhalos have  indeed been formed due to tidal forces caused by galactic interactions (see Section~\ref{sec:formation_TDGCs}, Appendix \ref{sec:appendix_Images_TDGCs_DMCDGs}, and the movies in the supplementary information).
	
	Tidal dwarf galaxies lack dark matter due to the physics of their formation. 
	We point out that apart from galactic interactions, efficient cooling processes provided by the implemented galaxy-formation models of the Illustris simulation can also artificially trigger the formation of DMF stellar objects. Jeans instabilities depend on the mass and temperature of the molecular gas cloud. The collapse of a cloud is supported by an increase of the mass (at a given temperature) or a decrease of the temperature (at a given mass) \citep{Jeans_1902,Coles_2003}. Efficient cooling of great baryonic matter accumulations allows for the collapse of these structures without the need for high amounts of nonbaryonic matter. Cold accretion of gas clumps onto halos might also perhaps produce such DMF objects. The agreement of the properties of the TDGCs formed in Illustris-1 with independent work reporting the formation of TDGs \citep{Barnes_1992, Bournaud_2006,Wetzstein_2007,Fouquet_2012,Yang_2014,Ploeckinger_2018}, the shown formation scenarios in Section \ref{sec:formation_TDGCs}, and the applied $6$D phase-space halo finder on selected TDGCs (see Appendix \ref{sec:appendix_energy}) all together suggest that the TDGCs formed in Illustris-1 are physical. 
	
	By extracting the formation time of the oldest stellar particle within a dwarf galaxy identified at redshift $z=0$, we have shown that TDGCs and DM-poor DGs are typically younger than DM-rich DGs. This underlines that DM-rich DGs are formed in the early universe in contrast to TDGCs as expected from their different formation scenarios, since TDGs are being formed from the expelled gas from massive galaxies triggered by galactic encounters and interactions. Furthermore, gas-rich TDGCs are typically younger than gas-free TDGCs.
	
	We have shown that TDGCs with $M_{\mathrm{gas}}>5 \times 10^{7} \, \rm{M_{\odot}}$ and with at least one stellar particle (sample B) are typically more phase-space-correlated than DMC DGs. TDGCs with $M_{\mathrm{stellar}}>5 \times 10^{7} \, \rm{M_{\odot}}$ (sample A) are less phase-space-correlated than sample B but are still more so than DMC DGs. The difference is qualitatively consistent with sample A (gas-poor TDGCs) being older than sample B (gas-rich TDGCs). Gas-poor TDGs would have been stripped of their gas or would have consumed it and their orbits are likely perturbed due to later mergers of the hosting galaxy which are likely to destroy phase-space-correlated populations in the dark matter-based cosmological models \cite[see also][]{Kroupa_2015}. However, the small number of galactic systems of sample A hosting more than one TDGC  requires further study of this issue in order to produce any statistically robust conclusions about their phase-space correlation (see Section \ref{Results:angularmomentum}). In a more detailed analysis we also have to investigate if and how an initial phase-space correlation is affected by further galactic encounters and mergers. In the local Universe a large if not dominant fraction of the dwarf galaxies surrounding the MW, M31, and NGC 5128 (Centaurus A) are significantly phase-space-correlated \citep{Kroupa_2005,Metz_2007,Ibata_2013,Pawlowksi_2013,Ibata_2014,Mueller_2018, Pawlowski_2018}. This observed ubiquitous occurrence of disks or planes of satellites \citep{Ibata_2014, Pawlowski_2018} may thus imply an absence of such encounters in the real Universe.
	
	\subsection{Gas masses and star formation rates of TDGCs}
	\label{sec:discussion_gas_sfr}
	TDGCs are likely formed out of the stellar and gas reservoir of their host galaxies. We have shown that the amount of gas depends strongly on the applied selection criteria. Our main sample (sample A) includes $97$ TDGCs with $M_{\mathrm{stellar}}> 5 \times 10^{7} \, \rm{M_{\odot}}$, such that around $89$~percent are completely gas-free suggesting that a significant fraction of TDGCs have already converted their gas content to stars. The large fraction of gas-free TDGCs has a direct consequence on the star formation rate such that $90$~percent have no star formation. However, when we apply selection criteria similar to \cite{Ploeckinger_2018} we find a larger number of TDGCs (sample B, see Table~\ref{tab:TDGCs}). These are young and gas-rich TDGCs which have recently formed out of the gaseous disk of their host galaxies (see Sections \ref{sec:formation_TDGCs} and \ref{sec:Physical properties of TDGCs and DMC DGs}).
	
	TDGCs and DM-poor DGs ($M_{\mathrm{dm}}/M_{\mathrm{baryonic}} < 1$) are often more metal-rich and younger than DM-rich DGs ($M_{\mathrm{dm}}/M_{\mathrm{baryonic}} \geq 1$). This is consistent with the formation theory of TDGs and underlines that TDGs can also capture at least a small amount of dark matter particles (see Appendix \ref{sec:appendix_metallicity}).
	By back-tracing the particle identification numbers, one can decipher whether or not a TDGC in the Illustris simulation can indeed capture dark matter particles. Such events must be extremely rare given the weak gravitational potential of TDGCs, but it may be interesting to study this in the future. Indications of such a capture can be seen in Fig. \ref{fig:mass_evolution} in Section \ref{sec:formation_TDGCs}, but it is likely that these dark matter particles identified by the Subfind algorithm are just individual particles crossing the object.
	Nevertheless, the small number of DMC DGs with a dark-to-baryonic matter fraction smaller than one and their similar physical properties to TDGCs indicate that such DM-poor DGs are TDGs.  
	
	\subsection{Radius--mass relation} \label{sec:Discussion_Mass_radius_relation}
	According to the dual dwarf theorem, two different types of dwarf galaxies should exist in the mass range between about $10^{6} \, \rm{M_{\odot}}$ and $10^{10} \, \rm{M_{\odot}}$ \citep{Kroupa_2012,Dabringhausen_2013}. Although observed dEs and TDGs are indistinguishable in the radius--mass plane \citep{Dabringhausen_2013}, simulated TDGCs are clearly separated by being smaller than DMC DGs. By showing that TDGCs and DM-poor DGs are more compact than DM-rich DGs in the stellar mass range between $5 \times 10^{7} \, \rm{M_{\odot}}$ and $10^{9} \, \rm{M_{\odot}}$ we have verified the dual dwarf theorem  for the first time in a self-consistent $\Lambda$CDM simulation. The KS test underlines a statistically highly significant difference between the stellar-half mass distribution of TDGCs and DMC DGs. The P-value of the KS test is $<10^{-12}$. These results are consistent with the formation scenario of TDGs in the $\Lambda$CDM framework, which are understood to be formed naked without the help of a dark matter potential and ought to be therefore more compact than primordial dwarfs \citep{Kroupa_2012}. It is noteworthy that the observed physical stellar half-light radii more closely resemble simulated stellar half-mass radii of dark matter-free and -poor galaxies rather than of dark matter-dominated galaxies in the stellar mass regime of $5 \times 10^{7} \, \rm{M_{\odot}}$ and $10^{9} \, \rm{M_{\odot}}$ \citep{Dabringhausen_2013,Duc_2014}. Comparing the stellar half-mass radius distributions of dispersion-dominated TDGCs and DMC DGs from the Illustris-1 simulation with observed early-type galaxies gives a P-value of $0.209$ and $<10^{-12}$, respectively. The radii of TDGCs formed in the Illustris-1 simulation are confirmed by the independent simulations of \cite{Fouquet_2012}. The fact that the radius of TDGCs and DM-poor galaxies in Illustris-1 agree with the observed dE galaxies suggests that the latter are TDGs, as also concluded by \citet{Okazaki_2000} based on different arguments.
	
	The first results from the new Illustris TNG simulation have shown that a modification of the galactic wind model reduces the stellar half-mass radii by a factor of two for galaxies with $M_{\mathrm{stellar}}<10^{10} \, \rm{M_{\odot}}$ \citep{Pillepich_2018}. Nevertheless, we have shown in the present paper that even these current state-of-the-art cosmological hydrodynamical simulations cannot reproduce the observed galaxy sizes (see Section \ref{sec:MR_BTF}). 
	
	A further consistency test of the $\Lambda$CDM cosmology would be to study the positions on the baryonic Tully-Fisher relation (BTFR) of simulated dark matter-poor and dark matter-dominated galaxies. Dark matter-poor galaxies (i.e., TDGs) are thus expected to lie above the BTFR by having smaller rotation speeds than DMC galaxies of the same baryonic mass. The apparent absence of observed dwarf galaxies, some of which must be TDGs that lie above the BTFR, may pose a serious challenge for dark matter cosmology \citep{Kroupa_2012, Flores_2016}. This issue could be directly tested with confirmed old TDGs settled down to virial equilibrium as is likely with the observed TDGs identified by \cite{Duc_2014}. Observations show a very tight power-law correlation between the baryonic mass and the circular velocity for galaxies with baryonic masses between $10^{7} \, \rm{M_{\odot}}$ and  $5 \times 10^{11} \, \rm{M_{\odot}}$ \citep{McGaugh_2012,Lelli_2016}.
	However, a proper analysis requires the extraction of each model galaxy and the fitting of its rotation curve, meaning that this line of investigation needs to be postponed to a detailed analysis of the rotation curves of galaxies in the Illustris and EAGLE simulations.
	
	\section{Conclusion}\label{sec:Conclusion}
	
	We studied the physical properties of dwarf galaxies lacking dark matter in the $\Lambda$CDM Illustris simulation. In particular, we identified $3484$ stellar objects without any dark matter in the simulation volume of $(75 \, h^{-1} \, \rm{cMpc})^{3}$ at redshift $z=0$. After applying a minimum mass criterion, we separated them into substructures and TDGCs based on their separation to their host galaxies (see Section~\ref{sec:Methods_distancecriteria}). The minimum stellar mass of our main sample (sample A) is set to $5 \times 10^{7} \, \rm{M_{\odot}}$ and includes $97$ TDGCs corresponding to a co-moving number density of $2.3 \times 10^{-4} \, h^{3} \, \rm{cMpc^{-3}}$ in the Illustris-1 simulation. These galaxies have total masses up to $3.1 \times 10^{9} \, \rm{M_{\odot}}$, which is comparable to the mass of the LMC. \cite{Fouquet_2012} suggest that the observed Magellanic Clouds could be TDGs (see Section \ref{sec:Physical properties of TDGCs and DMC DGs}). We present movies of the formation scenarios of TDGCs confirming their tidal dwarf nature. In particular, TDGCs are formed through galactic interactions and in the ram-pressure-stripped gas clouds of the host galaxy (see Section~\ref{sec:formation_TDGCs}, Appendix \ref{sec:appendix_Images_TDGCs_DMCDGs} and the movies in the supplementary information). However, TDGCs may conceivably also be formed in other scenarios such as cold accretion of gas clumps onto halos (see also Section~\ref{sec:discussion_DMF_galaxies}) which has not been addressed in this work, and has also never been reported to actually occur. 
	
	Dwarf galaxies lacking dark matter are mostly found in massive halos, in the vicinity of a massive galaxy, and are typically younger than DM-rich DGs. These results support the theory that these objects are TDGs formed through galactic interactions and mergers (see Sections \ref{sec:Methods_distancecriteria} and \ref{sec:Physical properties of TDGCs and DMC DGs}). 
	TDGCs and DMC DGs with a small dark matter mass are often metal-rich (see Appendix \ref{sec:appendix_metallicity}), which indicates that these satellite galaxies are possibly TDGs formed from a chemically enriched host galaxy \citep{Recchi_2015}.
	Almost all of the TDGCs of sample A are gas depleted and dispersion-dominated, which suggest that these are older TDGs which could have already lost their gas content by feedback processes and suffered from an angular momentum loss.  
	The density distribution of four selected massive TDGCs can be described by a Plummer model and an exponential profile (see Figs. \ref{fig:density_distribution} and \ref{fig:density_SUBFIND_3SIGMA_plots}).
	
	Analyzing the orbital angular momenta of TDGCs which have a minimum gas mass of $5 \times 10^{7} \, \rm{M_{\odot}}$ and at least one stellar particle (sample B) yields that these objects are significantly phase-space-correlated as observed in the Local Group \citep{Kroupa_2005,Pawlowski_2018} and Centaurus A \citep{Mueller_2018}. Moreover, simulated TDGCs are significantly more phase-space-correlated than DMC DGs (see Section \ref{Results:angularmomentum}). 
	
	Throughout the paper we have identified TDGCs based on the Subfind algorithm, which is a position-space subhalo finder. We also apply a $2 \sigma$-clipping scheme as a $6$D phase-space halo finder to the surroundings of gas-free Subfind TDGCs of sample A and we show that $92$~percent of these objects are gravitationally self-bound by including the velocity-space information of stellar particles (see Appendix \ref{sec:appendix_energy}).
	
	We quantified the probability of finding a NGC 1052-DF2-like galaxy in the Illustris-1 simulation at redshift $z=0$. While finding a few similar galaxies (Table \ref{table_NGC1052-DF2_probability}), there is also a non-detection of gas-free TDGCs with $M_{\mathrm{stellar}} \geq 0.8 \times \left(2 \times 10^{8} \right) \, \rm{M_{\odot}}$
	and $R_{\mathrm{0.5 \, stellar}} \geq 0.8 \times 2.7 \, \rm{kpc}$. Thus, such dwarf galaxies appear to be extremely rare in the Illustris-1 simulation. However, we note that this analysis does not consider the peculiar velocity of the observed NGC 1052-DF2 dwarf galaxy (see Section \ref{sec:Results_NGC1052-DF2}).
	
	We reported for the first time that the dual dwarf theorem is fulfilled in the self-consistent $\Lambda$CDM cosmological Illustris-1 simulation. TDGCs and DM-poor DGs populate regions in the radius--mass plane that are different from those populated by DM-rich DGs. In the stellar mass range between $5 \times 10^{7} \, \rm{M_{\odot}}$ and $10^{9} \, \rm{M_{\odot}}$ galaxies which are dark matter-poor have smaller stellar half-mass radii than dark matter-dominated galaxies as predicted by \cite{Kroupa_2012}. In particular, the KS test showed that the probability that simulated TDGCs and DMC DGs follow the same stellar half-mass radius distributions is less than $10^{-12}$. The independent simulations by \cite{Fouquet_2012} lead to TDGs which have radii in agreement with those of TDGCs in the Illustris simulation. However, the work of \cite{Dabringhausen_2013} has shown that observed TDGs occupy the same region in the radius--mass diagram as elliptical dwarf galaxies. This region agrees with our model TDGCs,  causing a major conflict with $\Lambda$CDM cosmology, because based on Illustris-1 some splitting is expected for galaxies with such different origins  (see Section \ref{sec:MR_BTF}). However, we note that only $0.17$~percent of all galaxies with $M_{\mathrm{stellar}} = 5 \times 10^{7}-10^{9} \, \rm{M_{\odot}}$ are TDGCs or DM-poor DGs in the Illustris-1 simulation.  
	
	It is therefore essentially important to conduct dedicated dwarf galaxy surveys \citep[such as][]{Merritt_2016, Javanmardi_2016} to find dwarf galaxies which are lacking dark matter, as this would confirm the dual dwarf galaxy theorem (it would be established that dwarfs with and without dark matter do exist in the real Universe), thus supporting $\Lambda$CDM cosmology. \cite{vDokkum_2018b, vDokkum_2019} have indeed found two candidates for such dwarf galaxies in the group of NGC 1052 (i.e., NGC 1052-DF2 and -DF4). However, the different positions of the Illustris TDGCs and DMC DGs in the radius--mass plane are in significant tension with the fact that there is only one population of observed dwarf galaxies. That the TDGCs in Illustris-1 have radii in agreement with the observed radii of dE galaxies suggests that the latter are TDGs, as also concluded independently by \citet{Okazaki_2000}.
	
	\begin{acknowledgements}
		We thank the anonymous referee for her/his constructive comments and suggested improvements. IB is supported by an Alexander von Humboldt research fellowship. We thank the DAAD-Ostpartnerschaftsprogramm für 2018 at the University of Bonn for funding exchange visits between Charles University in Prague and Bonn University. We would like to thank the staff of the Illustris project, especially Dylan Nelson and Vicente Rodriguez-Gomez, for providing the Illustris data, the $\kappa_{\mathrm{rot}}$ morphology parameter, example scripts, and useful suggestions. We thank Paul Torrey for explanations about the Illustris Galaxy Observatory. 
	\end{acknowledgements}
	
	\bibliography{library.bib}
	\bibliographystyle{aa}
	
	\appendix
	
	\section{Stellar luminosities and gas densities of TDGCs and their host galaxies}
	\label{sec:appendix_Images_TDGCs_DMCDGs}
\begin{figure*}
		\centering
		
		\includegraphics[width=91mm,trim={0.0cm 0.0cm 0.0cm 0.0cm},clip]{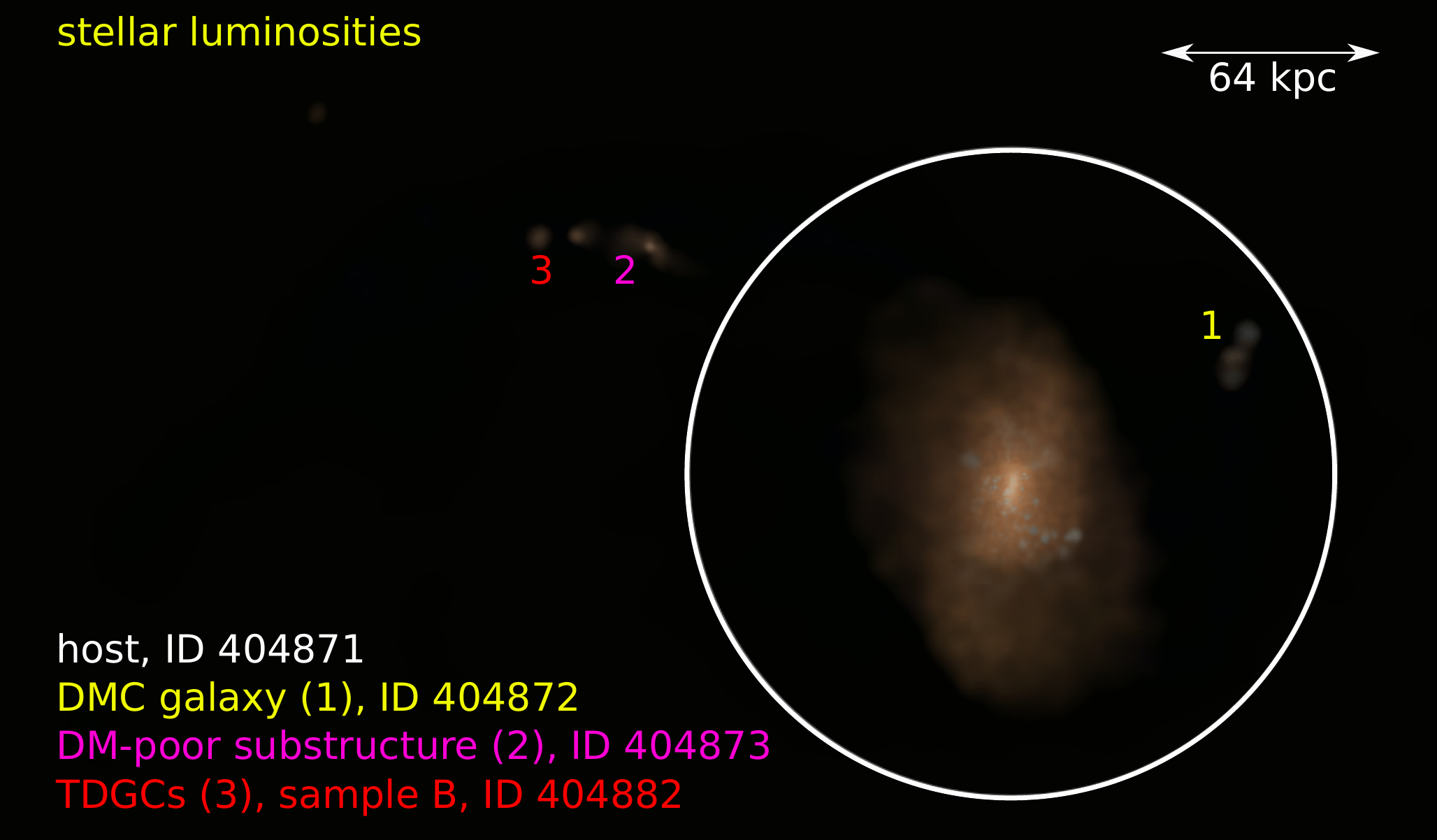}
		\includegraphics[width=91mm,trim={0.0cm 0.0cm 0.0cm 0.0cm},clip]{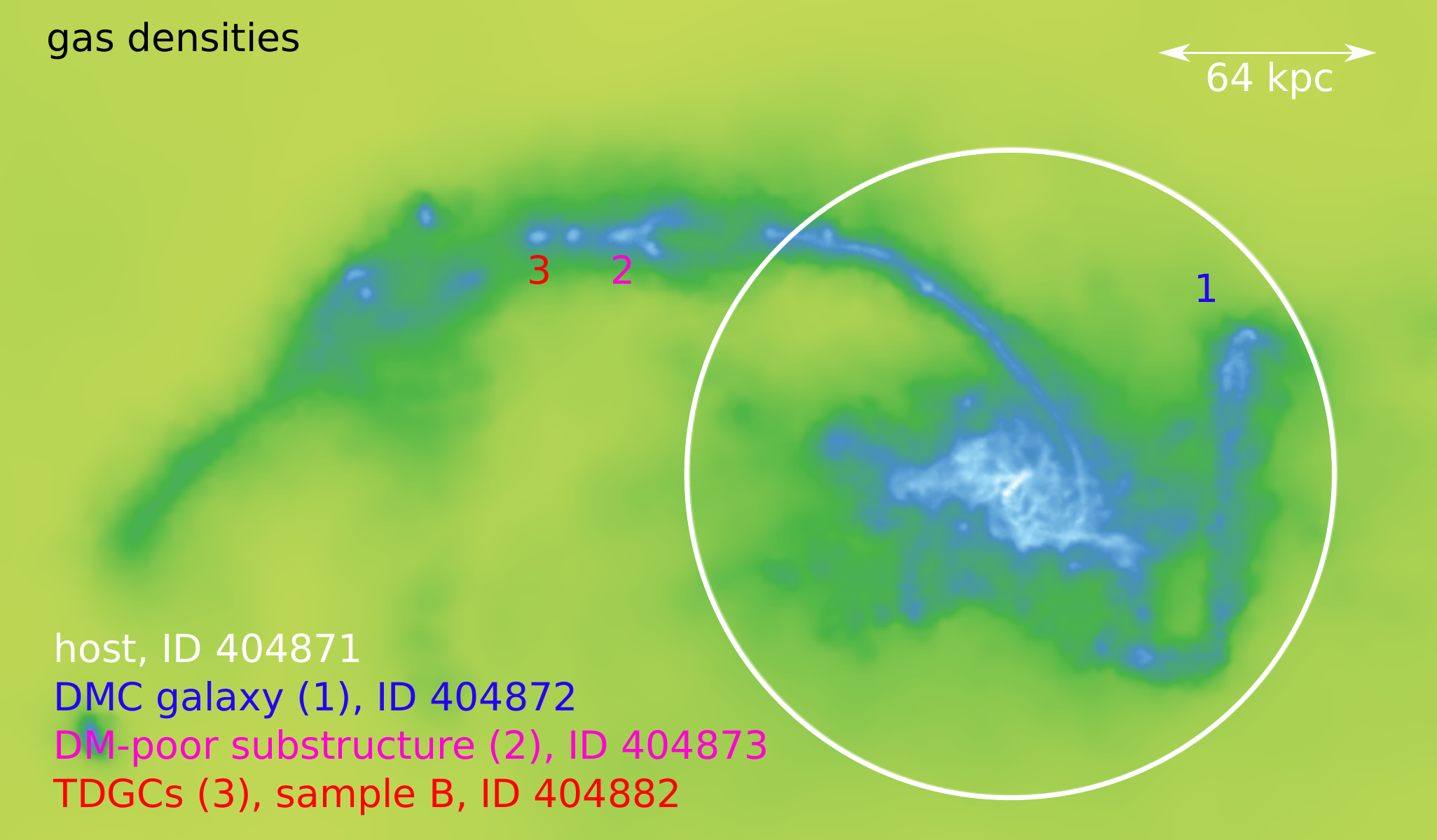}
		
		\vspace{0.2cm}
		
		\includegraphics[width=91mm,trim={0.0cm 0.0cm 0.0cm 0.0cm},clip]{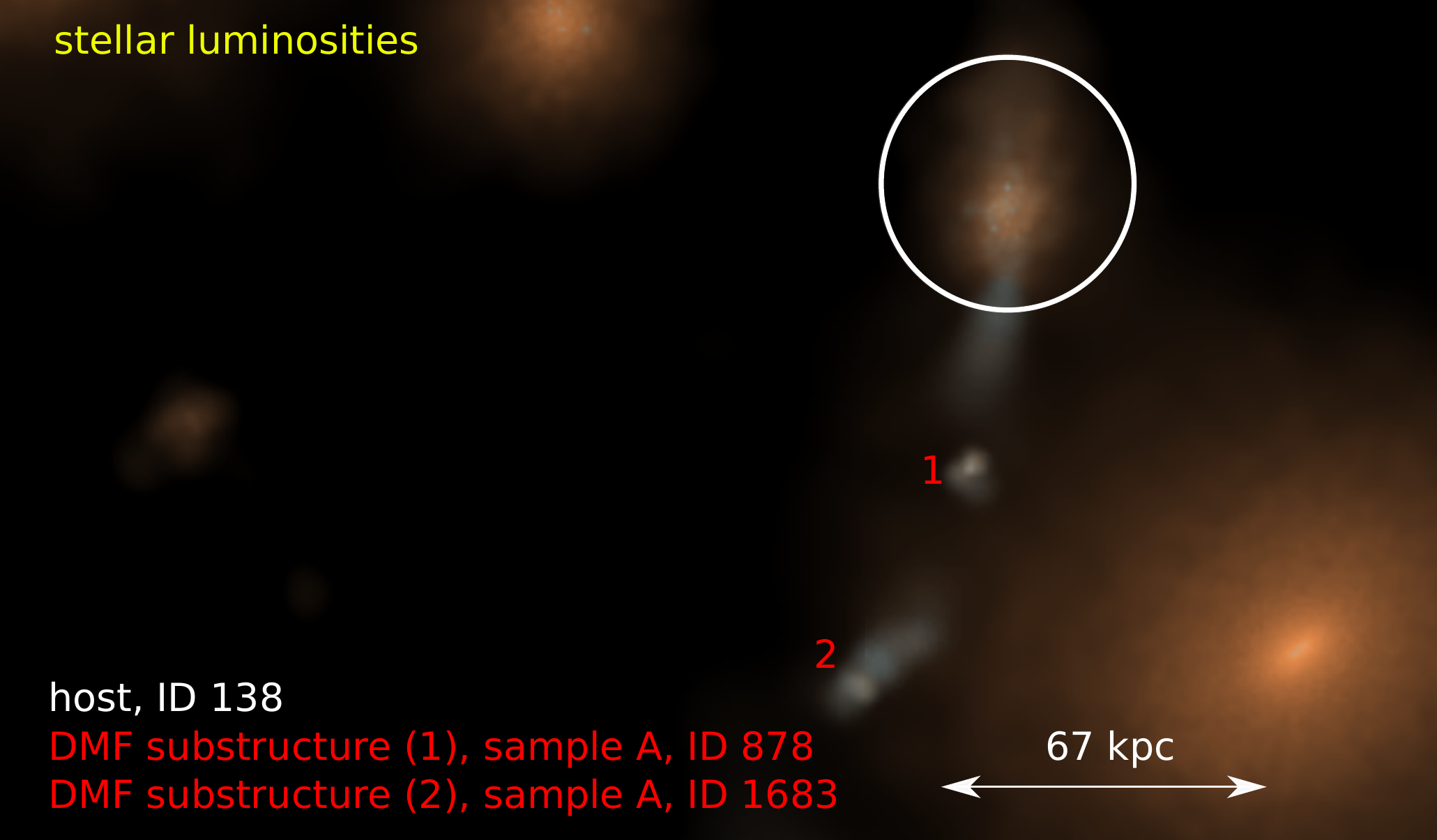}
		\includegraphics[width=91mm,trim={0.0cm 0.0cm 0.0cm 0.0cm},clip]{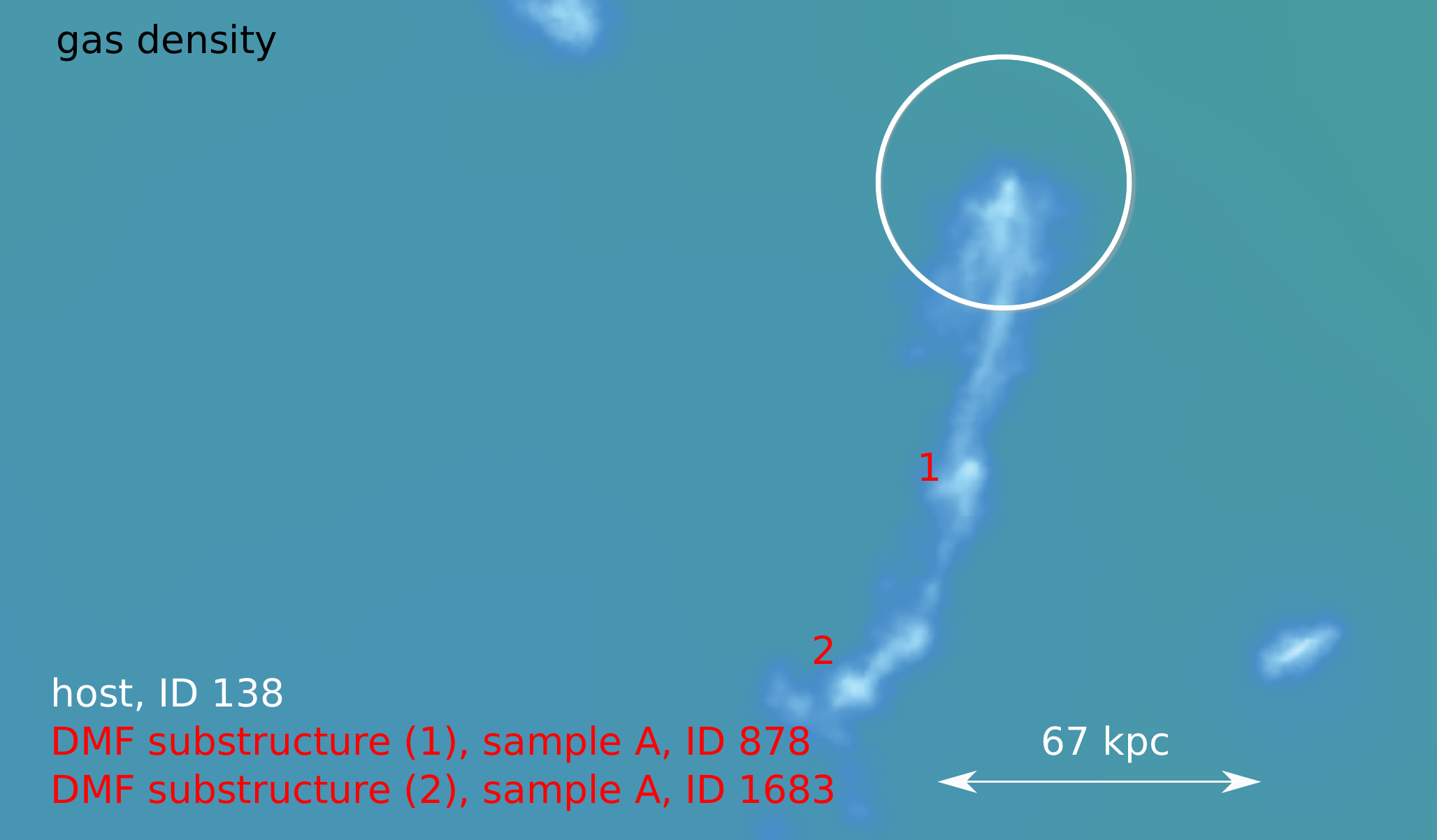}
		\caption{Projected stellar luminosity (left panels) and gas density (right panels) of the environment of the host galaxy ID $404871$ (top panels) and ID $138$ (bottom panels) at redshift $z = 0$. Top panels: The stellar objects (1) - (3) close to ID $404871$ (the white circle marks the host galaxy) are explained in the figure legend. Bottom panels: The stellar objects (1) and (2) are identified as DMF substructures of the host galaxy ID $138$ (the white circle as above). The luminous galaxy in the lower-right corner is a foreground galaxy, which has a 3D separation of $1626 \, \rm{kpc}$ to the galaxy ID $138$. Credits: \url{http://www.illustris-project.org/explorer/} [13.08.2018].}
		\label{fig:appendix_Images_TDGCs_DMCDGs}
	\end{figure*}

	Using the Illustris Explorer online tool\footnote{\url{http://www.illustris-project.org/explorer/}}, we visualize in Fig.~\ref{fig:appendix_Images_TDGCs_DMCDGs} the projected gas densities and stellar luminosities of TDGCs and DM-poor objects of the studied host galaxies ID $138$ and ID $404871$ in Section~\ref{sec:formation_TDGCs} at redshift $z = 0$. The identification number and physical parameters of the stellar objects shown in Fig.~\ref{fig:appendix_Images_TDGCs_DMCDGs} are listed in Table~\ref{tab:TDGCs_DMCDGs}. 
	
	\begin{table*}
		\centering
		\caption{Properties of the discussed and depicted stellar objects identified at redshift $z=0$ and shown in Fig.~\ref{fig:appendix_Images_TDGCs_DMCDGs}.}
		\label{tab:TDGCs_DMCDGs}
		\small
		\begin{tabular}{lllllllllll}
			\hline
			galaxy (ID) & host (ID) & $M_{\mathrm{stellar}} \, [\rm{M_{\odot}}]$ &  $M_{\mathrm{gas}} \,  \, [\rm{M_{\odot}}]$ & $M_{\mathrm{dm}}/M_{\mathrm{baryonic}}$  & $\kappa_{\mathrm{rot}}$ & $s_{\mathrm{host}} \,[\rm{kpc}]$ & $s_{\mathrm{host}}/R_{\mathrm{0.5 \, stellar}}^{\mathrm{host}}$ & $\widehat{\vec{L}}_{\mathrm{orbit}}$ \\
			\hline \hline
			$878$    &  $138$   & $6.7\times 10^{7}$ & $1.4 \times 10^{9}$  & $0.0$ & $0.47$ & $94$ & $9.7$ & $(+0.812, -0.0896, +0.576)$  \\ 
			$1683$       & $138$   & $5.7 \times 10^{7}$ & $5.6 \times 10^{8}$ & $0.0$ & $0.39$ & $49$ & $5.1$ & $(+0.936, -0.0479, +0.350)$ \\ \hline 
			$404872$ & $404871$ & $1.1 \times 10^{9}$ & $2.1 \times 10^{9}$ & $4.5$ & $0.66$ & $115$ & $7.0$ & $(-0.712, +0.225, +0.665)$ \\
			$404873$ & $404871$ & $8.0 \times 10^{6}$ & $1.8 \times 10^{9}$ & $0.020$ & $0.39$ & $157$ & $9.5$ & $(+0.0415, -0.0105, -0.994)$ \\
			$404879$ & $404871$ & $2.2 \times 10^{7}$ & $3.0 \times 10^{8}$ & $0.019$ & $0.37$ & $182$ & $11$ & $(+0.0831, -0.0204,  -0.996)$ \\
			$404882$ & $404871$ & $1.7 \times 10^{7}$ & $2.5 \times 10^{8}$ & $0.0$ &  $0.41$ & $193$ & $12$ & $(-0.0283, -0.158, -0.987)$ \\
			\hline 
		\end{tabular}
		\tablefoot{Listed are the identification number of the stellar object, the identification number of its host galaxy, the stellar mass, $M_{\mathrm{stellar}}$, gas mass, $M_{\mathrm{gas}}$, the fraction of dark matter-to-baryonic matter, $M_{\mathrm{dm}}/(M_{\mathrm{gas}}+M_{\mathrm{stellar}}) = M_{\mathrm{dm}}/M_{\mathrm{baryonic}}$, the morphological parameter, $\kappa_{\mathrm{rot}}$, the 3D separation between the stellar object and its host galaxy, $s_{\mathrm{host}}$, the fraction of this separation to the stellar half-mass radius of the host galaxy, $s_{\mathrm{host}}/R_{\mathrm{0.5 \, stellar}}^{\mathrm{host}}$, and the normalized specific orbital angular momentum, $\widehat{\vec{L}}_{\mathrm{orbit}}$ (eq. \ref{eq:orbital_angular_momentum}).}
	\end{table*}

	\section{Gas and stellar metallicity of TDGCs and DMC DGs}  \label{sec:appendix_metallicity}
	
\begin{figure}
		\centering
		\includegraphics[width=\columnwidth,trim={0cm 0.0cm 0 0.0cm},clip]{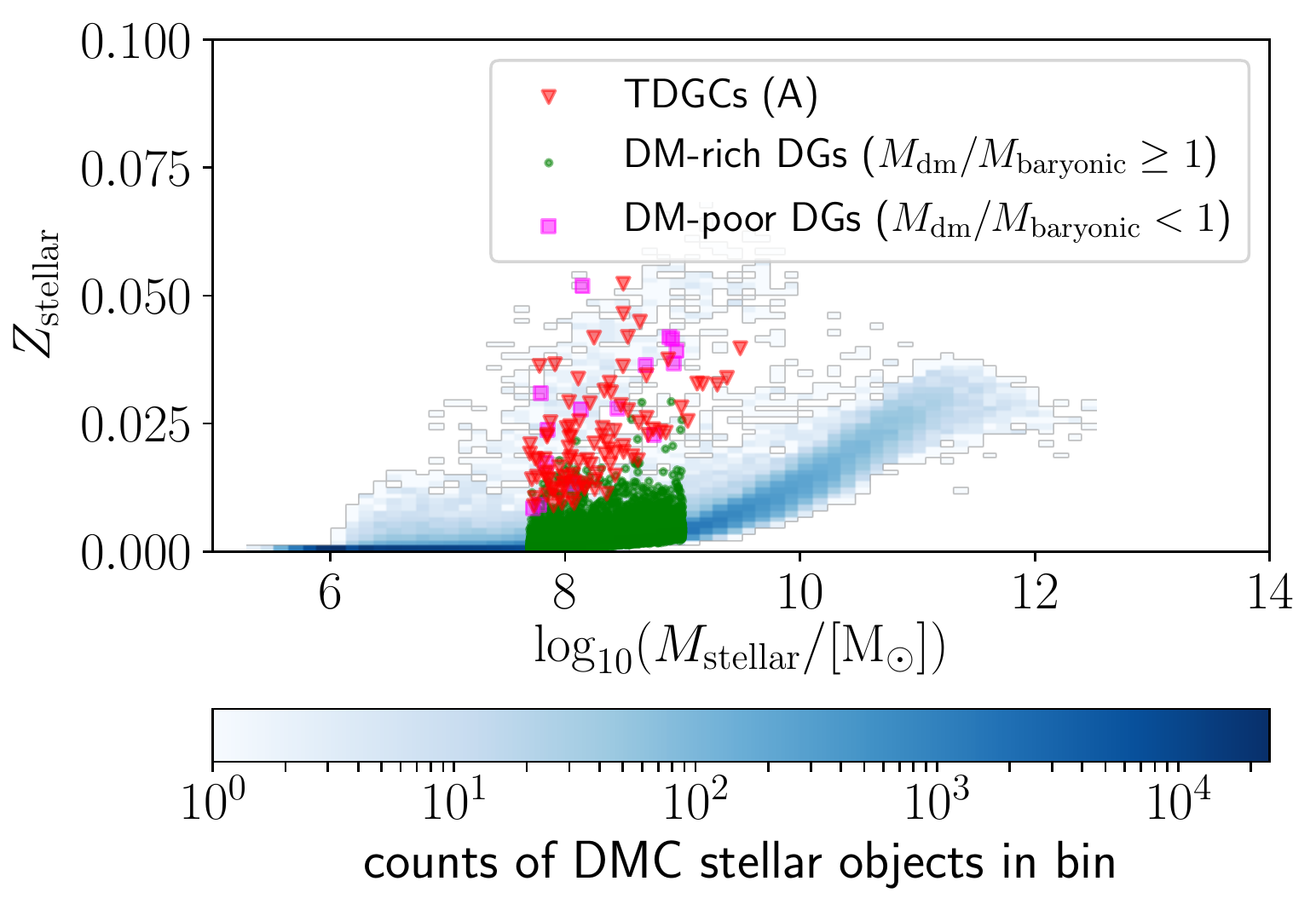}
		\includegraphics[width=\columnwidth,trim={0cm 0.0cm 0 0.0cm},clip]{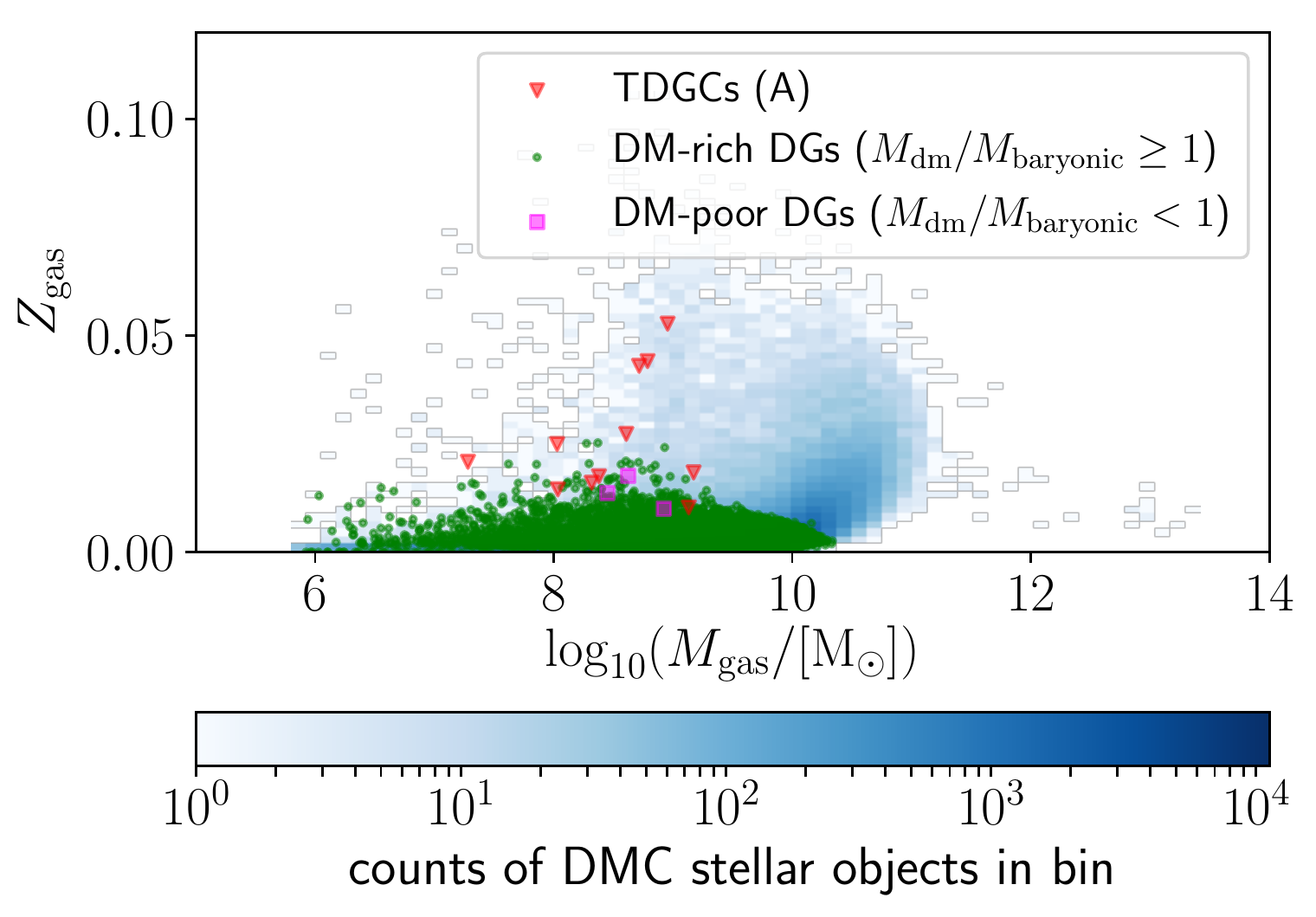}
		\caption{Stellar (top) and gas (bottom) mass-weighted average metallicity, $Z_{\mathrm{stellar}}$ and $Z_{\mathrm{gas}}$, respectively, in dependence of the stellar mass, $M_{\mathrm{stellar}}$, and gas mass, $M_{\mathrm{gas}}$. The stellar and gas mass-weighted average metallicities, $Z_{\mathrm{stellar}}$ and $Z_{\mathrm{gas}}$, are defined by Eq. \ref{equation_metallicity}. Blue bins are DMC stellar objects, green dots are DM-rich DGs ($M_{\mathrm{dm}}/M_{\mathrm{baryonic}} \geq 1$), magenta squares are DM-poor DGs ($M_{\mathrm{dm}}/M_{\mathrm{baryonic}}<1$), and red triangles are TDGCs of sample A.}
		\label{metallcity_subhalo}
	\end{figure}

	Figure~\ref{metallcity_subhalo} shows the stellar and gas metallicity in dependence of the stellar and gas mass, respectively, for DMC stellar objects, DMC DGs, and TDGCs (sample A). The stellar and gas mass-weighted average metallicities are defined as
	\begin{equation}
	\begin{aligned}
	&Z_{\mathrm{stellar}} \equiv \bigg( \frac{M_{\mathrm{>He}}}{M_{\mathrm{tot}}} \bigg)_{\mathrm{stellar}} \, , \\
	&Z_{\mathrm{gas}} \equiv \bigg( \frac{M_{\mathrm{>He}}}{M_{\mathrm{tot}}} \bigg)_{\mathrm{gas}} \, , \\
	\end{aligned}
	\label{equation_metallicity}
	\end{equation}
	where $M_{\mathrm{tot}}$ is the total mass and $M_{\mathrm{>He}}$ is the mass of all elements above Helium (He).\footnote{Only cells within twice the stellar half-mass radius are considered.} 
	DM-poor DGs and TDGCs are often significantly more metal-rich than DM-rich DGs, which is consistent with these galaxies being TDGs formed from chemically enriched host galaxies in the Illustris model \citep{Recchi_2015}. Dark matter-containing stellar objects in the region of TDGCs and DM-poor DGs in Fig.~\ref{metallcity_subhalo} (top) are typically metal-enriched substructures of massive subhalos. Therefore, DM-poor DGs can have captured dark matter particles. 
	
	\section{Internal structures and kinematics of TDGCs}
	\label{sec:appendix_energy}
	In the following we discuss the internal structures and kinematics of TDGCs, which includes an analysis of their energy contents, density and dispersion distributions, and rotation curves. Different halo finders running on the same simulation can provide different results \citep{Knebe_2011}. Therefore  a $\sigma$-clipping procedure as a $6$D phase-space halo finder is applied to the surroundings of the Subfind TDGCs to examine whether or not these objects are gravitationally bound by including their particle velocity data.
	
	\subsection{Virial equilibrium of TDGCs}
	The kinetic energy (including the thermal energy of the gas), $E_{\mathrm{kin}}$, and the potential energy, $E_{\mathrm{pot}}$, of subhalos at redshift $z=0$ are taken from a supplementary catalog prepared by \cite{Zjupa_2017} in which the energies of the subhalos are calculated by the particles identified by the Subfind algorithm \citep{Springel_2001}. 
	
\begin{figure}
		\centering
		\includegraphics[width=\columnwidth,trim={0cm 0.0cm 0 0.0cm},clip]{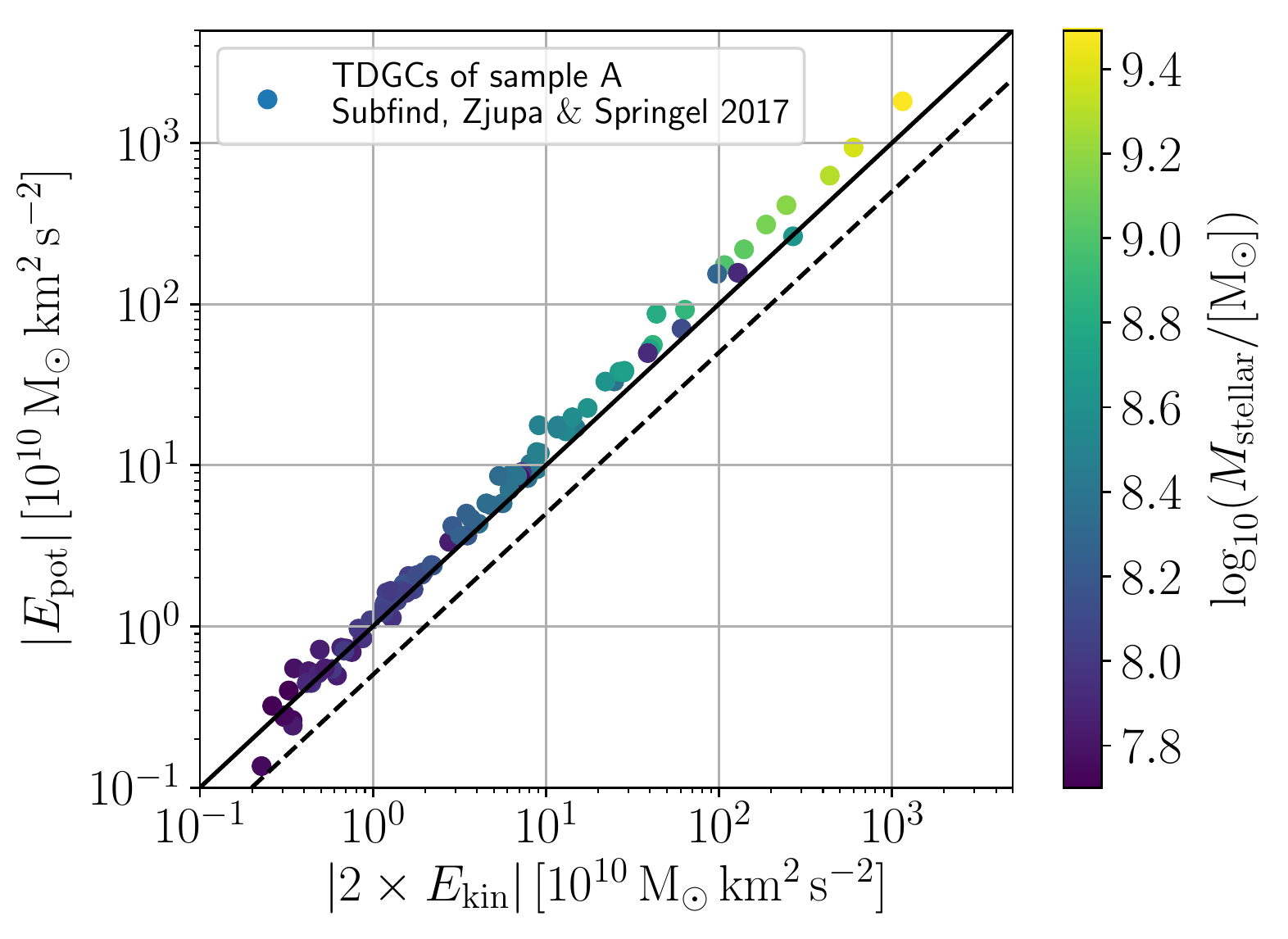}
		\caption{Relation between the potential and kinetic energy for all TDGCs of sample A. The colorbar presents the stellar mass, $M_{\mathrm{stellar}}$, of the TDGCs. The black solid lines highlight the condition for virial equilibrium, i.e., where the virial ratio becomes $q=1$ (see Eq. \ref{eq:virial_equilbrium}). All objects above the black dashed line are gravitationally bound ($q<2$).}
		\label{fig:virial_equilbrium}
	\end{figure}

	All identified Subfind TDGCs of sample A and sample B fulfill the condition $ \lvert E_{\mathrm{pot}} \rvert > \lvert E_{\mathrm{kin}} \rvert$, which demonstrates that these subhalos are gravitationally bound. Here the virial ratio is defined by
	\begin{equation}
	\begin{aligned}
	q \equiv \bigg \lvert \frac{2 \times E_{\mathrm{kin}}}{E_{\mathrm{pot}}} \bigg \rvert \, , \\
	\end{aligned}
	\label{eq:virial_equilbrium}
	\end{equation}
	
	such that for a self-gravitating object virial equilibrium is achieved if $q=1$ and the object is gravitationally bound if $q<2$. Figure~\ref{fig:virial_equilbrium} shows that TDGCs of sample A only slightly deviate from virial equilibrium for larger stellar masses or smaller stellar half-mass radii in the sense that they become dominated by potential energy ($q<1$). The mean and the standard deviation of the virial ratio of all TDGCs of sample A is $0.85$ and $0.18$, respectively. Therefore, the condition for virial equilibrium lies roughly within the $1 \sigma$ range of the virial ratio distribution. The statistics of the virial ratio for different dwarf galaxy samples are given in Table~\ref{tab:virial_equilibrium}.
	
	\begin{table}
		\centering
		\caption{Virial ratio, $q$ (eq. \ref{eq:virial_equilbrium}), for TDGC samples and DMC DGs at redshift $z=0$ as resulting from the Subfind algorithm.}
		\label{tab:virial_equilibrium}
		\begin{tabular}{lllll} \hline
			Sample & Mean & Median & Std. & $16^{\mathrm{th}}-84^{\mathrm{th}}$ \\
			& & & & percentile \\ \hline \hline
			TDGCs (sample A) & $0.85$ & $0.82$ & $0.18$ & $0.69-0.97$ \\
			TDGCs (sample B) & $0.59$ & $0.58$ & $0.25$ & $0.34-0.86$ \\ 
			DM-poor DGs      & $0.95$ & $0.92$ & $0.28$ & $0.73-1.2$ \\
			DM-rich DGs      & $1.0$ & $1.0$ & $0.080$ & $0.98-1.1$ \\  \hline 
		\end{tabular}
		\tablefoot{Listed are the mean, the median, the standard deviation (std.), and the $16^{\mathrm{th}}-84^{\mathrm{th}}$ percentile of the virial ratio, $q$, for dwarf galaxies.}
	\end{table}
	
	\subsection{Density distribution, velocity distribution, and rotation curves of TDGCs}
\begin{figure*}
		\centering
		\includegraphics[width=\linewidth,trim={1.5cm 1.0cm 3.0cm 2.0cm},clip]{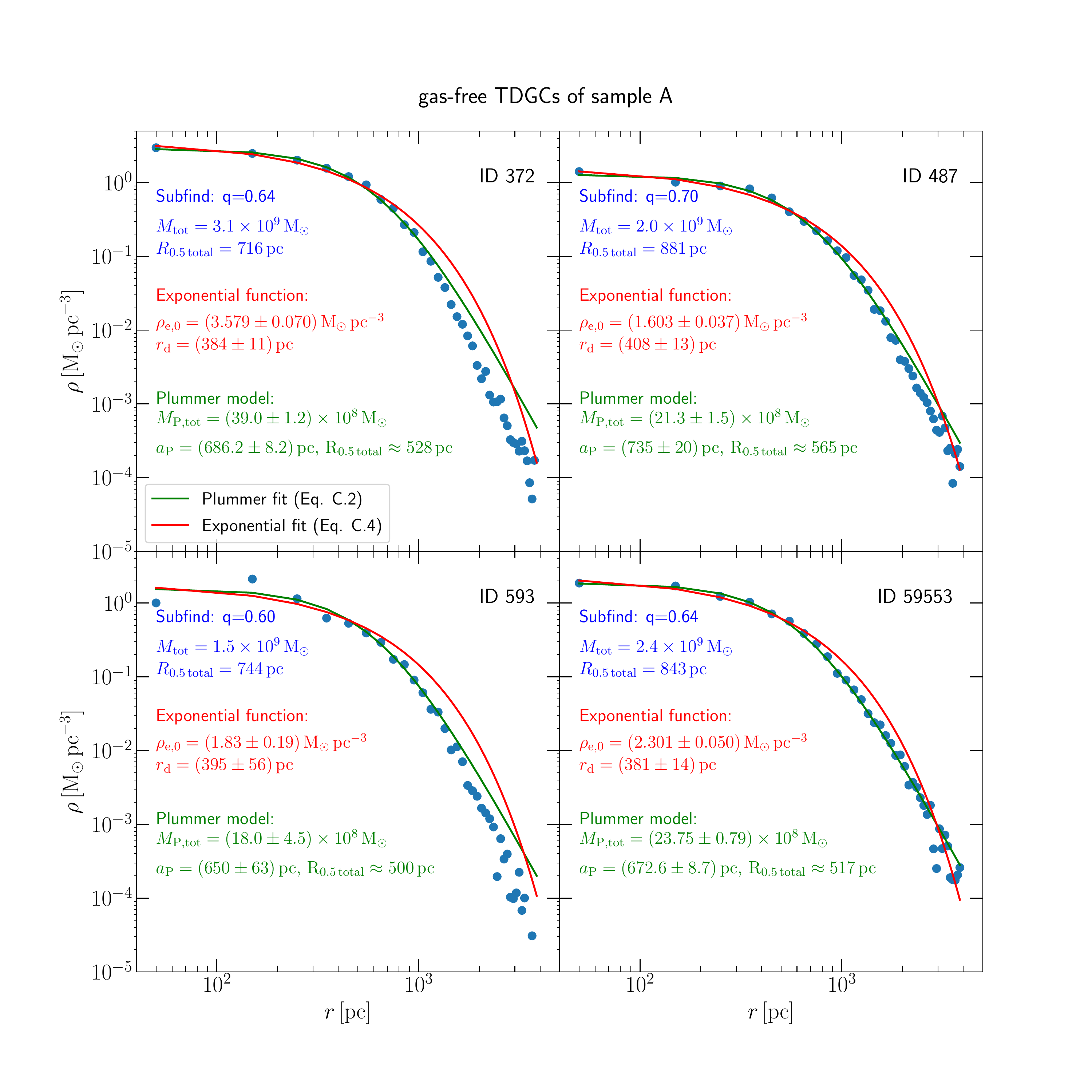}
		\caption{Radial density distributions of four selected gas-free TDGCs of sample A (IDs $372$, $487$, $593$, and $59553$) fitted with a Plummer model (green; Eq. \ref{eq:fit_plummer}) and an exponential function (red; eq. \ref{eq:fit_exponential}). The radial bins have a width of $100 \, \rm{pc}$. The identification number, ID, their corresponding virial ratios, $q$ (eq. \ref{eq:virial_equilbrium}), their total masses and stellar half-mass radii obtained from the Subfind algorithm, and their fitting parameters are given in the panels. According to the particles identified by the Subfind algorithm, the TDGCs shown here are not fully in virial equilibrium. The rotation curves and $3$D velocity dispersion of the TDGCs discussed here are shown in Fig.~\ref{fig:velocitydispersion_rotationcurve_time_distribution_1} and Fig.~\ref{fig:velocitydispersion_rotationcurve_time_distribution_2}.}
		\label{fig:density_distribution}
	\end{figure*}

	In Fig.~\ref{fig:density_distribution} we study the internal structure of four massive gas-free TDGCs of sample A which are among the most deviant ones from virial equilibrium (i.e., with $q = 0.60 - 0.70$) by fitting their density distributions with a Plummer model and an exponential function. The $3$D density profile for a Plummer model \citep[e.g.,][]{heggie2003gravitational,Kroupa_2008} is given by
	
	\begin{equation}
	\begin{aligned}
	\rho_{\mathrm{P}}(r) = \frac{3 M_{\mathrm{tot}}}{4 \pi a_{\mathrm{P}}^{3}} \bigg( 1 +\frac{r^{2}}{a_{\mathrm{P}}^{2}} \bigg)^{-5/2} \, , \\
	\end{aligned}
	\label{eq:fit_plummer}
	\end{equation}
	where $M_{\mathrm{tot}}$ is the total mass of the object and $a_{\mathrm{P}}$ is the Plummer radius which is related to the half-mass radius, $R_{\mathrm{0.5 \, total}}$, by
	\begin{equation}
	\begin{aligned}
	a_{\mathrm{P}} \approx 1.3 R_{\mathrm{0.5 \, total}}. \\
	\end{aligned}
	\label{eq:plummer_radius_radius}
	\end{equation}
	The exponential function describes the observed dwarf satellite galaxies well \citep{Binggeli_1984} and has the form, 
	\begin{equation}
	\begin{aligned}
	\rho_{\mathrm{e}}(r) = \rho_{\mathrm{e,0}} \exp (-r/r_{\mathrm{d}}) \, , \\
	\end{aligned}
	\label{eq:fit_exponential}
	\end{equation}
	where $\rho_{\mathrm{e,0}}$ is the central density and $r_{d}$ is the scale radius. Both fitting functions can reasonably describe the density profile of the shown TDGCs, that is, their total masses derived from the fits are comparable to the total masses obtained from the Subfind algorithm. The half-mass radius calculated from the Plummer fit for all the four TDGCs is smaller than the radii given by the Subfind algorithm. The fitting parameters and their uncertainties and a quantitative comparison with the total mass and half-mass radii obtained from the Subfind algorithm are labeled in the panels of Fig.~\ref{fig:density_distribution}. These demonstrate that even the density distributions of simulated TDGCs which are not fully in virial equilibrium follow a profile which is expected from observed dwarf galaxies. 
	
\begin{figure*}
		\centering
		
		\includegraphics[width=91mm,trim={0.0cm 0.0cm 0.0cm 0.0cm},clip]{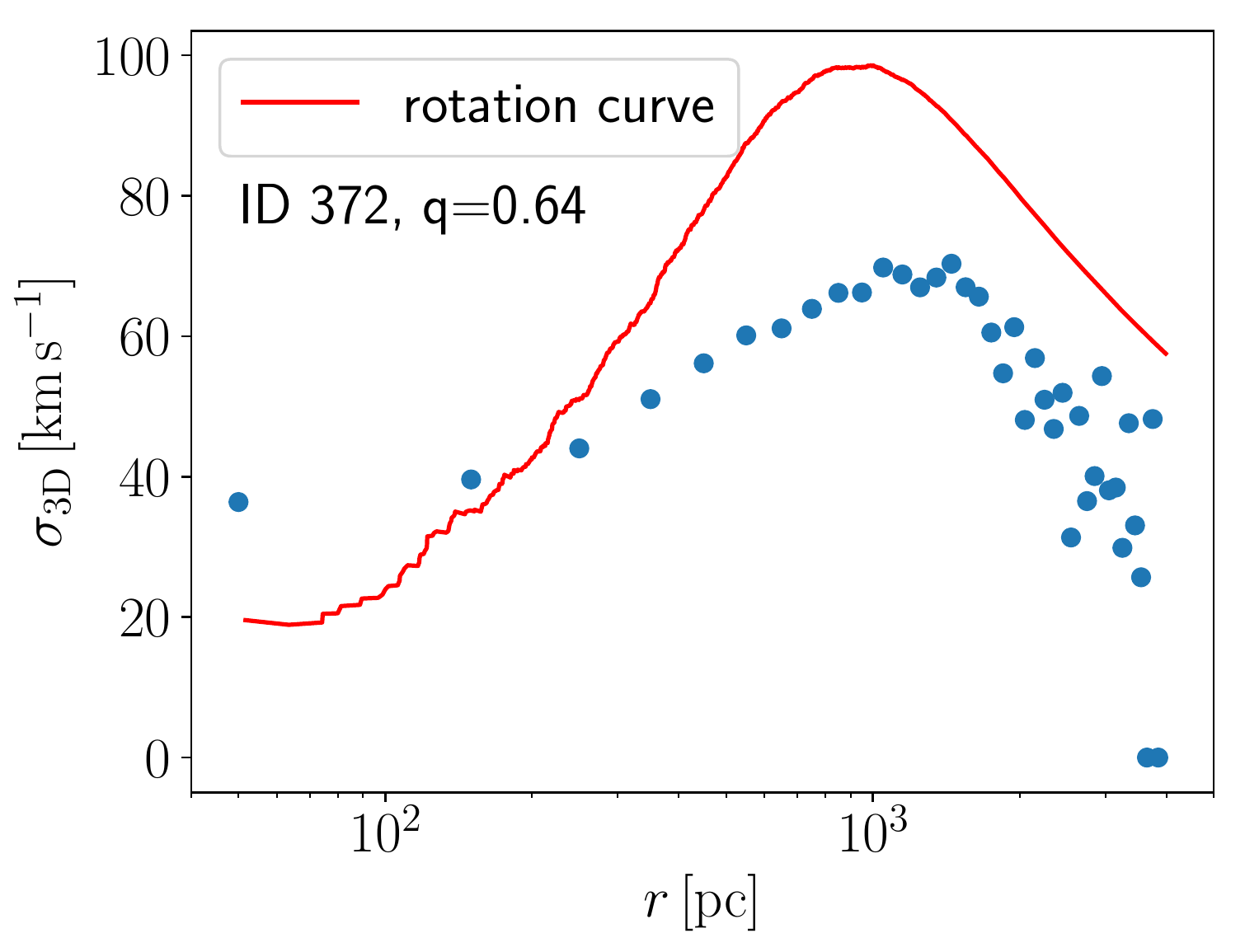}
		\includegraphics[width=91mm,trim={0.0cm 0.0cm 0.0cm 0.0cm},clip]{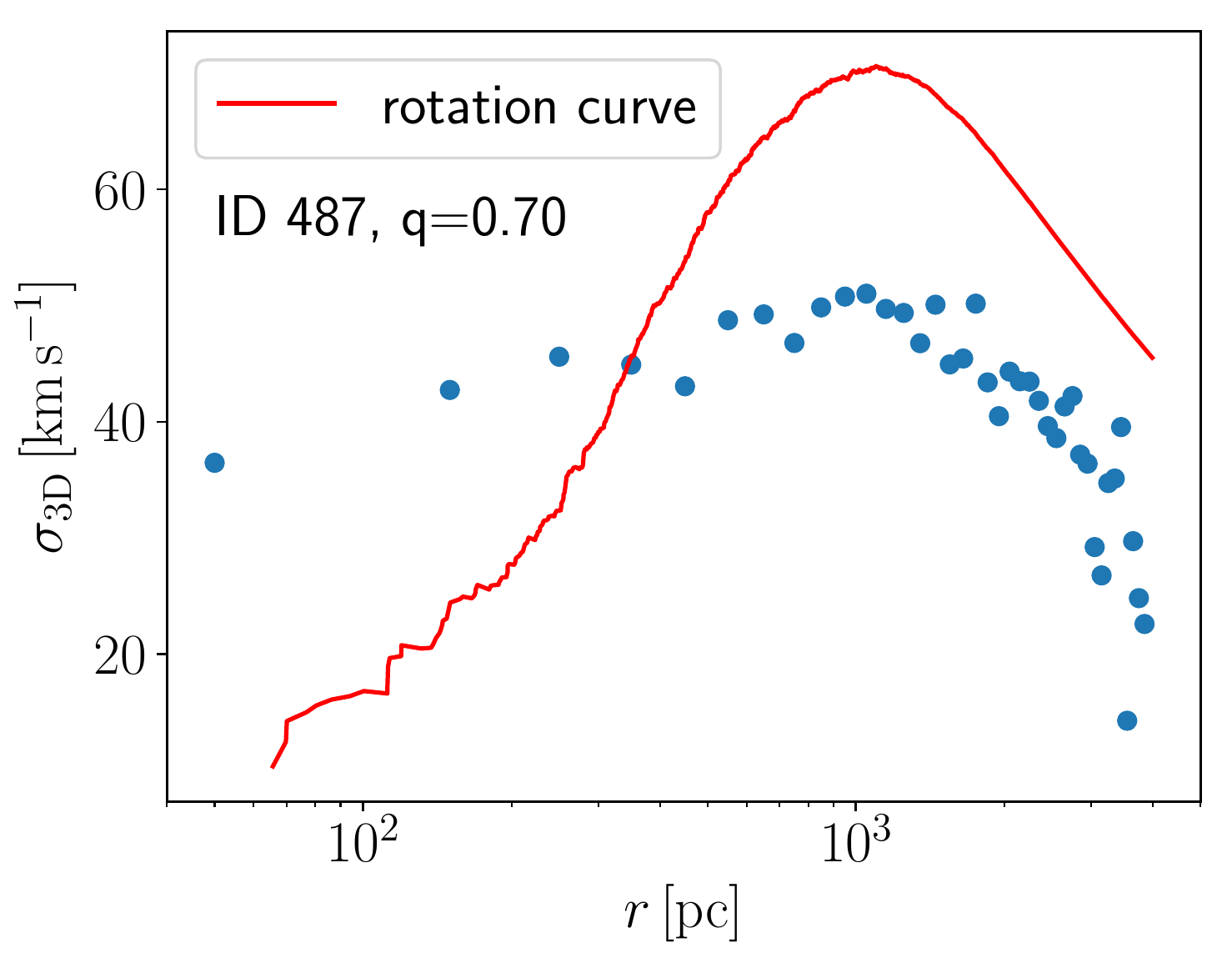}
		
		\includegraphics[width=91mm,trim={0.0cm 0.0cm 0.0cm 0.0cm},clip]{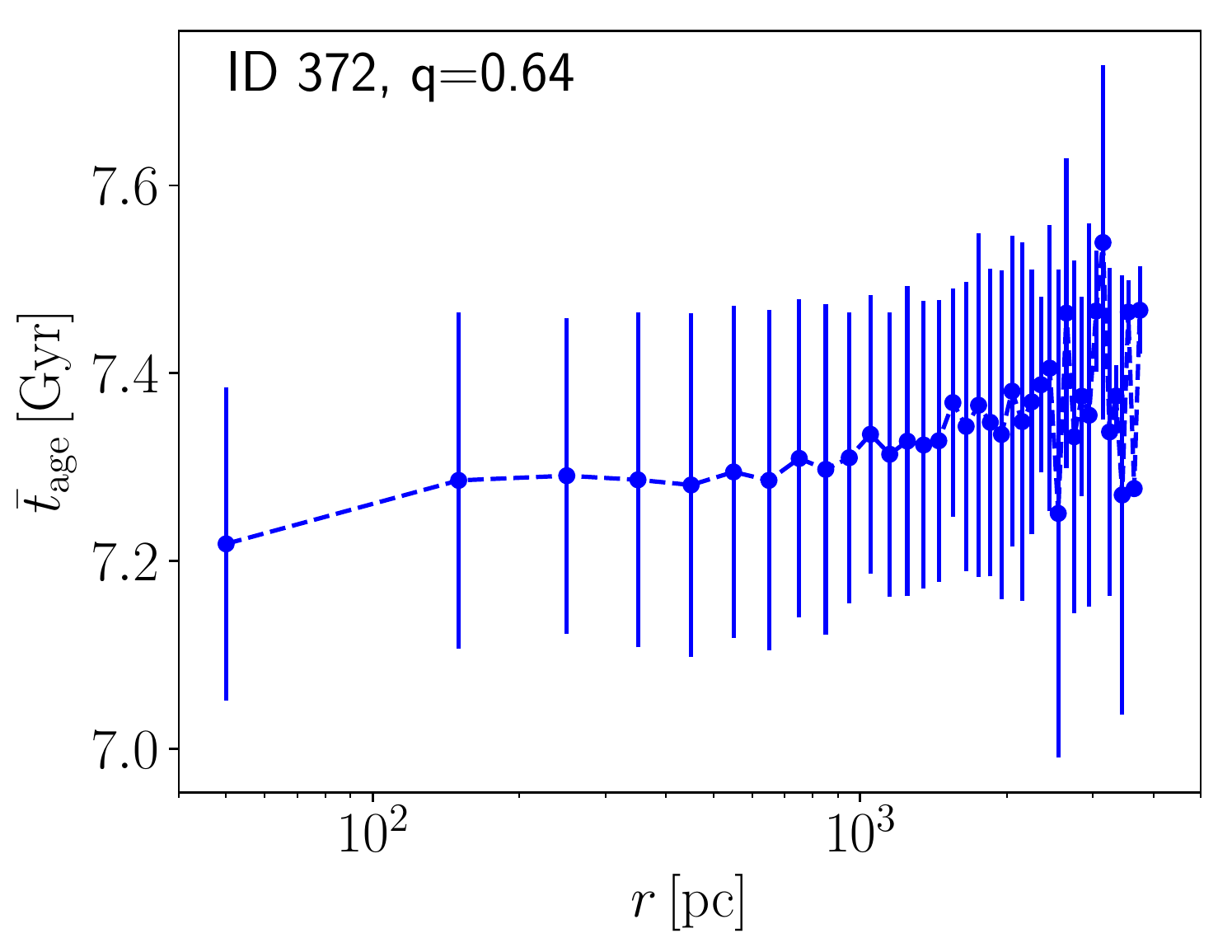}
		\includegraphics[width=91mm,trim={0.0cm 0.0cm 0.0cm 0.0cm},clip]{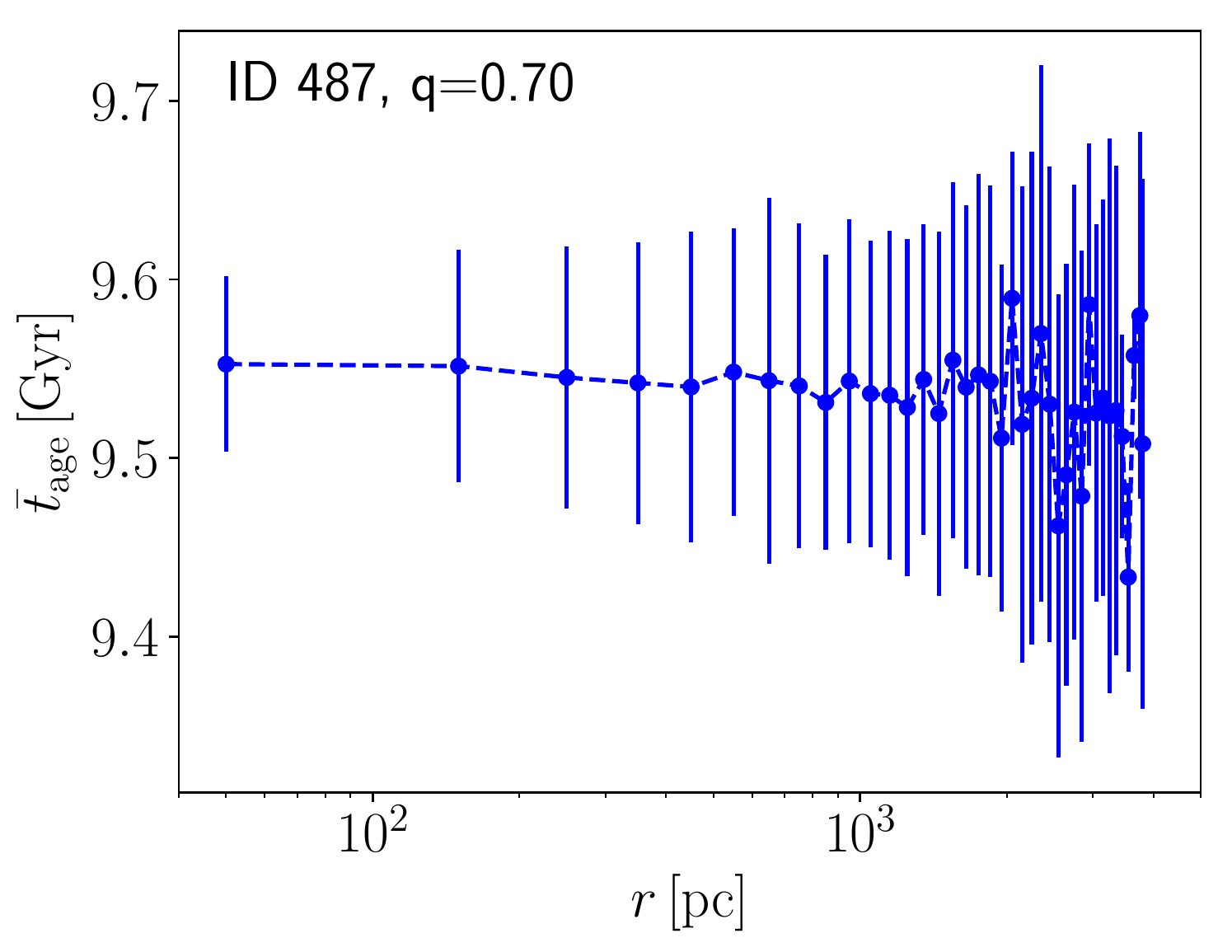}
		\caption{Rotation curves and radial $3$D velocity dispersion distributions (top), and the radial distribution of stellar particle ages (bottom) of gas-free TDGCs of sample A with the ID $372$ and ID $487$. The radial bins have a width of $100 \, \rm{pc}$. The error bars in the bottom panels correspond to the standard deviation of the stellar particle age within a radial bin. The identification number, ID, and their corresponding virial ratios, $q$ (Eq.~\ref{eq:virial_equilbrium}), are given in the panels.  The radial density distributions of the TDGCs discussed here are shown in Fig.~\ref{fig:density_distribution}.}
		\label{fig:velocitydispersion_rotationcurve_time_distribution_1}
	\end{figure*}

\begin{figure*}
		\centering
		
		\includegraphics[width=91mm,trim={0.0cm 0.0cm 0.0cm 0.0cm},clip]{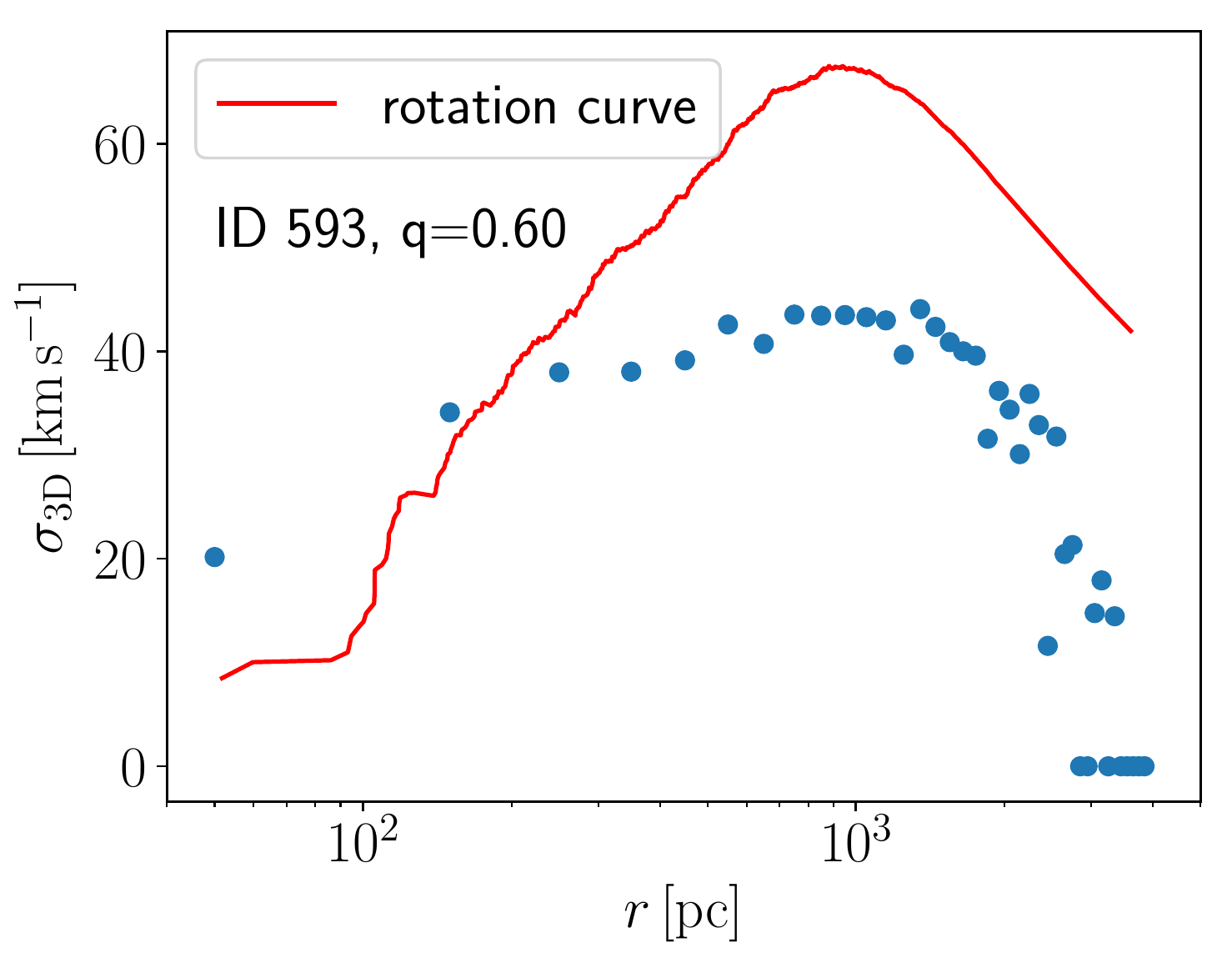}
		\includegraphics[width=91mm,trim={0.0cm 0.0cm 0.0cm 0.0cm},clip]{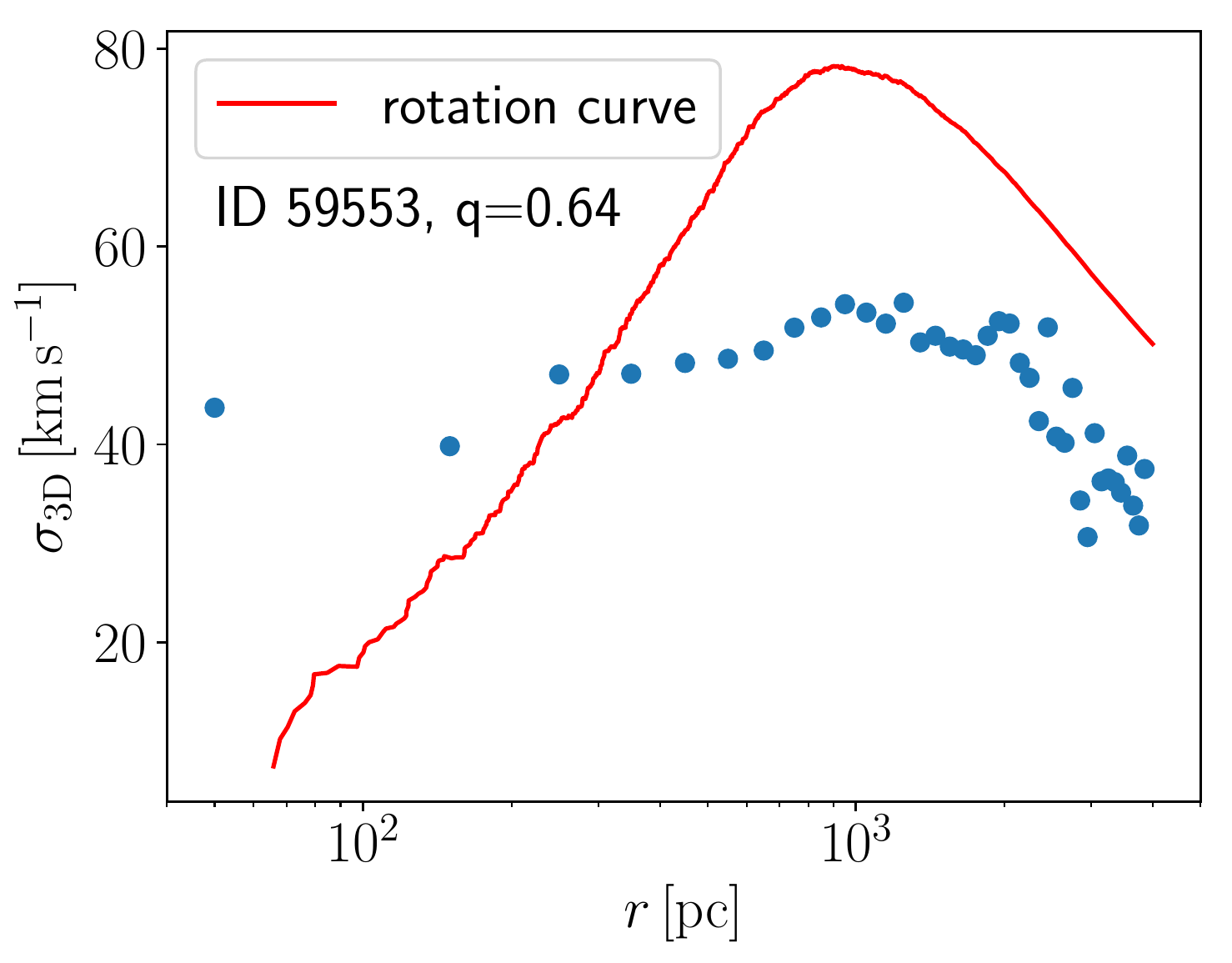}
		
		\includegraphics[width=91mm,trim={0.0cm 0.0cm 0.0cm 0.0cm},clip]{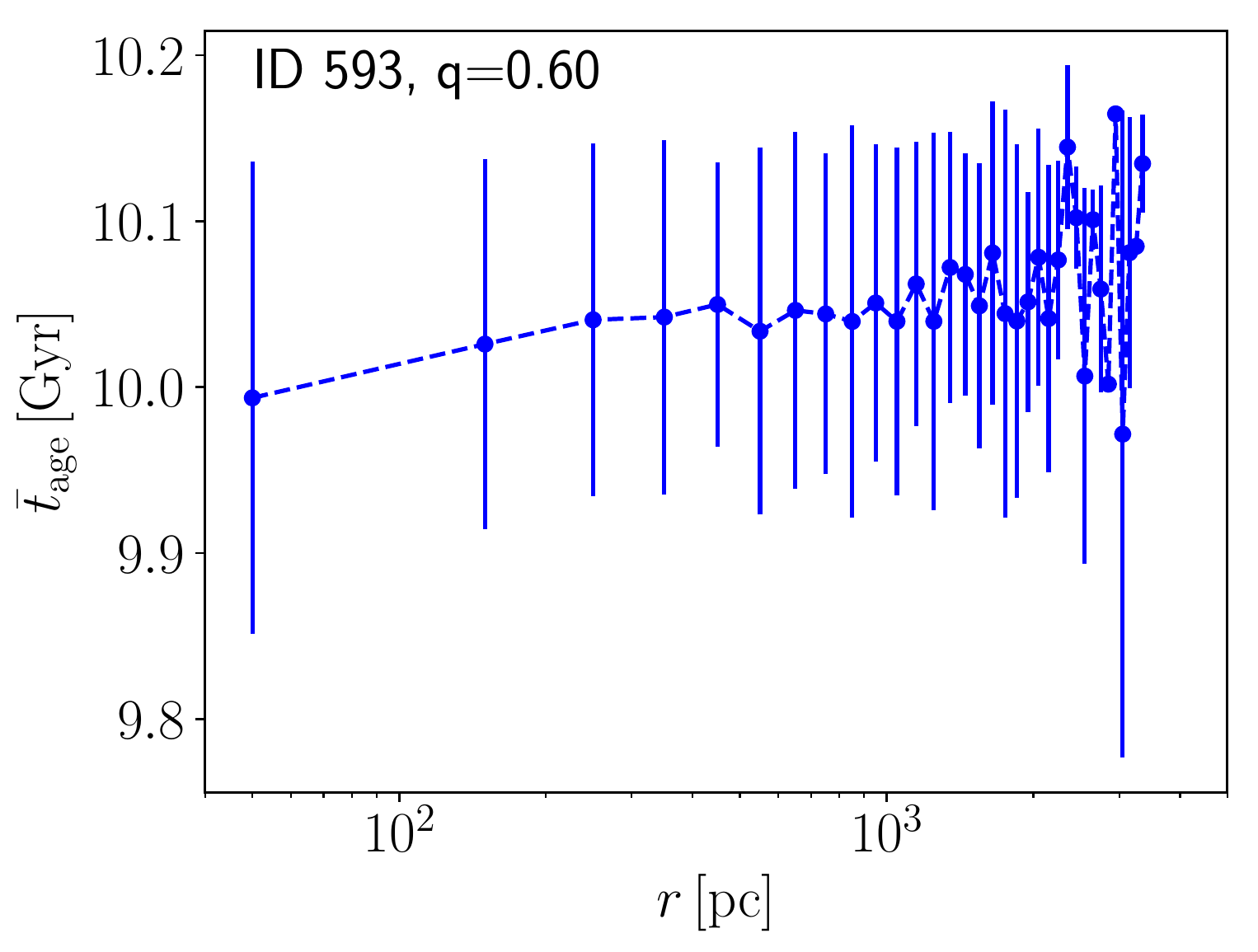}
		\includegraphics[width=91mm,trim={0.0cm 0.0cm 0.0cm 0.0cm},clip]{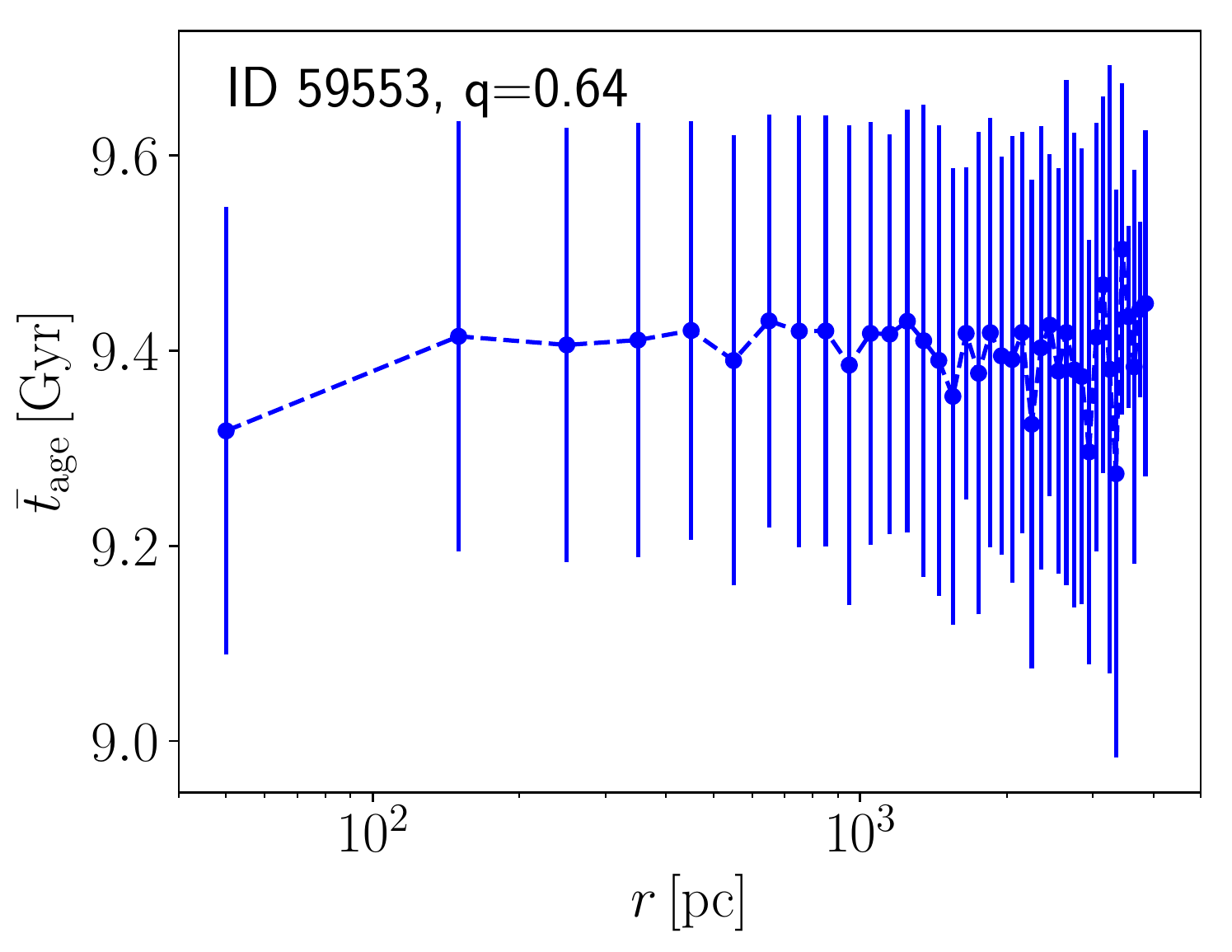}
		\caption{As in Fig.~\ref{fig:velocitydispersion_rotationcurve_time_distribution_1} but for gas-free TDGCs of sample A with the ID $593$ and ID $59553$. The radial density distributions of the here discussed TDGCs are shown in Fig.~\ref{fig:density_distribution}.}
		\label{fig:velocitydispersion_rotationcurve_time_distribution_2}
		
	\end{figure*}

	The internal kinematics of the above discussed gas-free TDGCs are analyzed in Figs. \ref{fig:velocitydispersion_rotationcurve_time_distribution_1} and  \ref{fig:velocitydispersion_rotationcurve_time_distribution_2} in which their rotation curves and the $3$D velocity dispersion (top panels) are plotted. By assuming that the mass distribution of these TDGCs is spherical we can calculate the circularity velocity by
	\begin{equation}
	\begin{aligned}
	v_{\mathrm{c}}(r) = \sqrt{\frac{G M_{\mathrm{tot}}(<r)}{r}} \, , \\
	\end{aligned}
	\label{eq:rotation_curve}
	\end{equation}
	where $G$ is the gravitational constant and $M_{\mathrm{tot}}(<r)$ is the total mass within the radius $r$. The rotation curves of TDGCs decline for larger radii because of the absence of dark matter particles. The $3$D velocity dispersion for the TDGCs with the ID $372$, ID $487$, and ID $593$ increases with increasing radii up until reaching a maximum value, which could indicate that their stellar mass grew during formation inside-out. In order to address this formation theory, we plot in Figs. \ref{fig:velocitydispersion_rotationcurve_time_distribution_1} and  \ref{fig:velocitydispersion_rotationcurve_time_distribution_2} (bottom panels) the mean and the standard deviation as error bars of the stellar particle age in dependence of the distance to the center of the TDGC. The age gradient is almost zero across the whole galaxy, which does not confirm an inside-out formation scenario of these TDGCs. 
	
	\subsection{Mass-weighted $\sigma$-clipping (outlier-rejection)}
	Throughout this study the analysis relies on the Subfind algorithm, which is a position-space subhalo finder \citep{Springel_2001}. We apply a mass-weighted $\sigma$-clipping scheme in the $6$D phase-space as an outlier-rejection method in order to cross-check if the TDGCs identified by the Subfind algorithm are indeed gravitationally bound objects. For simplicity we only analyze TDGCs that  are gas-free according to the Subfind algorithm  (i.e., $86$ out of $97$ TDGCs of sample A), such that the $\sigma$-clipping procedure is only applied to the stellar particles within the Illustris-1 box.\footnote{The Illustris-1 simulation box includes $1820^{3}$ dark matter particles and $1820^{3}$ gas tracer particles (gas cells and stellar particles). Therefore operating on the gas cells and dark matter particles would be extremely storage consuming such that we focus here purely on the stellar particles.} 
	
	The $\sigma$-clipping scheme for a single gas-free TDGC which was originally identified by the Subfind algorithm works as follows: In the first step, we load the center and the stellar half-mass radius of the considered TDGC provided by the Subfind algorithm. Secondly, we extract all stellar particles within a sphere centered at the position of the considered TDGCs with a radius of $20$ times the stellar half-mass radius. Subsequently, the $\sigma$-clipping procedure starts by calculating the stellar center-of-mass position and velocity in the first iteration by
	\begin{equation}
	\begin{aligned}
	&\vec{r}_{\mathrm{com}} \equiv \frac{\sum_{i=1}^{N} m_{i} \vec{r}_{i}}{M} \, , \\
	&\vec{v}_{\mathrm{com}} \equiv \frac{\sum_{i=1}^{N} m_{i} \vec{v}_{i}}{M} \, , \\
	\end{aligned}
	\label{eq:sigma_clipping_center_of_mass}
	\end{equation}
	where $N$ is the number of all selected stellar particles within the sphere, $m_{i}$ is the mass, $\vec{r}_{i}$ is the position vector, $\vec{v}_{i}$ is the velocity vector of the i-th stellar particle, and $M$ is the mass of all selected stellar particles. Each of the selected stellar particles has to fulfill the following two conditions
	\begin{equation}
	\begin{aligned}
	&\vec{r}_{\mathrm{rel},i}^{2} \equiv (\vec{r}_{i}-\vec{r}_{\mathrm{com}})^{2} < \eta^{2} \sigma_{r}^{2} \, , \\
	&\vec{v}_{\mathrm{rel},i}^{2} \equiv (\vec{v}_{i}-\vec{v}_{\mathrm{com}})^{2} < \eta^{2} \sigma_{v}^{2} \, , \\
	\end{aligned}
	\label{eq:sigma_clipping_single_particle}
	\end{equation}
	where $\eta$ is the $\sigma$-threshold, and $\sigma_{\mathrm{r}}$ and $\sigma_{\mathrm{v}}$ are the standard deviations in position and velocity space given by
	\begin{equation}
	\begin{aligned}
	&\sigma_{\mathrm{r}}^{2} = \frac{\sum_{i}^{N} m_{i} (\vec{r}_{i} - \vec{r}_{\mathrm{com}})^{2} }{M} \, , \\
	&\sigma_{\mathrm{v}}^{2} = \frac{\sum_{i}^{N} m_{i} (\vec{v}_{i} - \vec{v}_{\mathrm{com}})^{2} }{M} \, , \\
	\end{aligned}
	\label{eq:sigma_clipping_convergence}
	\end{equation}
	whereby $\sigma_{\mathrm{r}}$ and $\sigma_{\mathrm{v}}$ are set to be zero in the first iteration. The stellar particles which do not fulfill the condition \ref{eq:sigma_clipping_single_particle} are rejected and the center-of-mass position and velocity are re-calculated for the accepted stellar particles (see Eq. \ref{eq:sigma_clipping_center_of_mass}, where $N$ and $M$ refer to only the accepted stellar particles). This procedure is iteratively repeated until the algorithm converges such that the following three conditions are fulfilled
	\begin{equation}
	\begin{aligned}
	& \frac{\lvert \vec{r} - \vec{r}_{\mathrm{old}} \rvert^{2} }{\sigma_{\mathrm{r}}^{2}} < \delta^{2} \, , \\
	& \frac{\lvert \vec{v} - \vec{v}_{\mathrm{old}} \rvert^{2} }{\sigma_{\mathrm{v}}^{2}} < \delta^{2} \, , \\
	&\bigg \lvert \bigg( \frac{\sigma_{\mathrm{v}}}{\sigma_{\mathrm{v}}^{\mathrm{old}}} \bigg)^{2} -1 \bigg \rvert < \delta^{2} \, ,
	\end{aligned}
	\label{eq:sigma_clipping_sigma}
	\end{equation}
	where $\delta$ is the convergence threshold and $\sigma_{\mathrm{v}}^{\mathrm{old}}$ is the velocity dispersion of the previous iteration. We chose for the following analysis a $\sigma$-threshold of $\eta = 2$ or $ \eta = 3$ and a convergence threshold of $ \delta = 0.01$.
	
	Finally, the virial ratio is calculated using the stellar particles accepted by the $\sigma$-clipping scheme. The kinetic and potential energy are given, respectively, by
	\begin{equation}
	\begin{aligned}
	&E_{\mathrm{kin}} = \frac{1}{2} \sum_{i=1}^{N} m_{i} (\vec{v}_{i} - \vec{v}_{\mathrm{com}})^{2} \, , \\
	\end{aligned}
	\label{eq:Ekin}
	\end{equation}
	
	\begin{equation}
	\begin{aligned}
	&E_{\mathrm{pot}} = - \sum_{i=1}^{N} \sum_{j = i+1}^{N} \frac{G m_{i} m_{j}}{r_{i,j}^{2}+\epsilon_{\mathrm{baryonic}}^{2}} \, , \\
	\end{aligned}
	\label{eq:Epot_Plummer}
	\end{equation}
	where $N$ is the number of all accepted stellar particles based on the $\sigma$-clipping scheme, $G$ is the gravitational constant, and $\epsilon_{\mathrm{baryonic}}$ is the softening length, for which we assume for simplicity a fixed value of $710 \, \rm{pc}$ corresponding to the resolution of gravitational dynamics in Illustris-1.
	
\begin{figure}
		\centering
		\includegraphics[width=\columnwidth,trim={0cm 0.0cm 0 0.0cm},clip]{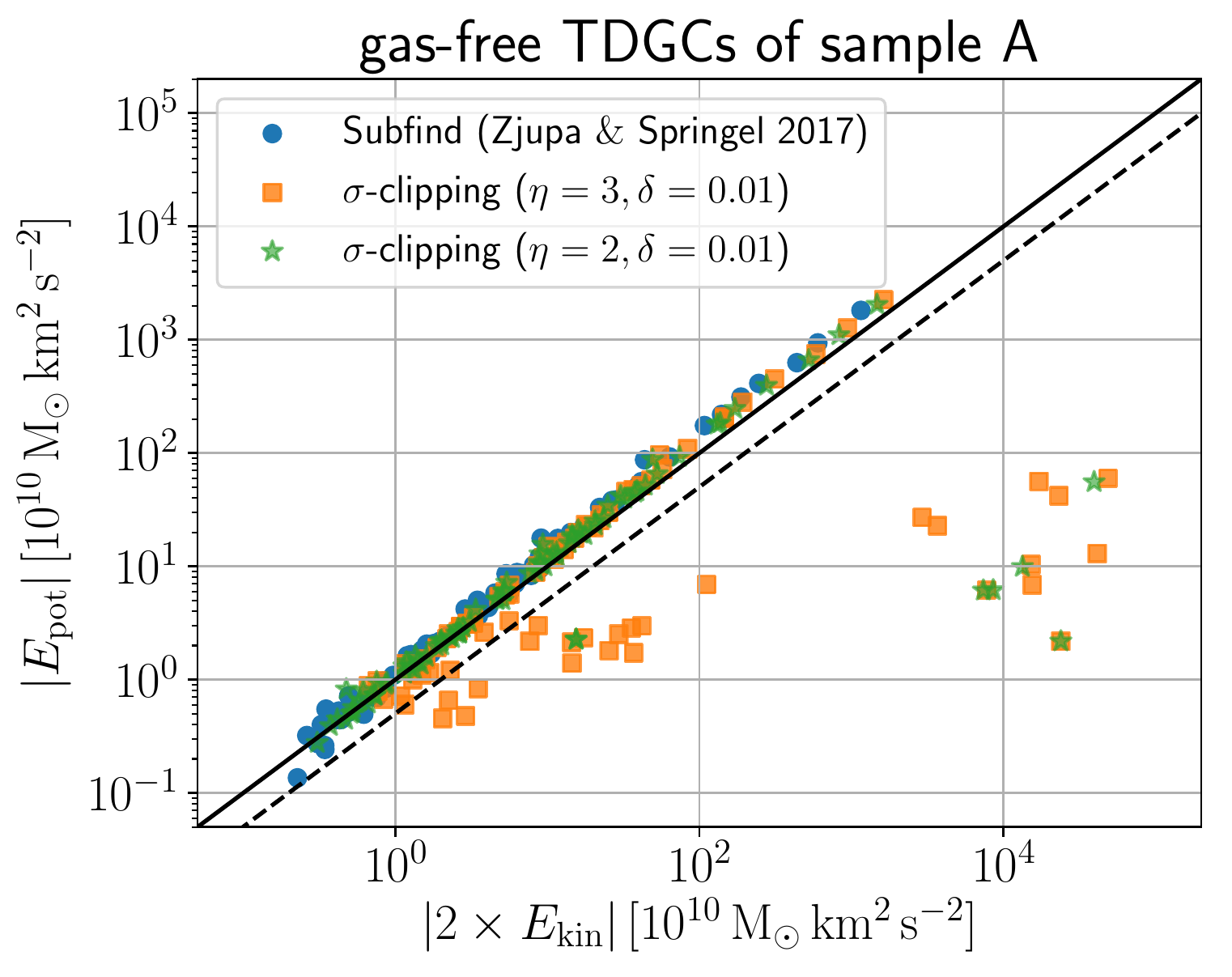}
		\caption{Relation between the potential and kinetic energy for gas-free TDGCs of sample A calculated by the Subfind and $\sigma$-clipping ($\eta =2$, $3$; $ \delta = 0.01$) algorithms. The black solid line highlights the condition for virial equilibrium, i.e., where the virial ratio becomes $q=1$ (see Eq. \ref{eq:virial_equilbrium}). All objects above the black dashed line are gravitationally bound ($q<2$) . The kinetic and potential energy for the Subfind algorithm are taken form \citet{Zjupa_2017} and the calculation for the potential energy using particles identified by $\sigma$-clipping assumes a fixed softening length in Eq. \ref{eq:Epot_Plummer}.}
		\label{fig:virial_equilbrium_SUBFIND_SIGMA}
	\end{figure}

\begin{figure*}
		\centering
		\includegraphics[width=72mm,trim={0.0cm 0.0cm 0.0cm 0.0cm},clip]{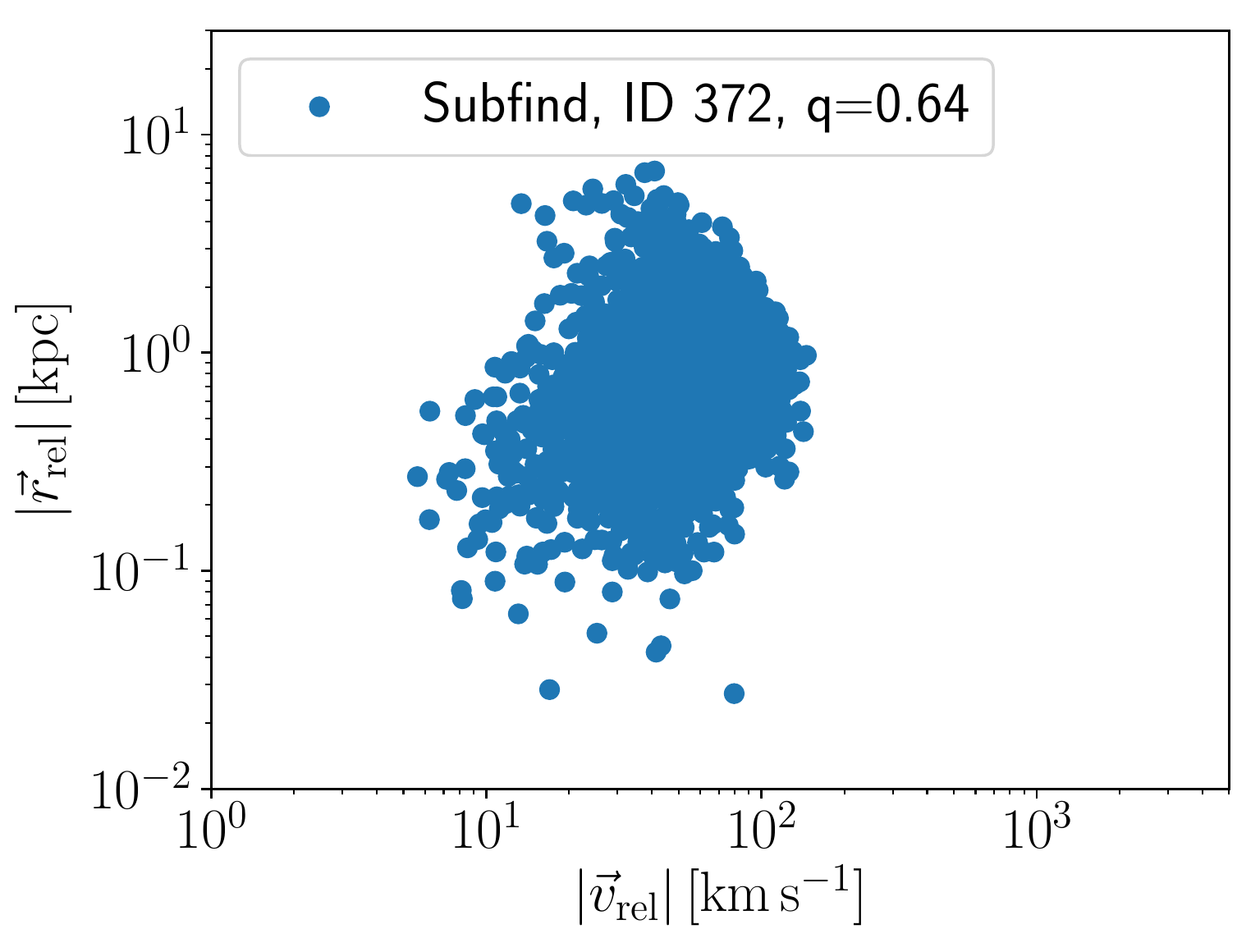}
		\includegraphics[width=72mm,trim={0.0cm 0.0cm 0.0cm 0.0cm},clip]{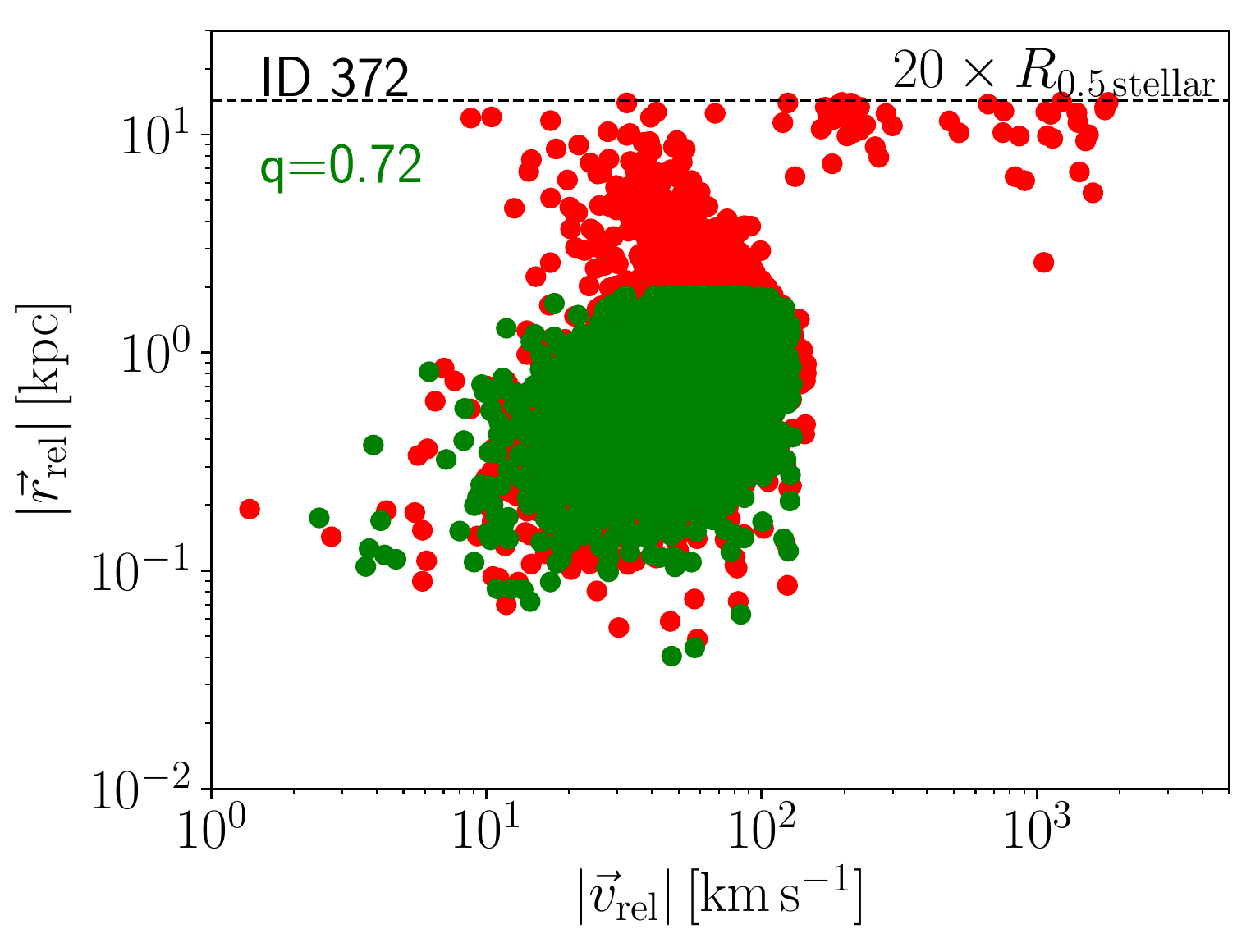}
		
		\includegraphics[width=72mm,trim={0.0cm 0.0cm 0.0cm 0.0cm},clip]{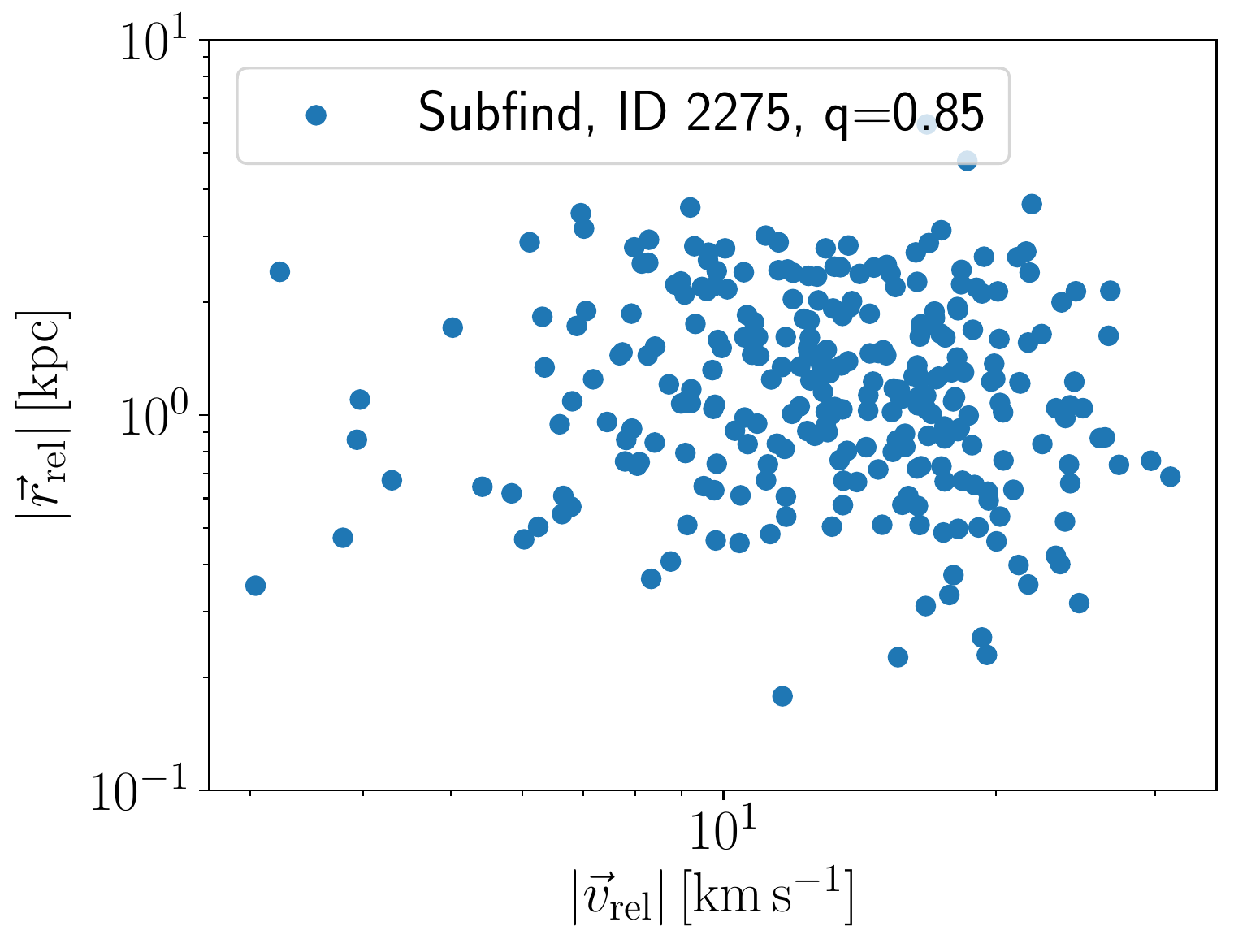}
		\includegraphics[width=72mm,trim={0.0cm 0.0cm 0.0cm 0.0cm},clip]{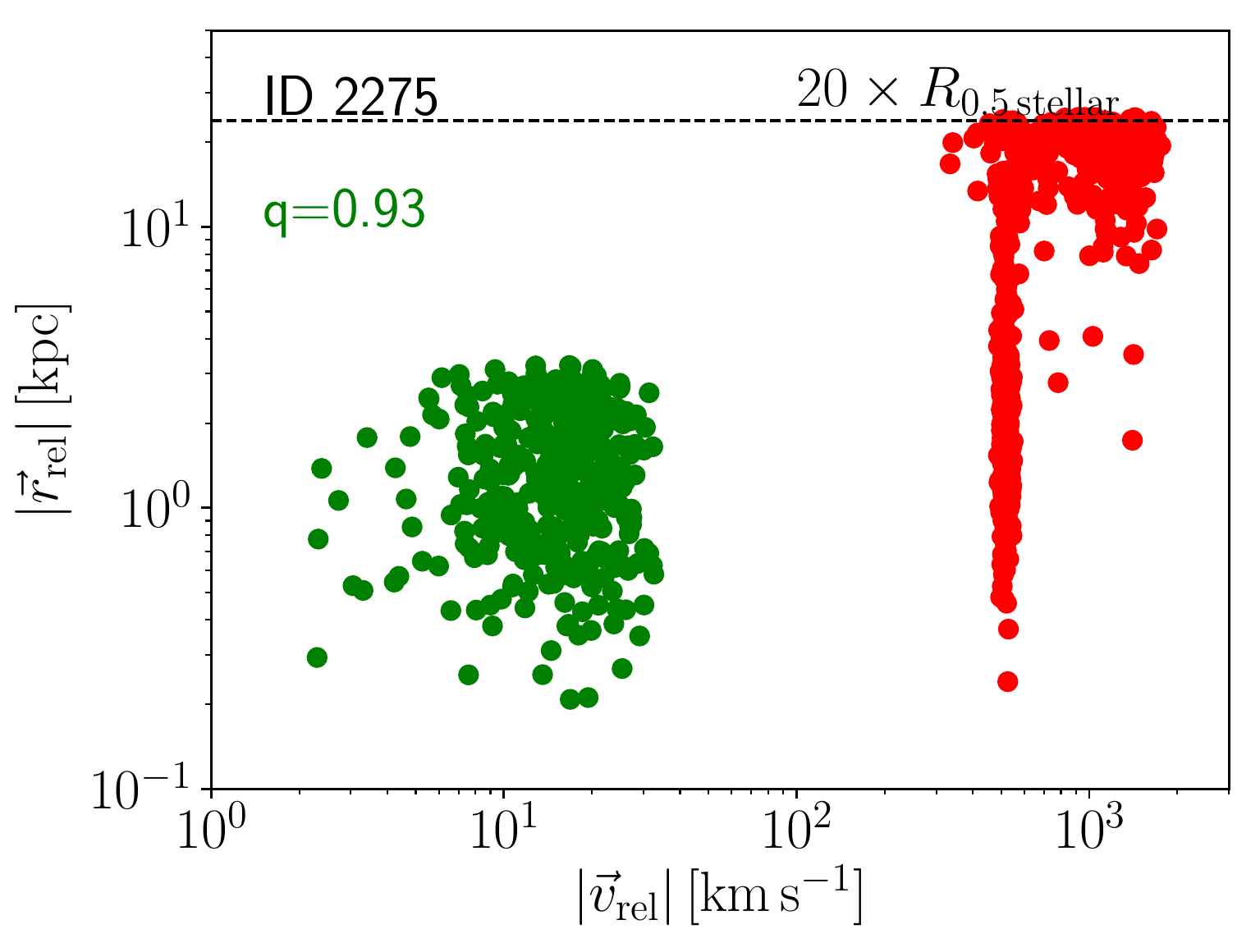}
		
		\includegraphics[width=72mm,trim={0.0cm 0.0cm 0.0cm 0.0cm},clip]{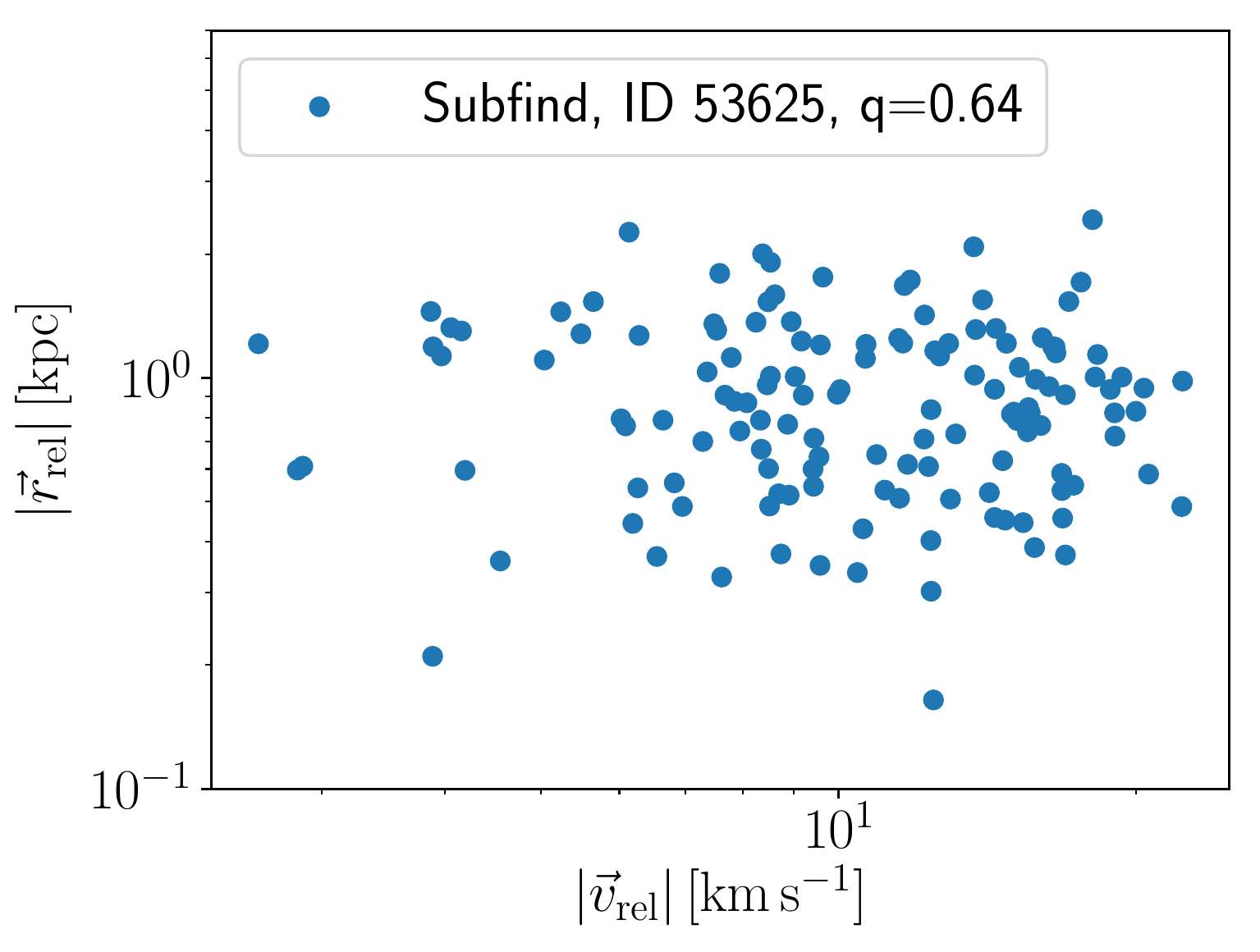}
		\includegraphics[width=72mm,trim={0.0cm 0.0cm 0.0cm 0.0cm},clip]{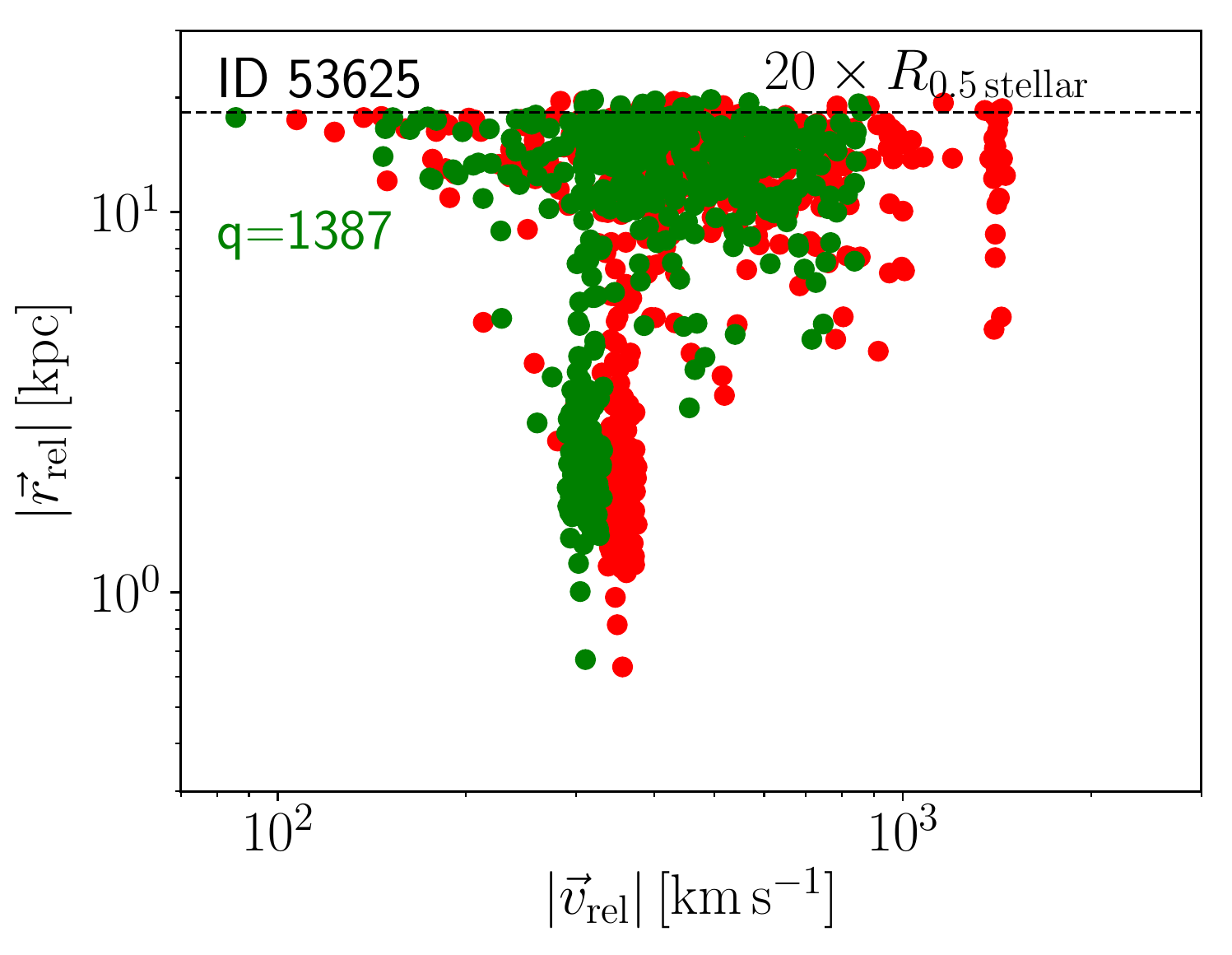}
		
		\includegraphics[width=72mm,trim={0.0cm 0.0cm 0.0cm 0.0cm},clip]{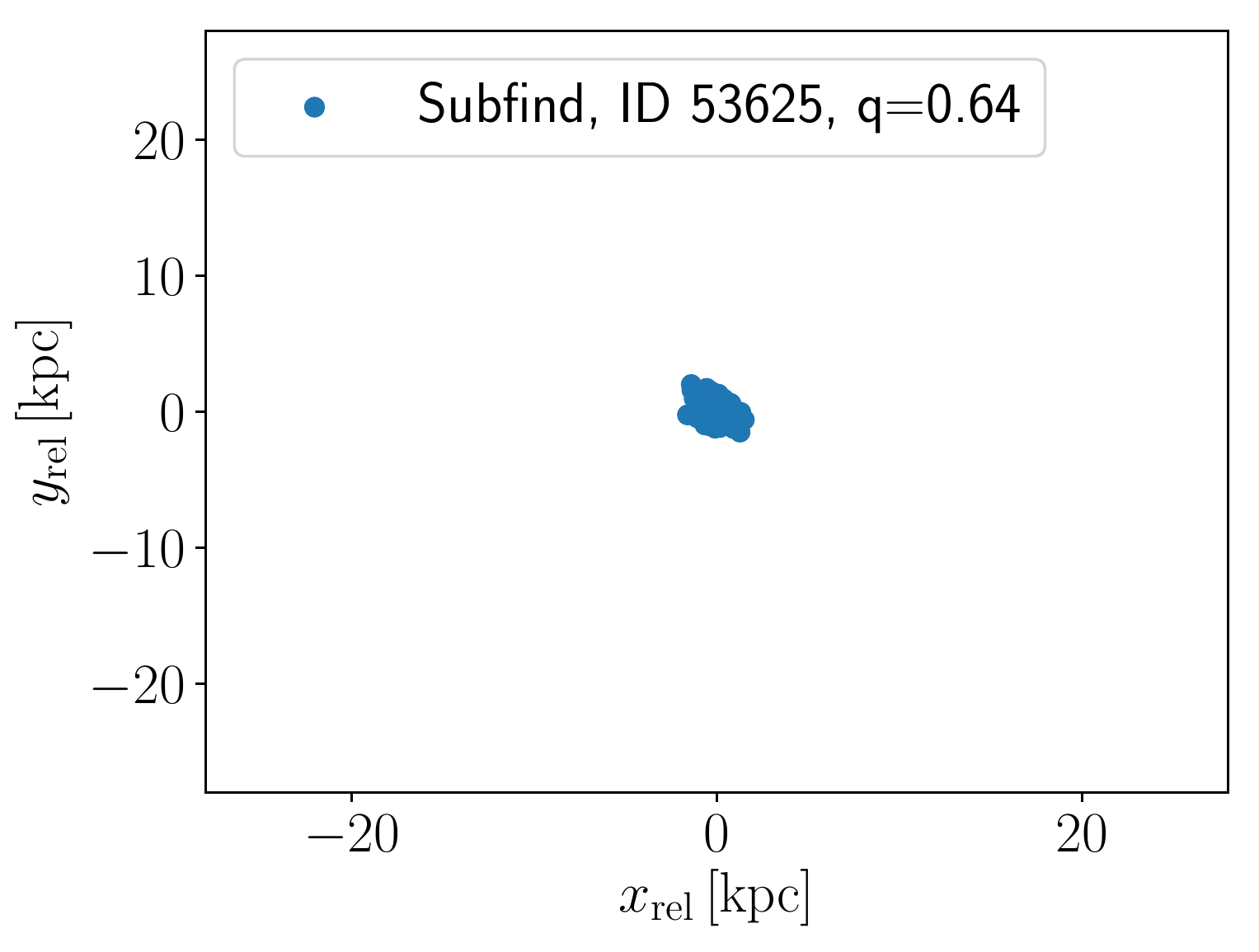}
		\includegraphics[width=72mm,trim={0.0cm 0.0cm 0.0cm 0.0cm},clip]{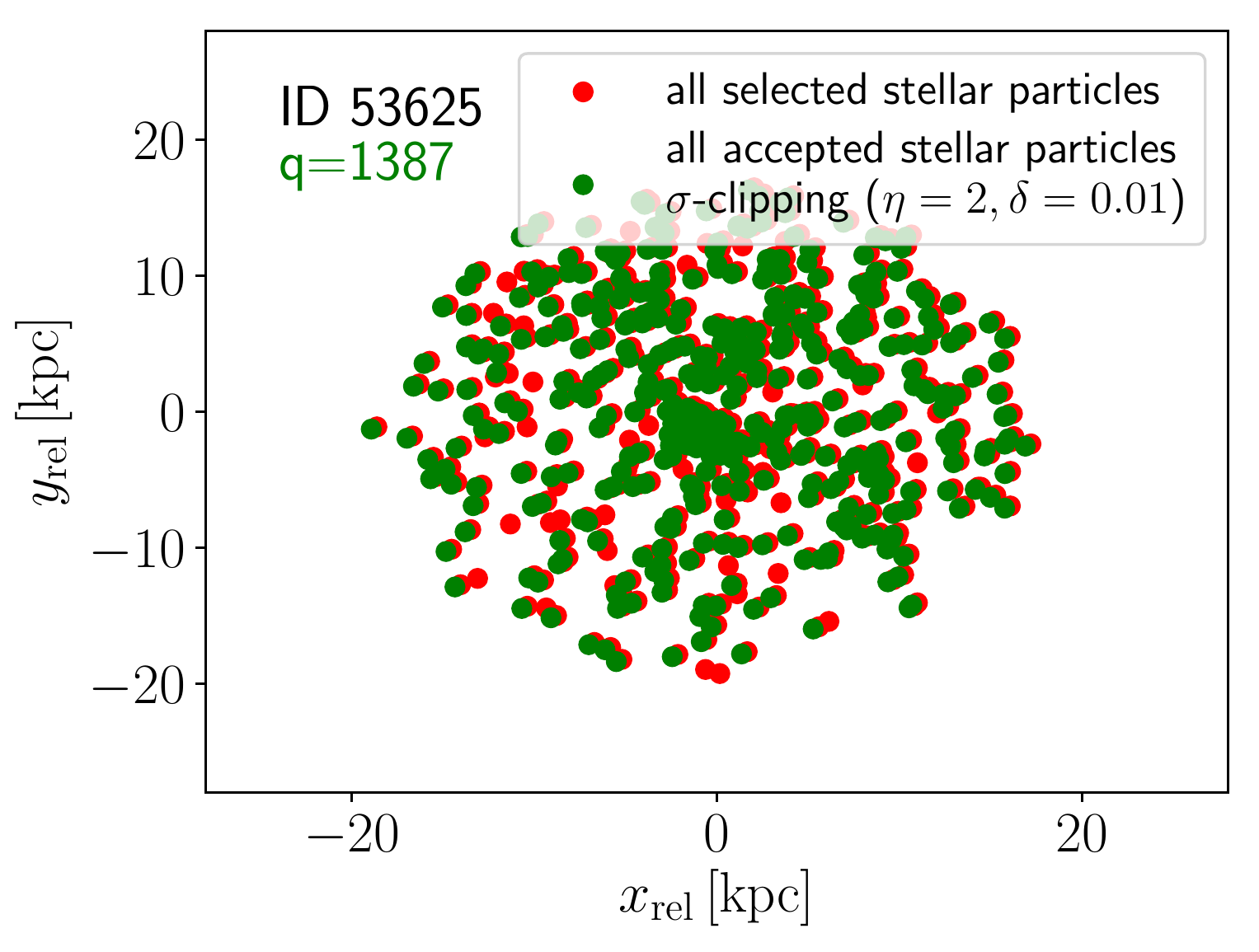}
		
		\caption{Top three panels: Modulus of the position and velocity vectors with respect to the center-of-mass for stellar particles (see Eq. \ref{eq:sigma_clipping_single_particle}). Blue dots are stellar particles identified by the Subfind algorithm (left panels), red dots are all selected stellar particles within a sphere with radius $20$ times the stellar half-mass radius (highlighted by the dashed solid black line; right panels) and centered around the considered Subfind TDGCs, and green dots are all stellar particles which become accepted by the $\sigma$-clipping algorithm with $\eta = 2$ and $\delta = 0.01$ (right panels). \newline 
			Bottom: Projected stellar particle distribution with positions relative to the center-of-mass for ID $53625$. For the $\sigma$-clipping, the center-of-mass position and velocity change in each iteration and also for the final accepted stellar particles (see text).}
		\label{fig:outliers_2sigma}
	\end{figure*}

\begin{figure}
		\centering
		\includegraphics[width=\linewidth]{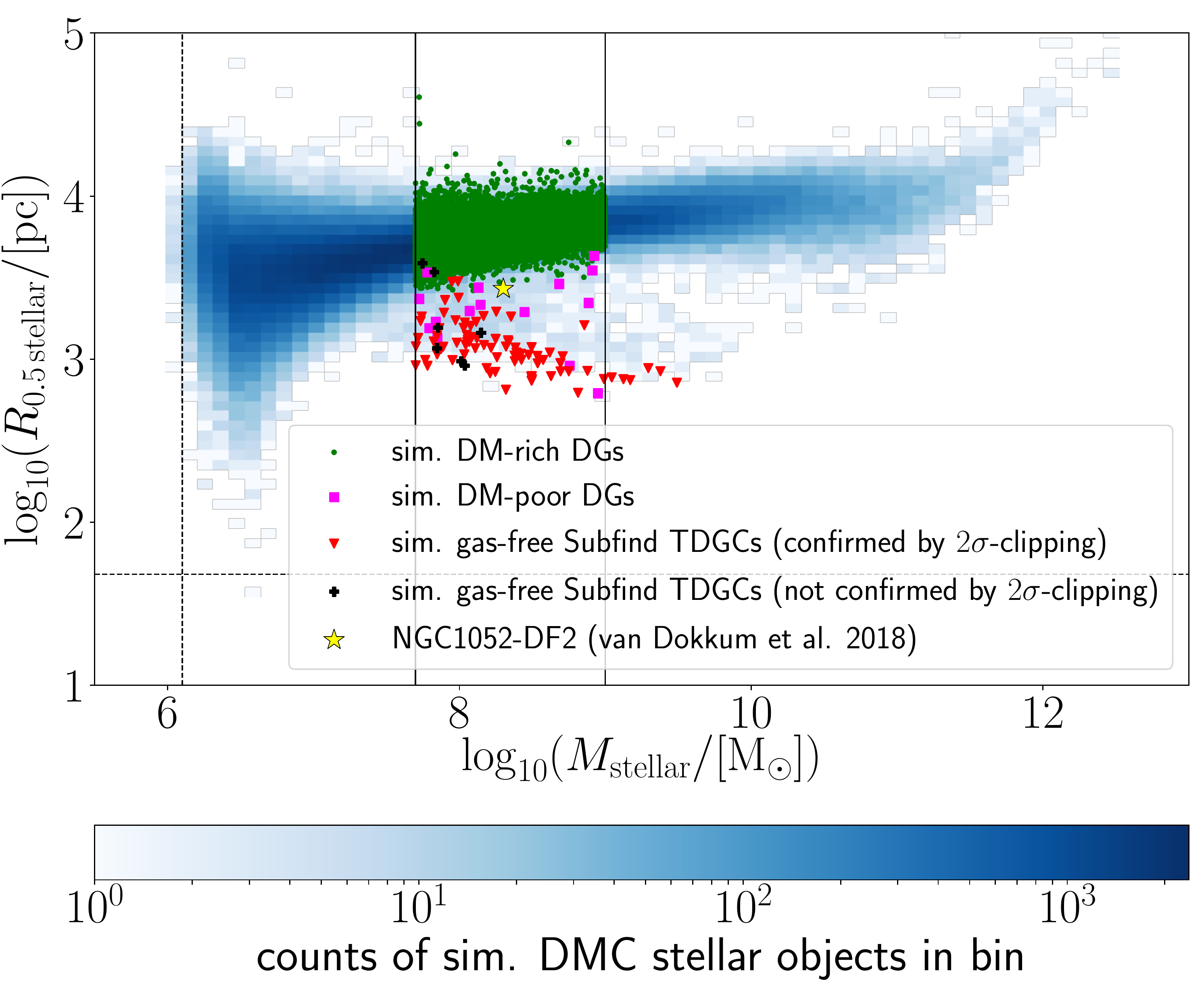}
		\caption{As in Fig.~\ref{S_massinhalfrad_stellar_vs_S_halfmassrad_stellar_DMCDGs_TDGCs} (top) but for gas-free TDGCs analyzed by a $\sigma$-clipping scheme ($\eta = 2$, $\delta=0.01$). Red triangles are gas-free TDGCs confirmed by $2\sigma$-clipping and black crosses are gas-free TDGCs which are according to $2\sigma$-clipping not gravitationally bound objects.}
		\label{fig:S_massinhalfrad_stellar_vs_S_halfmassrad_stellar_DMCDGs_TDGCs_SIGMA}
	\end{figure}

	Figure~\ref{fig:virial_equilbrium_SUBFIND_SIGMA} and Table~\ref{tab:virial_equilibrium_SUBFIND_SIGMA} compare the virial ratio of gas-free TDGCs of sample A calculated by the Subfind and $\sigma$-clipping algorithms. As expected, the percentage of gravitationally bound objects depends on the $\sigma$-threshold. Choosing $\eta = 2$ in the $\sigma$-clipping scheme we find that $92$~percent of our gas-free TDGCs of sample A identified by the Subfind algorithm fulfill $\vert E_{\mathrm{pot}} \rvert > \vert E_{\mathrm{kin}} \rvert$ and are thus also gravitationally bound when they become selected by a $6$D phase-space halo finder. Objects that are not gravitationally bound have $q > 2$. In Fig.~\ref{fig:outliers_2sigma} we illustrate for three TDGCs the $\sigma$-clipping scheme by plotting selected and accepted stellar particles in a position--velocity diagram and compare them with the particle distribution obtained by the Subfind algorithm. The TDGCs ID $372$ (top panels) and ID $2275$ (panels in the second row) are examples that are also gravitationally bound objects according to a $\sigma$-clipping scheme with $\eta = 2$. Despite the unusual appearance in the right panel of ID $2275$ the velocity--radius distribution of accepted particles is rather similar to the distribution of the Subfind particles. The $\sigma$-clipping algorithm calculates the center-of-mass position and velocity by using the accepted stellar particles resulting in a shift in the velocity-position diagram (see ID $2275$ in the left panel in the second row of Fig.~\ref{fig:outliers_2sigma}). The TDGC ID $53625$ is an example which is gravitationally bound according to the Subfind algorithm ($q = 0.78$), but cannot be identified as a gravitationally bound object by using the $\sigma$-clipping scheme as a $6$D phase-space halo finder, that is, the virial ratio calculated by all accepted stellar particles is $q = 1387$ and thus $\lvert E_{\mathrm{kin}} \rvert \gg \lvert E_{\mathrm{pot}} \rvert$.
	The Subfind TDGCs which are not gravitationally bound according to the $\sigma$-clipping scheme typically have lower stellar masses in the range of $5.6 \times 10^{7} - 1.4 \times 10^{8} \, \rm{M_{\odot}}$ and their positions in the radius--mass diagram are highlighted in Fig.~\ref{fig:S_massinhalfrad_stellar_vs_S_halfmassrad_stellar_DMCDGs_TDGCs_SIGMA}. 
	
\begin{figure*}
		\centering
		\includegraphics[width=\linewidth,trim={1.5cm 1.0cm 3.0cm 2.0cm},clip]{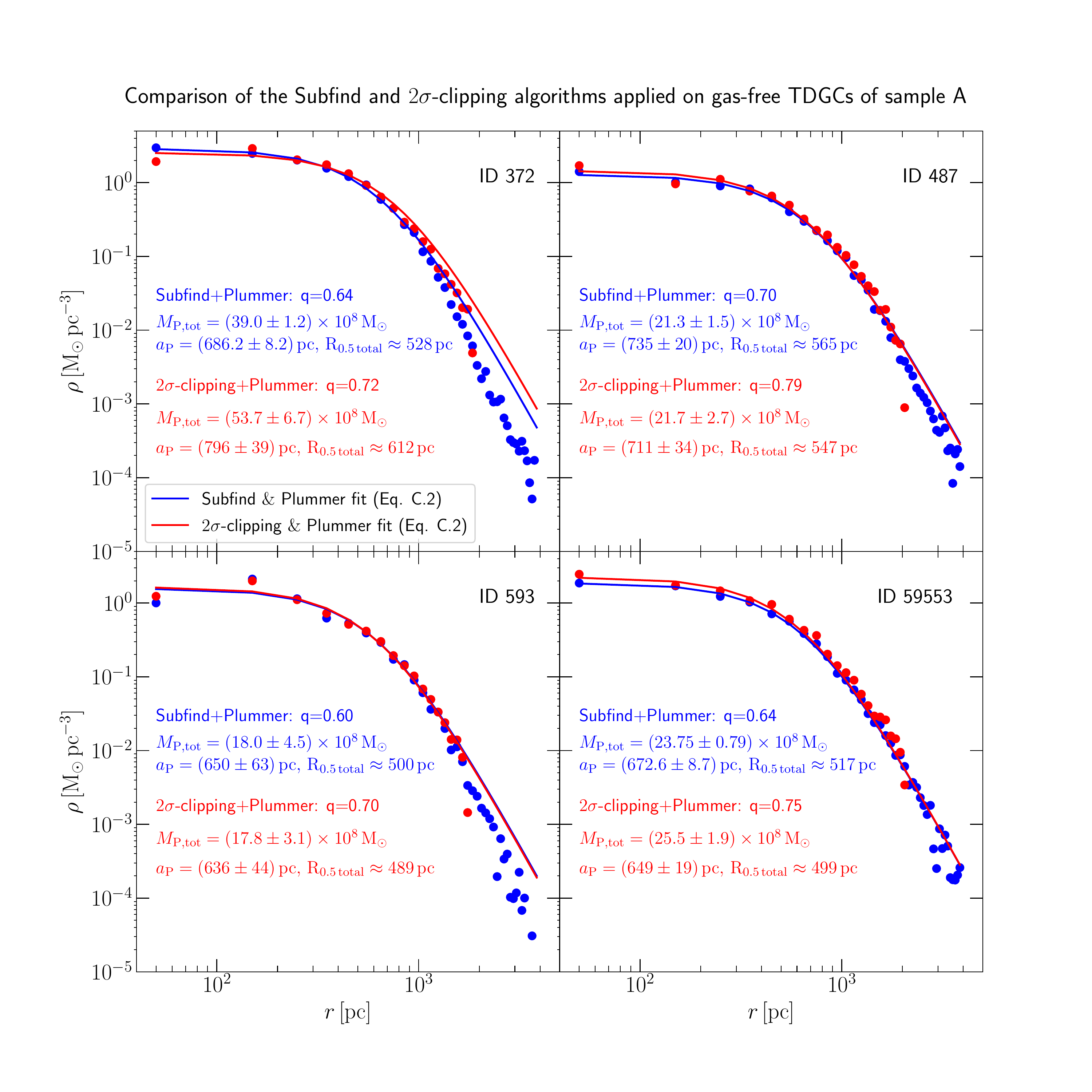}
		\caption{Radial density distributions of four selected gas-free TDGCs of sample A (IDs $372$, $487$, $593$, and $59553$) calculated by the particles identified by the Subfind (blue) and $\sigma$-clipping ($ \eta = 2$ and $ \delta = 0.01$; red) algorithms and fitted with a Plummer model (eq. \ref{eq:fit_plummer}). The radial bins have a width of $100 \, \rm{pc}$. The identification number, ID, their corresponding virial ratios, $q$ (Eq. \ref{eq:virial_equilbrium}), and their fitting parameters are given in the panels. The kinetic and potential energy for the Subfind algorithm are taken form \citet{Zjupa_2017} and the calculation for the potential energy using particles identified by $\sigma$-clipping assumes a fixed softening length in Eq. \ref{eq:Epot_Plummer}. We note that $q$ is computed solely for the particles selected using Subfind or $2\sigma$-clipping.}
		\label{fig:density_SUBFIND_3SIGMA_plots}
	\end{figure*}

	We also compare the Subfind algorithm and $\sigma$-clipping scheme with a $\sigma$-threshold of $\eta = 2$ for the density distribution of the TDGCs discussed in Fig.~\ref{fig:density_distribution}. In 
	Fig.~\ref{fig:density_SUBFIND_3SIGMA_plots} the density distributions of the TDGCs are fitted with a Plummer model (see Eq. \ref{eq:fit_plummer}) and the fitting parameters are qualitatively compared in the figure panels. The two different halo finders (i.e., $3$D position-space and $6$D phase-space algorithms) give these TDGCs approximately the same density distributions as those pointed out by the fitting parameters of the Plummer model.
	
	Summarizing, a small number of TDGCs identified with the Subfind algorithm are not confirmed by a $\sigma$-clipping scheme as gravitationally self-bound objects. These TDGCs might either not be real or cannot be identified as bound objects given the limitations of the $\sigma$-clipping scheme. Therefore, the frequency of Subfind TDGCs is $\approx 10$~percent too high and this does not affect our results significantly.
	
	\begin{table*}
		\centering
		\caption{Virial ratio, $q$ (Eq. \ref{eq:virial_equilbrium}), for gas-free TDGCs of sample A calculated by the stellar particles identified by the Subfind and $\sigma$-clipping algorithms.}
		\label{tab:virial_equilibrium_SUBFIND_SIGMA}
		\begin{tabular}{llllllll} \hline
			Algorithm & $\#$ & $\#$ of $\lvert E_{\mathrm{pot}} \rvert > \lvert E_{\mathrm{kin}} \rvert$ & Mean & Median & Std. & $16^{\mathrm{th}}-84^{\mathrm{th}}$ percentile \\ \hline \hline
			Subfind \citep{Zjupa_2017} & $86$ & $86 \, (100 \, \mathrm{percent})$ & $0.85$ & $0.83$ & $0.19$ & $0.69-0.97$ \\ \hline 
			$2\sigma$-clipping (+fixed softening length) & $86$ & $79 \, (92 \, \mathrm{percent})$ & $0.88$ & $0.88$ & $0.12$ & $0.76-1.0$ \\
			$3 \sigma$-clipping (+fixed softening length) & $86$ & $61 \, (71 \, \mathrm{percent})$ & $0.95$ & $0.87$ & $0.28$ & $0.76-1.0$ \\ \hline 
		\end{tabular}
		\tablefoot{Listed are the number of gas-free TDGCs of sample A, the number of gas-free TDGCs which fulfill $\lvert E_{\mathrm{pot}} \rvert > \lvert E_{\mathrm{kin}} \rvert$, the mean, the median, the standard deviation (std.), and the $16^{\mathrm{th}}-84^{\mathrm{th}}$ percentile of the virial ratio, $q$, for gas-free TDGCs with $\lvert E_{\mathrm{pot}} \rvert > \lvert E_{\mathrm{kin}} \rvert$, i.e., with $q<2$.}
	\end{table*}
	
\end{document}